\documentclass[a4paper,11pt]{article}
\usepackage{jheppub}
\usepackage{needspace}
\usepackage{bm,booktabs,array,longtable,placeins,capt-of,xcolor}
\usepackage{fontspec}
\hypersetup{citecolor=black,linkcolor=black,urlcolor=black,anchorcolor=black,filecolor=black,menucolor=black,pdftitle={Electroweak Incarnations of Instantons at Inert Higgs Doublet Interfaces},pdfauthor={Muneto Nitta}}
\newcommand{\dd}{\mathrm d}\newcommand{\Tr}{\operatorname{tr}}
\newcommand{\bmn}{\boldsymbol n}\newcommand{\bmN}{\boldsymbol N}

\numberwithin{equation}{section}\allowdisplaybreaks[2]
\title{Electroweak Incarnations of Instantons at Inert Higgs Doublet Interfaces}
\author[a,b]{Muneto Nitta}
\affiliation[a]{Department of Physics \& Research and Education Center for Natural Sciences, Keio University,\\4-1-1 Hiyoshi, Yokohama, Kanagawa 223-8521, Japan}
\affiliation[b]{International Institute for Sustainability with Knotted Chiral Meta Matter (WPI-SKCM$^2$), Hiroshima University, 1-3-1 Kagamiyama, Higashi-Hiroshima, Hiroshima 739-8531, Japan}
\emailAdd{mune.nitta@gmail.com}
\abstract{The inert doublet model is a two-Higgs-doublet extension of the Standard Model. Its scalar potential can support both an inert vacuum and an inert-like vacuum, distinguished by which doublet condenses. Using the degenerate limit of these vacua as a reference, we study how electroweak instantons are realized on their interface and on the defects it hosts. An explicit interpolation of the non-Abelian gauge connection relates the interface Skyrmion number to the bulk Yang--Mills instanton charge. At vanishing hypercharge coupling, a full turn of the internal phase of a confined or deconfined Nambu monopole provides a unit-charge Yang--Mills instanton configuration. When a Nambu monopole joins two half-unit $Z$ strings, this instanton appears as a baby Skyrmion on their Euclidean world sheet. On superconducting electroweak strings, a unit-charge Yang--Mills instanton can be realized as a pair of complementary merons. In the mixed $Z$--$W$ string regime, their geometric charges and actions are unequal, yet the magnetic-flux contribution gives each meron half a unit of Yang--Mills instanton charge for the specified bulk interpolation. On a superconducting interface, we derive a double sine-Gordon theory for the neutral phase in the core of a local vortex. Degenerate core states at nonzero soft mixing support separate events with unequal fractional charges, whereas a higher-energy intermediate core binds the constituents into a two-peak unit-charge event. The two peaks persist when the vortex cross section is relaxed in the ungauged interface theory. We also compute a half-instanton profile in the local gauged interface theory and derive the geometric charge fractions of global-monopole constituents on loci with spherical interface vacua.}
\begin{document}
\maketitle
\raggedbottom

\FloatBarrier\section{Introduction}\label{sec:intro}\suppressfloats[t]
Instantons reveal how the topology of gauge fields enters quantum dynamics beyond perturbation theory. With the usual vacuum boundary conditions, their finite-action Euclidean configurations carry an integer Yang--Mills instanton charge and provide semiclassical contributions associated with tunneling between sectors of different winding~\cite{BPST}. Their importance extends to the determination of nonperturbative interactions and vacuum structure. Supersymmetric gauge theories make this role particularly explicit: instanton calculations determine nonperturbative superpotentials in calculable regimes of $\mathcal N=1$ theories~\cite{ADS}, constrain quantum modifications of the vacuum moduli space~\cite{SeibergVacua}, and supply the instanton expansion of the low-energy prepotential in $\mathcal N=2$ theories~\cite{SW,Nekrasov}. Understanding the configurations that carry instanton charge is therefore an essential part of understanding the nonperturbative dynamics of gauge theories.

The realization of instantons depends on the background in which they are embedded. In particular, topological solitons can support lower-dimensional effective theories whose instantons encode instanton effects of the parent theory, revealing connections between nonperturbative physics in different dimensions. A prominent example is provided by non-Abelian vortex strings in supersymmetric gauge theories~\cite{HananyTong2003,Auzzi2003,EtoModuli2006}. Yang--Mills instantons inside a vortex appear as sigma-model lumps on its Euclidean world sheet~\cite{higgsinstantons,HananyTong2004,ShifmanYung2004}, while confined magnetic monopoles are described as world-sheet kinks~\cite{TongMonopoles2003,NittaVinci2011}. This correspondence extends to nonperturbative quantum physics, including exact Bogomolny--Prasad--Sommerfield (BPS) spectra and partition functions~\cite{HananyTong2004,ShifmanYung2004,Dorey1998,DoreyHollowoodTong1999,DoreyHollowoodLee2011,ChenDorey2011,FujimoriCounting2012,FujimoriKimuraNittaOhashi2015}. More generally, solitons can support internal degrees of freedom whose nontrivial configurations carry bulk instanton charge. A domain wall provides a principal chiral effective theory in which the instanton appears as a Skyrmion~\cite{eto2005}; non-Abelian walls, localized modes, and Josephson junctions furnish related settings~\cite{earlywall,cloud,jwall,josephson}. Instantons also occur in intersecting vortex sheets and webs, whose geometry can be described using amoeba and tropical geometry~\cite{amoeba2008}. Further localization on monopole strings realizes instantons as sine-Gordon kinks in the effective theory of their internal phase~\cite{nittaMonopoleStrings2013,incarnations,relations2022}. These relations allow the same bulk topological charge to be studied through successive effective theories on its host solitons.

In the electroweak theory, instanton topology has a direct physical consequence. Anomalous nonperturbative processes violate baryon and lepton numbers, although these numbers are conserved in perturbation theory~\cite{tHooft}. The Higgs condensate changes the instanton size dynamics and motivates the constrained-instanton treatment of the broken phase~\cite{affleck1981}. The Higgs background is thus part of the nonperturbative problem. The Standard Model also admits $Z$-flux strings and their Nambu-monopole endpoints~\cite{nambu,vachaspati,review}, but its bulk vacuum does not topologically protect either defect. The familiar electroweak $Z$ string is also unstable at physical Standard Model parameters~\cite{james1992,james,goodband}. Its persistence is consequently an important issue for proposals involving electroweak strings in baryogenesis~\cite{baryogenesis} and primordial magnetic fields and helicity~\cite{magnetic,helicity}. Attempts to stabilize the strings have examined bound fields~\cite{vachaspatiwatkins}, external magnetic fields~\cite{garriga}, thermal effects~\cite{nagasawa}, additional scalar condensates~\cite{forgacs}, and fermionic contributions~\cite{naculich,etohamada2024}. These studies motivate examining how the Higgs environment can supply persistent electroweak defects. Modifications of the Higgs potential can also substantially extend the lifetime of electroweak oscillons~\cite{DrogoszOscillons2026}, illustrating the sensitivity of localized field configurations to the scalar sector.

Two-Higgs-doublet models provide a simple extension of that environment~\cite{2hdmreview}. Their additional relative phases and possible discrete symmetries enrich the vacuum topology~\cite{vacuumtopology}, allowing flux tubes~\cite{earlyvortex,twohiggsvortex} and fractional non-Abelian electroweak strings~\cite{fluxtube,ekn,eknstrings2018}. Symmetries can relate two fractional strings and balance their tensions at a Nambu-monopole junction~\cite{topmonopole,monodynamics,nambudyon}, while binding fractional strings into a molecule can stabilize a flux tube~\cite{molecule}. Domain walls provide another setting in which the relative Higgs fields become active: their profiles can involve CP-violating or charged condensates~\cite{membrane,cpsoliton,chargedwall}, and their solutions, evolution, and cosmological constraints have been investigated~\cite{wallsim,wallconstraint,collapse,wallclassification}. Superconducting strings and walls, wall-bound condensates, vortons, and global-monopole structures further illustrate the defect dynamics opened by the enlarged Higgs sector~\cite{superconducting2hdm,percolation,kinkyvortons,globalmonopoles,hopffibration}. The additional doublet therefore changes both the possible host solitons and the internal degrees of freedom available on them.

The inert doublet model offers a particularly simple framework in which to exploit this possibility. A discrete symmetry distinguishes the second doublet and forbids its Yukawa couplings~\cite{deshpande,barbieri}. The standard inert vacuum has an expectation value only for the first doublet and permits a dark-matter candidate~\cite{idmdm}. The additional scalar sector is tested by collider and precision-Higgs observables~\cite{aokicollider,idmprecision,idmlimits} and can modify the electroweak transition~\cite{idmewpt,gil,idmperturbativity}, providing a setting for the broader connection between extended Higgs sectors and electroweak baryogenesis~\cite{kanemuraewbg,aks,aksewbg}. For the present purpose, its essential feature is the possibility of a competing inert-like minimum, in which the second doublet condenses instead of the first. Vacuum coexistence and multistep thermal histories involving these minima have been studied extensively~\cite{ferreira,ginzburg,aokimultistep,aoki,fabian}, including intermediate charge-breaking phases~\cite{aokicharged}. Coexisting inert and inert-like regions admit an interface across which the identity of the condensing doublet changes. In contrast to conventional two-Higgs-doublet models, the topology and associated soliton structures of the inert doublet model had remained largely unexplored.

\begin{figure}[p]\centering
\begin{minipage}{\textwidth}
\begingroup
\centering\small
\begin{tabular}{cl>{\raggedright\arraybackslash}p{7.5cm}}
\toprule
Phase & Conditions & Vacuum set\\\midrule
A & generic $a<h$, $a-b<h$ & $u_0=1$, $u_3=c=0$; one point\\
B & $b<0$, $a-b>h$ & $(u_0,u_3,c)=\left(\frac{h}{a-b},\ \pm\sqrt{1-\frac{h^2}{(a-b)^2}},\ 0\right)$; 2 points\\
C & $b>0$, $a>h$ & $u_0=h/a$, $u_3=0$, $|c|^2=1-(h/a)^2$; charged $S^1$\\
D & $h=0$, $a<\min(b,0)$ & $u_0=\pm1$, $u_3=c=0$\\
E & $h=0$, $a=b<0$ & $c=0$, $u_0^2+u_3^2=1$; neutral $S^1$\\
F & $b=0$, $a>h$ & $u_0=h/a$, $u_3^2+|c|^2=1-(h/a)^2$; $S^2$\\
G & $h=a=0$, $b>0$ & $u_3=0$, $u_0^2+|c|^2=1$; $S^2$\\
H & $a=b=h=0$ & $u_0^2+u_3^2+|c|^2=1$; $S^3$\\\bottomrule
\end{tabular}
\captionof{table}{Interface vacuum sets. We use $h\geq0$. Special phase loci D--H are listed separately. The row for phase A excludes those special degeneracies. Critical boundaries require their own limiting analysis.}\label{tab:vacua}

\endgroup\vspace{3mm}

\centering\includegraphics[width=.89\textwidth]{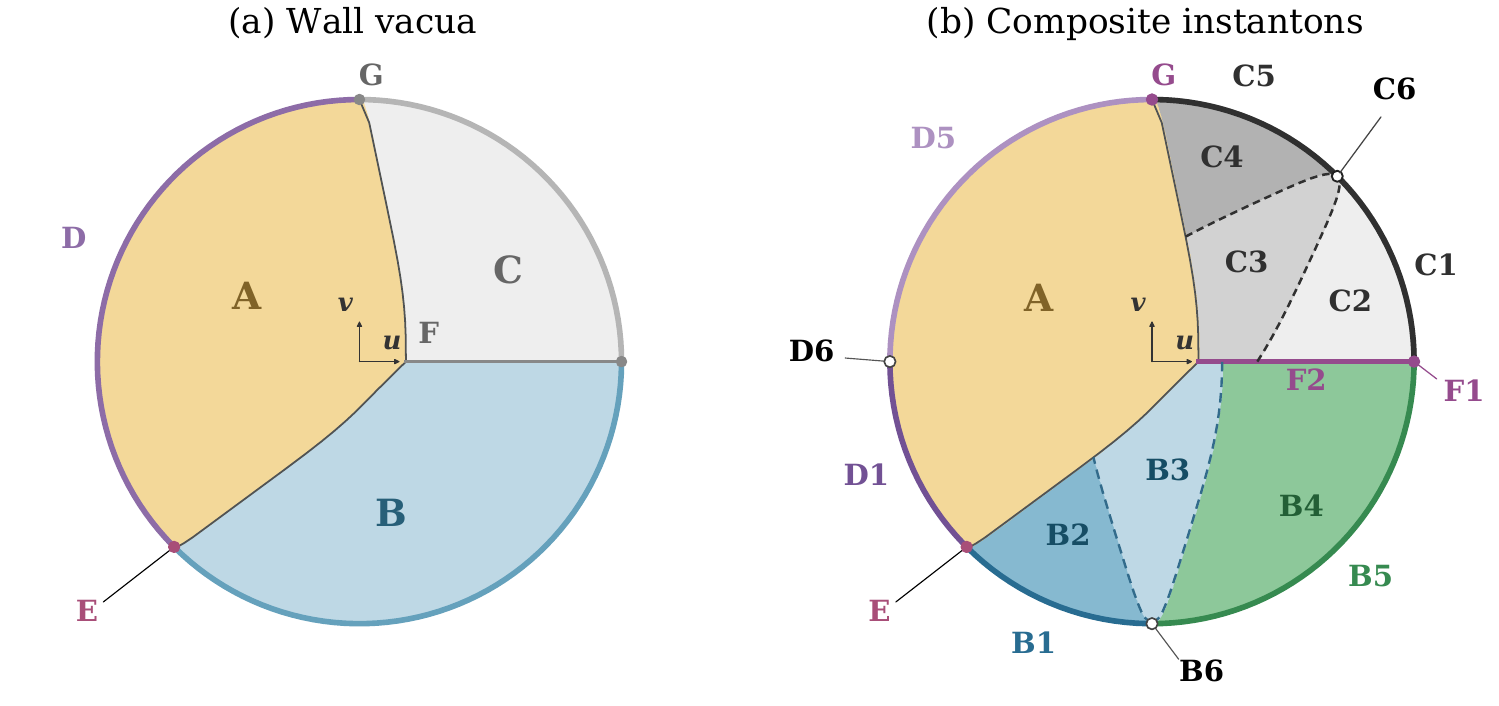}
\caption{\footnotesize Interface vacua and composite-instanton structures. (a) Vacuum classification of the interface theory. (b) Realizations of electroweak instantons in the corresponding phases. The radial coordinate is $\rho=\log[1/(1-\alpha r)]/\log[1/(1-\alpha)]$, where $r=\sqrt{a^2+b^2}/\sqrt{a^2+b^2+h^2}$, $\theta=\arg(a+ib)$, and $\alpha=1-10^{-3}$. The plotted coordinates are $u=\rho\cos\theta$ and $v=\rho\sin\theta$, with $u^2+v^2\leq1$ and $-1\leq u,v\leq1$. The rim corresponds to $h=0$; phase H denotes the separate point $a=b=h=0$. Ordinary interface Skyrmions are admitted in all phases (section~\ref{sub:Skyrme}); the diagram highlights additional composite realizations. In phases B1 and D1, a full winding of the internal phase of a confined Nambu monopole along Euclidean time forms a unit lump on the host string (section~\ref{sub:B1}). In phases B2 and B3, instantons appear as baby Skyrmions on strings (section~\ref{sub:B23}). In phases B4, B5, and D5, two complementary merons carry unit total instanton charge. Their geometric charge contributions are unequal, $q$ and $1-q$, in phase B4, and equal, $1/2$ and $1/2$, in phases B5 and D5 (section~\ref{sec:merons}). In phase B, blue denotes fractional $Z$-string regimes, while green denotes mixed $Z$--$W$ and pure $W$-string regimes, following Ref.~\cite{interface2026}. Phase E admits a complementary construction through phase winding in the charged core of a deconfined Nambu monopole (section~\ref{sub:E}). In phases C1--C5, instantons are kinks of the neutral vortex-core phase along Euclidean time. The leading double sine-Gordon (DSG) theory gives deconfined symmetric half instantons in phases C1 and C5, deconfined asymmetric fractional instantons in phase C2, single unit instantons in phase C3, and confined fractional-instanton molecules in phase C4 (sections~\ref{sec:DSG}--\ref{sec:Cclassification}). The dashed boundaries within phase C extrapolate the weak-deformation core theory. In phases F and G, the interface Skyrmion charge can be shared by two global-monopole constituents (section~\ref{sec:FG}). The open circles B6 and D6 have full $S^2$ orientational degeneracy. At C6 the neutral vortex-core phase is free on $S^1_{\rm core}$. The diagram classifies composite structures; full four-dimensional stability is not assessed.}\label{fig:phase}

\end{minipage}
\end{figure}

Our previous work showed that this interface supports locally stable electroweak $Z$ strings and topologically protected Nambu monopoles within a controlled effective theory~\cite{interface2026}. We used degenerate inert and inert-like vacua as a solvable reference. In the overlap region of the two Higgs condensates, their relative orientation becomes a normalizable field localized on the interface. Near a symmetric reference point, this field takes values in $SU(2)\simeq S^3$. Small scalar couplings generate an orientation potential, whose vacuum structure determines which defects the interface supports. The two bulk vacua and the vacua of this interface theory play different roles: the former define the host interface, whereas the latter classify its internal phases. Table~\ref{tab:vacua} lists the interface vacuum sets, and Figure~\ref{fig:phase}(a) displays their organization in parameter space. There are three generic phases. Phase A has a unique interface vacuum and, in part of its parameter region, supports locally stable unit-flux $Z$ strings represented by sine-Gordon solitons in the interface theory. Phase B has two disconnected interface vacua, and a topological kink interpolating between them represents a string localized on the interface. Phase C has a charged Higgs condensate and a circle of vacuum orientations before electromagnetic identification; it is a superconducting interface phase. The string branches within phase B are further distinguished by their internal orientation: they include fractional $Z$ strings in subphases B1--B3, mixed $Z$--$W$ strings in phase B4, and pure $W$ strings in phase B5. The mixed and pure $W$ branches have charged core condensates and are superconducting strings. Their phase degrees of freedom will be important for the instanton constituents studied below. The parameters and phase boundaries used in Table~\ref{tab:vacua} and Figure~\ref{fig:phase} are defined in Sec.~\ref{sec:setup}.

Special parameter loci give additional vacuum structures. Phase D has two interface vacua related to those of phase B by an exchange of orientation components; it includes the corresponding $Z$-string and pure $W$-string branches in phases D1 and D5, which meet at the orientationally degenerate point D6. Phase E has a neutral vacuum circle, phases F and G have vacuum spheres, and phase H has the full $S^3$ vacuum manifold. These loci connect configurations that otherwise have different localized descriptions. In particular, phases B1, D1, and E exhibit distinct realizations of Nambu monopoles. In phases B1 and D1, a Nambu monopole joins two symmetry-related, equal-tension half-unit $Z$ strings. The strings lie on opposite sides of the monopole, and its $Z$ flux is concentrated into them. A discrete symmetry exchanges the string branches and enforces the tension balance. On the string, the monopole is a second kink, now in the string orientation theory. In phase E, the monopole instead becomes a global vortex of the neutral interface vacuum circle. It has no attached $Z$ strings: its $Z$ flux spreads radially through the interface, while the monopole itself remains bound to the interface. As phase E is approached, the fractional strings broaden and their tension vanishes, so the monopole description passes from a kink on a string to a vortex in the interface. Across this confinement transition, the outward electromagnetic magnetic charge remains unchanged, and the electromagnetic field extends outside the interface in both regimes~\cite{interface2026}. These interface defects provide a setting in which to investigate how electroweak instanton charge is realized through their localized degrees of freedom.

In this paper we establish a common description of these configurations using the interface relative-orientation field. An explicit gauge interpolation relates its Euclidean Skyrmion number to the bulk Yang--Mills instanton charge. In phase B, we follow this charge into the string and monopole effective theories. A full turn of the monopole phase gives a unit lump on the string, while other orientation potentials give baby Skyrmions or meron constituents. We derive the string kinetic and baby Skyrme coefficients from the interface action. For the mixed $Z$--$W$ and pure $W$ branches, we identify the two meron constituents that together carry unit instanton charge. We study symmetric and asymmetric constituents in a local gauged effective theory, including non-BPS couplings, and distinguish their geometric charges from the magnetic-flux contributions to the gauge-completed charge. The right panel of Figure~\ref{fig:phase} summarizes the instanton structures investigated here, extending the vacuum and defect classification of Ref.~\cite{interface2026}.

In phase C, the superconducting interface provides a complementary realization. A spatial vortex winds the charged condensate, while a neutral condensate fills its core. The neutral core phase is normalizable and generally acquires a double sine-Gordon potential from the scalar couplings. Kinks of the neutral core phase along the Euclidean time direction represent symmetric half instantons, asymmetric fractional instantons, single unit instantons, and confined fractional-instanton molecules. We show why degenerate core states permit deconfined fractional events even at nonzero soft mixing, whereas a higher-energy intermediate core confines the constituents into a molecule. We derive the core coefficients and compare the reduced descriptions with configurations in which the vortex cross section also relaxes, including a local planar gauge model.

The special vacuum manifolds provide further connections. The deconfined monopole in phase E gives a complementary vortex-core construction in which the neutral and charged sectors exchange roles relative to phase C. In phases F and G, the interface Skyrmion admits a decomposition into fractional constituents carrying global-monopole structure~\cite{fractional2015} in the Euclidean interface theory. Throughout, we distinguish the charge correspondence from the dynamical localization of a representative. Both localized reduced-theory solutions and configurations with spreading directions are included, with their boundary conditions and effective descriptions stated explicitly. Thus, Yang--Mills instantons admit a variety of composite-soliton realizations at the interface, appearing as solitons of different dimensions in the effective theories on their hosts. These solitonic incarnations provide a unified picture of how instanton charge is localized, fractionalized, and confined.

This paper is organized as follows. Section~\ref{sec:setup} introduces the model, interface theory, phase diagram, and common charge relation. Section~\ref{sec:BD} treats strings and Nambu monopoles in the B, D, and E phases. Section~\ref{sec:C} derives and tests the vortex-core instantons of the C phase. Section~\ref{sec:FG} discusses fractional instanton molecules in the F and G phases. We conclude in Sec.~\ref{sec:discussion}. Appendix~\ref{app:coefficients} derives the coefficients of the electroweak string effective theory and computes the four-derivative correction from relaxation of its transverse width. Appendix~\ref{app:rotor} compares free-phase winding on a Euclidean time circle and on the line for Nambu monopoles and C6 vortices, and gives a constrained solution of prescribed width. Appendix~\ref{app:charge} derives the magnetic-flux contribution to the charge of a gauged meron and shows how it cancels the latitude dependence of the geometric charge in phase B4 for the specified bulk connection. Appendix~\ref{app:numerics} gives the finite-soft-mixing charge formula and summarizes the numerical methods and convergence checks.

\FloatBarrier\section{Inert Higgs interfaces and electroweak instantons}\label{sec:setup}\suppressfloats[t]
This section establishes the common interface description used throughout the paper. Section~\ref{sub:model} introduces the scalar model and reference interface, and section~\ref{sub:action} derives its orientation action and vacuum phases. Section~\ref{sec:charge} relates the Euclidean interface winding to the bulk Yang--Mills instanton charge. Section~\ref{sub:Skyrme} derives the four-derivative term, describes interface Skyrmions, and introduces the hierarchy of nested solitons.

\subsection{The model and the reference interface}\label{sub:model}
We restrict our analysis to the bosonic Higgs and gauge sectors. The inclusion of fermions and their effects on the interface and its solitons is left for future work.
We consider two scalar doublets $H_1,H_2$ of $SU(2)_L$, each with hypercharge $1/2$. We take the quartic scalar interactions to respect a $Z_2$ symmetry, $H_1\to H_1$ and $H_2\to-H_2$, which excludes the quartic couplings $\lambda_6$ and $\lambda_7$. The real scalar potential is
\begin{align}
 V={}&-m_1^2r_1-m_2^2r_2+\frac{\lambda_1}{2}r_1^2+
 \frac{\lambda_2}{2}r_2^2+\lambda_3r_1r_2+\lambda_4|\zeta|^2+\frac{\lambda_5}{2}(\zeta^2+\zeta^{*2})-m_{12}^2(\zeta+\zeta^*),\notag\\
 &r_i\mathrel{\equiv} H_i^\dagger H_i,\qquad \zeta\mathrel{\equiv}H_1^\dagger H_2.
 \label{eq:UVpotential}
\end{align}
We allow soft breaking through the $m_{12}^2$ term.
The scalar symmetry assignments can be organized using the classifications of Refs.~\cite{symmetry2012,symmetry2020}.
For the reference interface, we choose $m_1^2=m_2^2\equiv m^2>0$, $\lambda_1=\lambda_2\equiv\lambda>0$, and $m_{12}^2=0$. With quartic couplings that favor condensation of only one doublet, the exchange symmetry relates the two degenerate vacua
\begin{equation}
 (H_1,H_2)=(v\xi_0,0),\qquad (H_1,H_2)=(0,v\xi_0),
 \qquad \xi_0=(0,1)^T.\label{eq:bulkvac}
\end{equation}
Here $v=m/\sqrt\lambda$ and $\xi_0$ fixes the Higgs direction in the doublet. In the usual inert-doublet assignment, only $H_1$ couples to fermions: the first vacuum is then inert and the second inert-like. The exchange symmetry of the reference scalar potential need not extend to the Yukawa interactions.

An analytic wall is available at $g'=0$, $\lambda_4=\lambda_5=m_{12}^2=0$, and $\lambda_3=3\lambda$. Here $g'$ denotes the hypercharge coupling; the $SU(2)_L$ gauge coupling is denoted by $g$. Rescaling the fields by $m/\sqrt\lambda$ and coordinates by $m^{-1}$ gives the reference units $m^2=\lambda=1$. In these units the potential is
\begin{equation}
 V_0=-(r_1+r_2)+\tfrac12(r_1^2+r_2^2)+3r_1r_2.
 \label{eq:referencepotential}
\end{equation}
Its two vacua have $(r_1,r_2)=(1,0)$ and $(0,1)$. A domain wall perpendicular to the $z$ axis and interpolating between these two vacua is given by
\begin{equation}
 H_1=f_-(z)\xi_0,\qquad H_2=f_+(z)U\xi_0,
 \qquad f_\mp(z)=\frac{1\mp\tanh(z/\sqrt2)}2 .\label{eq:wall}
\end{equation}
The relative orientation $U\in SU(2)$ parametrizes the internal moduli of the reference wall and is constant along each static planar member of this family. Small symmetry-breaking scalar couplings generate a potential for $U$, turning these moduli into pseudo-moduli. The reference wall tension is
\begin{equation}
 T_0=\frac{2\sqrt2}{3}.
 \label{eq:walltension}
\end{equation}

We parameterize the orientation by
\begin{equation}
 U=\begin{pmatrix}u_0-iu_3&c\\-c^*&u_0+iu_3\end{pmatrix},\qquad
 \eta\mathrel{\equiv}u_0+iu_3,\qquad |\eta|^2+|c|^2=1.\label{eq:U}
\end{equation}
The real fields $u_0,u_3$ and complex field $c$ thus describe a unit $S^3$.
The overlap $H_1^\dagger H_2=f_-f_+\eta$ is neutral. The residual electromagnetic transformation acts as
\begin{equation}
 c\longmapsto e^{i\omega}c,\qquad \eta\longmapsto\eta,
 \qquad U\longmapsto e^{i\omega\sigma_3/2}Ue^{-i\omega\sigma_3/2}.\label{eq:EM}
\end{equation}
Here $\omega$ is the electromagnetic gauge parameter and $\sigma_i$ are Pauli matrices. A nonzero $c$ therefore describes a charged condensate localized where the two Higgs profiles overlap.

\subsection{The interface action and its phases}\label{sub:action}
We work in Euclidean signature and denote the bulk coordinates by $x^\mu=(\tau,x,y,z)$, $\mu=0,1,2,3$. The interface coordinates are $x^a=(\tau,x,y)$, $a=0,1,2$, with $\tau$ the Euclidean time and $z$ transverse to the wall. Repeated interface indices are summed with the Euclidean metric.

The symmetry responsible for the reference moduli is explicit in the matrices $\Phi_i=(i\sigma_2H_i^*,H_i)$. At $g'=0$ and the reference scalar couplings, they transform as $\Phi_i\mapsto L\Phi_iR_i^\dagger$, where $L\in SU(2)_L$ is gauged and $R_i\in SU(2)_{R_i}$ are global right rotations. In a vacuum with one Higgs condensate, the diagonal left--right subgroup is the custodial $SU(2)_V$. On the reference wall the relative right orientation gives the $SU(2)$ modulus $U$. The independent right rotations make its leading metric the round metric on $S^3$. This is the symmetry used here within the scalar classifications of Refs.~\cite{maxsym,symmetry2012,symmetry2020}.

To derive the kinetic coefficient, we promote $U$ to a slowly varying field $U(x^a)$, keeping $f_\pm(z)$ fixed in the moduli approximation~\cite{manton,solitoneft}. The induced gauge field has transverse profile $q_g(z)$ multiplying $-\partial_aU\,U^{-1}$. Minimizing the quadratic gradient energy over this profile defines $K_g$:
\begin{equation}
 K_g=\min_{q_g}\int\dd z\left[f_-^2q_g^2+f_+^2(1-q_g)^2+
 \frac{2}{g^2}(q_g')^2\right].\label{eq:Kg}
\end{equation}
The minimizing function obeys
\begin{equation}
 -\frac2{g^2}q_g''+(f_-^2+f_+^2)q_g=f_+^2,
 \quad q_g(-\infty)=0,\quad q_g(+\infty)=1.\label{eq:qg}
\end{equation}
Primes in these equations denote $z$ derivatives. The overlap integrals needed for the scalar potential are
\begin{equation}
 I=\int\dd z\,f_-^2f_+^2=\frac{\sqrt2}{12},
 \qquad J=\int\dd z\,f_-f_+=\frac1{\sqrt2}.\label{eq:overlaps}
\end{equation}
Their values are in the reference units above. Restoring dimensions, $K_g$ has mass dimension one, while $I$ and $J$ have dimensions three and one. Integrating the transverse coordinate gives the leading two-derivative interface action
\begin{equation}
 S_w^{(2)}=\int\dd^3x\left[K_g\{\partial_a u_0\partial_a u_0+
 \partial_a u_3\partial_a u_3+\partial_a c^*\partial_a c\}+V_w\right].\label{eq:wallaction}
\end{equation}
Small real $\lambda_4,\lambda_5,m_{12}^2$ give the orientation potential
\begin{equation}
 V_w=I(au_0^2+bu_3^2-2hu_0),\qquad
 a\mathrel{\equiv}\lambda_4+\lambda_5,\quad b\mathrel{\equiv}\lambda_4-\lambda_5,\quad
 h\mathrel{\equiv}\frac JI m_{12}^2\geq0.\label{eq:wallpotential}
\end{equation}
The change of convention $U\to-U$ permits $h\geq0$. Equation~\eqref{eq:wallpotential} is the leading potential evaluated on the reference wall. At nonzero $m_{12}^2$, the bulk vacua shift away from Eq.~\eqref{eq:bulkvac}.

Minimizing $V_w$ on $u_0^2+u_3^2+|c|^2=1$ gives Table~\ref{tab:vacua}. The labels follow Ref.~\cite{interface2026}. For example, in phase B, $b<0$ favors the neutral direction $c=0$. Minimization then fixes $u_0=h/(a-b)$ and leaves the two choices
\begin{equation}
 (u_0,u_3,c)=\left(\frac{h}{a-b},\ \pm\sqrt{1-\frac{h^2}{(a-b)^2}},\ 0\right),
 \qquad b<0,\quad a-b>h.\label{eq:Bvacua}
\end{equation}
The strict inequality makes the two vacua distinct; at $a-b=h$ they merge at $(1,0,0)$. The table lists vacuum orientations before identifying points related by electromagnetic gauge transformations.

Subphases B1--B6 distinguish string orientations in phase B, and D1, D5, and D6 are the corresponding branches and meeting point in phase D (section~\ref{sec:BD}). Subphases C1--C6 classify the potential of the neutral phase in the core of a local vortex and its instanton configurations (section~\ref{sec:C}). Their boundaries are obtained from the leading potentials in the corresponding effective theories. The interface vacuum set depends only on the ratios of $a,b,h$: multiplying all three parameters by the same positive constant multiplies $V_w$ without changing its minima. Figure~\ref{fig:phase} therefore suppresses this overall scale.

\subsection{\texorpdfstring{Interface winding and bulk instanton charge}{Interface winding and bulk instanton charge}}\label{sec:charge}
Fix $U$ to one chosen interface vacuum at infinity in the Euclidean coordinates $(\tau,x,y)$. This compactifies the domain to $S^3$. Since $U\in SU(2)\simeq S^3$, the field carries a Skyrmion charge associated with $\pi_3(SU(2))\simeq\mathbb Z$:
\begin{equation}
 k_w\mathrel{\equiv}\frac1{24\pi^2}\int_{\mathbb R^3}\Tr(U^{-1}\dd U)^3\in\mathbb Z.\label{eq:kw}
\end{equation}
Here $\dd$ is the exterior derivative along the interface, the cube denotes a wedge product, and the trace is in the fundamental representation. To relate this charge to the bulk gauge field, define the anti-Hermitian one-form $L=-\dd U\,U^{-1}$ and use the transverse profile $q_g$ obtained above:
\begin{equation}
 L\mathrel{\equiv}-\dd U U^{-1},\qquad \mathcal A=q_g(z)L,\qquad \mathcal A_z=0.
 \label{eq:uplift}
\end{equation}
The bulk curvature is $\mathcal F=\dd\mathcal A+\mathcal A\wedge\mathcal A$, with the exterior derivative now acting on all four coordinates. It satisfies
\begin{align}
 \mathcal F&=q_g'\dd z\wedge L+q_g(q_g-1)L\wedge L,\\
 \Tr\mathcal F\wedge\mathcal F&=2q_g'q_g(q_g-1)\dd z\wedge\Tr L^3.
 \label{eq:FF}
\end{align}
The transverse integral is fixed by $q_g(-\infty)=0$ and $q_g(+\infty)=1$:
\begin{equation}
 2\int_0^1q(q-1)\dd q=-\frac13.
\end{equation}
With positive orientation $\dd z\wedge\dd\tau\wedge\dd x\wedge\dd y$, the Yang--Mills topological charge is
\begin{equation}
 Q_{\rm YM}=-\frac1{8\pi^2}\int\Tr\mathcal F\wedge\mathcal F=-k_w.
 \label{eq:QYM}
\end{equation}
Thus the interface Skyrmion charge equals the bulk Yang--Mills instanton charge in magnitude. The sign depends on the orientation convention. The identity follows from the displayed gauge field and its boundary values; it does not require self-duality.

\subsection{\texorpdfstring{The Skyrme term and interface Skyrmions}{The Skyrme term and interface Skyrmions}}\label{sub:Skyrme}
Within the moduli approximation, substituting Eq.~\eqref{eq:uplift} into the bulk Yang--Mills action induces a Skyrme term on the interface, as in the domain-wall construction of Ref.~\cite{eto2005}:
\begin{equation}
\begin{aligned}
 S_{4,w}&=\kappa_{\rm YM}\int\dd^3x\sum_{a,b=0}^2
 \{-\Tr[L_a,L_b]^2\},\\
 \kappa_{\rm YM}&\mathrel{\equiv}\frac1{2g^2}\int\dd z\,q_g^2(1-q_g)^2>0.
\end{aligned}\label{eq:Skyrme}
\end{equation}
Here $L_a$ are the interface components of $L$ in Eq.~\eqref{eq:uplift}. The sum counts ordered index pairs. The coefficient $\kappa_{\rm YM}$ has mass dimension minus one. Together with the two-derivative action (Eq.~\eqref{eq:wallaction}), this supplies the two gradient terms for the Derrick scaling argument below.

The target space remains $S^3$ in every phase. With a chosen vacuum fixed at infinity, all phases therefore admit ordinary interface Skyrmion configurations associated with $\pi_3(S^3)\simeq\mathbb Z$. The phase-dependent potential deforms these configurations. The subsequent sections describe their additional localization on defects or decomposition into constituents.

To estimate the size by Derrick scaling~\cite{derrick1964}, choose a smooth dimensionless profile $U_*(\boldsymbol y)$ of unit degree and define $U_{\lambda_{\mathrm s}}(\boldsymbol x)=U_*(\boldsymbol x/\lambda_{\mathrm s})$, with $\boldsymbol x=(\tau,x,y)$ and $\lambda_{\mathrm s}>0$. Factor the potential as $V_w-V_{\min}=\varepsilon\widehat V$, where $\varepsilon$ has mass dimension three and $\widehat V$ is dimensionless. In phase H we set $\varepsilon=0$. Define the dimensionless shape integrals
\begin{equation}\begin{aligned}
 A_2&\mathrel{\equiv}\int\dd^3y\left[(\partial_a u_{0*})^2+(\partial_a u_{3*})^2+|\partial_a c_*|^2\right],\\
 A_4&\mathrel{\equiv}\int\dd^3y\sum_{a,b=0}^2\{-\Tr[\ell_a,\ell_b]^2\},
 \quad \ell_a\mathrel{\equiv}-\partial_aU_*U_*^{-1},\\
 A_0&\mathrel{\equiv}\int\dd^3y\,\widehat V(U_*).
\end{aligned}\label{eq:shapeintegrals}
\end{equation}
The derivatives in these integrals act on $\boldsymbol y$; $u_{0*},u_{3*},c_*$ are the components of $U_*$. Let $\Delta S$ denote the interface action relative to the chosen uniform interface vacuum, evaluated in the same volume. The measure contributes $\lambda_{\mathrm s}^3$ and each derivative contributes $\lambda_{\mathrm s}^{-1}$, giving
\begin{equation}
 \Delta S(\lambda_{\mathrm s})=A_2K_g\lambda_{\mathrm s}+\frac{A_4\kappa_{\rm YM}}{\lambda_{\mathrm s}}+A_0\varepsilon\lambda_{\mathrm s}^3.
 \label{eq:Derrick}
\end{equation}
For a nontrivial profile $A_2,A_4>0$ and $A_0\geq0$. The two-derivative and potential terms decrease as the configuration shrinks, whereas the Skyrme term increases. At the phase H point, their balance gives $\lambda_{\mathrm s}^2=A_4\kappa_{\rm YM}/(A_2K_g)$. A nonzero potential changes this size balance. This fixed-shape variation tests the size direction only; it is not a fluctuation analysis of the full configuration.

For a quantitative use of the interface theory, its variation scale must exceed the wall thickness. If $M_\perp$ is the mass gap of the transverse modes omitted in the moduli approximation, this condition is $\lambda_{\mathrm s}M_\perp\gg1$.

\Needspace{6\baselineskip}
The effective theory on a defect can support a further soliton~\cite{relations2022}. We call the host defect the mother and the soliton it supports the daughter. For example, a half-unit $Z$ string is a daughter of the Higgs interface and can in turn be the mother of a Nambu monopole, which appears as a kink on the string. In the tables below, the rows identify the mothers, the columns identify the daughters, and each entry describes the daughter in the effective theory on the corresponding mother. We count the Higgs interface as the first generation and each additional nested soliton as one further generation, independently of codimension.

\FloatBarrier\section{Instantons on \texorpdfstring{electroweak}{electroweak} strings and Nambu monopoles: B, D, and E phases}\label{sec:BD}\suppressfloats[t]
Section~\ref{sub:string} derives the effective theory on an electroweak string. Section~\ref{sub:B23} uses it to estimate the size of the baby Skyrmion in phases B2 and B3. Sections~\ref{sub:B1} and~\ref{sub:E} describe phase winding on confined Nambu monopoles in phases B1 and D1 and on deconfined Nambu monopoles in phase E, respectively. Section~\ref{sec:merons} treats the meron constituents in phases B4, B5, and D5.

\subsection{\texorpdfstring{The effective theory on an electroweak string}{The effective theory on an electroweak string}}\label{sub:string}
We start with the reference wall in phase B at $a=h=0$, $b<0$, where the two interface vacua are $u_3=\pm1$. A kink between them varies across $x$ and extends along $y$, so it is a string inside the Higgs interface. Its internal orientation is a unit vector $\bmn=(n_1,n_2,n_3)$ on $S^2$. We write
\begin{equation}
 u_3=\cos\theta(x),\qquad
 (\Re c,\Im c,u_0)=\sin\theta(x)\bmn(\tau,y),\qquad \bmn^2=1,
 \label{eq:Bembedding}
\end{equation}
where $x_0$ denotes the string position and
\begin{equation}
 \theta=2\arctan e^{m_B(x-x_0)},\qquad m_B^2\mathrel{\equiv}\frac{I(-b)}{K_g}.
 \label{eq:theta}
\end{equation}
A constant change of target-space basis puts this embedding in the form $U=\cos\theta\,\mathbf1+i\sin\theta\,\bmn\cdot\boldsymbol\sigma$. This form is convenient for computing its winding. We now allow $\bmn$ to vary on the Euclidean string world sheet $(\tau,y)$. Indices $\alpha,\beta$ run over these two directions, and $\partial_\alpha\bmn\,\partial_\alpha\bmn=(\partial_\tau\bmn)^2+(\partial_y\bmn)^2$. Integrating over the transverse kink coordinate $x$ gives the world-sheet action of a baby Skyrme model, measured relative to a straight string in its vacuum orientation with the same world-sheet regulator
\begin{equation}
 S_{\rm str}=\int\dd\tau\dd y\left[
 G_B(\partial_\alpha\bmn)^2+
 \lambda_B\{\bmn\cdot(\partial_\tau\bmn\times\partial_y\bmn)\}^2
 +V_{\rm str}-V_{\min}+\cdots\right].\label{eq:stringaction}
\end{equation}
The coefficients obtained from Eqs.~\eqref{eq:wallaction} and~\eqref{eq:Skyrme} are
\begin{equation}
 G_B=\frac{2K_g}{m_B}+\frac{64}{3}\kappa_{\rm YM}m_B,
 \qquad \lambda_B=\frac{64\kappa_{\rm YM}}{3m_B}>0.\label{eq:GB}
\end{equation}
Here $G_B$ is the coefficient of the orientational kinetic term and $\lambda_B$ multiplies its four-derivative baby Skyrme term. The latter is the square of the $S^2$ area density. Both coefficients follow from the fixed transverse kink profile; Appendix~\ref{app:coefficients} gives the integrals.

{
With the string orientation fixed at world-sheet infinity, the configurations are classified by $\pi_2(S^2)\simeq\mathbb Z$. In the baby Skyrme model of Eq.~\eqref{eq:stringaction}, a unit-charge lump is a baby Skyrmion~\cite{pietteMultisolitons,pietteDynamics}. The lump charge is
\begin{equation}
 Q_{\bmn}\mathrel{\equiv}\frac{1}{4\pi}\int\dd y\,\dd\tau\,
 \bmn\cdot(\partial_y\bmn\times\partial_\tau\bmn).
 \label{eq:stringlump}
\end{equation}
In the reference embedding \eqref{eq:Bembedding}, the transverse kink profile $\theta(x)$ lifts this $S^2$ map to the $S^3$ interface field. Integrating over the transverse direction gives
\begin{equation}
 |k_w|=\left|\frac{1}{2\pi^2}\int_0^\pi\sin^2\theta\,\dd\theta
 \int\dd y\,\dd\tau\,\bmn\cdot(\partial_y\bmn\times\partial_\tau\bmn)\right|
 =|Q_{\bmn}|.
 \label{eq:suspensioncharge}
\end{equation}
The gauge interpolation in Eq.~\eqref{eq:uplift} then gives
$|Q_{\bmn}|=|k_w|=|Q_{\rm YM}|$ by Eq.~\eqref{eq:QYM}.
Thus a unit-charge baby Skyrmion on the string is a three-dimensional Skyrmion in the interface theory and realizes a unit-charge Yang--Mills instanton in the four-dimensional bulk. Viewed at the level of the world sheet and the bulk, this gives the lump--instanton correspondence familiar from non-Abelian vortices~\cite{HananyTong2004,ShifmanYung2004,higgsinstantons}.
}

\begin{figure}[!htbp]
\centering
\includegraphics[width=0.91\linewidth]{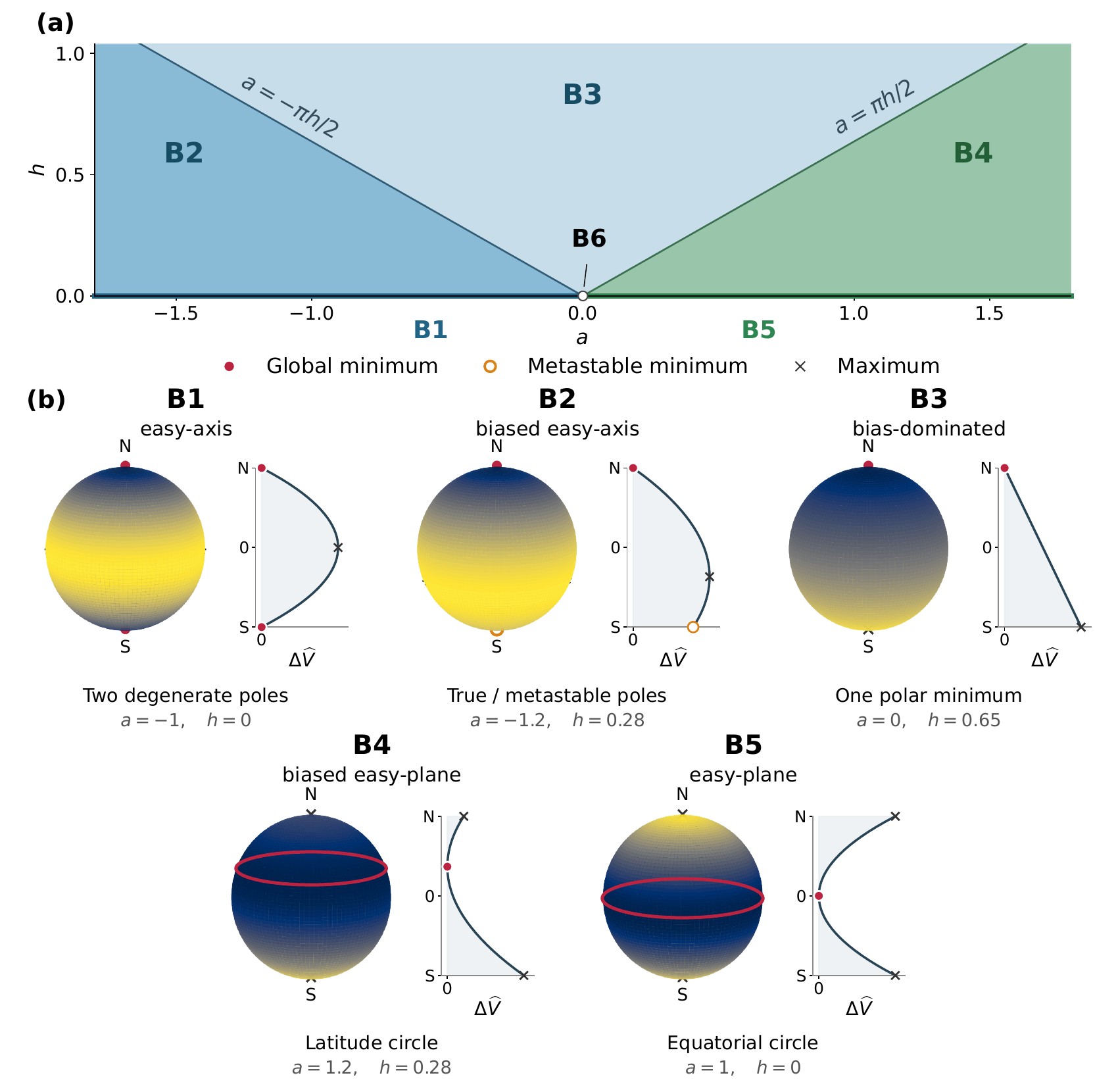}
\caption{String orientation potential in phase B. (a) Subphases of the reduced potential $\widehat V=a n_3^2-\pi h n_3$ at fixed $b<0$ and $h\geq0$. The boundaries are $a=\pm\pi h/2$, and the origin $a=h=0$ is phase B6, where the orientational potential vanishes identically, leaving the full $S^2$ degenerate. (b) Representative potentials on $S^2$ and profiles of $\widehat V-\widehat V_{\min}$ versus $n_3$ for phases B1--B5. Filled red circles, open orange circles, and crosses denote global minima, metastable minima, and maxima, respectively. The classification applies in the regime $|a|,h\ll-b$.}
\label{fig:Borientation}
\end{figure}

Small $a,h$ lift the orientational degeneracy. We write the resulting potential as
\begin{equation}
 V_{\rm str}=\mathrm{const}-D n_3^2-H n_3,
 \qquad H\mathrel{\equiv}\frac{2\pi Ih}{m_B}.\label{eq:orientpotential}
\end{equation}
Figure~\ref{fig:Borientation} shows the six orientational subphases B1--B6. For the reference scalar profile, $D=-2Ia/m_B$, and gauge corrections can shift its value. At $H=0$, the potential has two degenerate polar minima for $D>0$ (phase B1), an equatorial circle of minima for $D<0$ (phase B5), and a flat potential with full orientational $S^2$ degeneracy for $D=0$ (phase B6). Positive $H$ favors the north pole. For $D>H/2$, the south pole is a metastable minimum (phase B2). For $-H/2<D<H/2$, the north pole is the only minimum (phase B3). For $D<-H/2$, the minima form a latitude circle (phase B4). At $h=0$, exchanging $u_0\leftrightarrow u_3$ and $a\leftrightarrow b$ gives the D phase, with an easy-axis branch in phase D1 and an easy-plane branch in phase D5. They meet at phase D6 ($b=h=0$, $a<0$), where the exchanged orientational potential vanishes and the full $S^2$ is degenerate.

In the interface theory, the electroweak string is itself a domain wall interpolating between the two interface vacua. A Skyrmion trapped on this wall is therefore a three-dimensional domain-wall Skyrmion in the Euclidean interface theory. This correspondence has precedents in Skyrme-type models~\cite{nittaWallSkyrmion2013,nittaMatryoshka2013,gudnasonWallSkyrmions2014,incskyrmion,etoNittaNonAbelianSG2015} and in low-energy QCD~\cite{etoQCDwall,etoQCDchain,etoQCDrotation,amariQCDgauge,copingerFermionicWalls2026,copingerWallFermiGas2026}, where Skyrmions trapped on domain walls appear as lumps in the wall effective theory. On the electroweak string, the trapped instanton appears as a two-dimensional baby Skyrmion in the string effective theory~\cite{HananyTong2004,ShifmanYung2004,higgsinstantons}.

Throughout phase B, a unit-charge instanton on the electroweak string can be described as a baby Skyrmion in the string effective theory.

\subsection{\texorpdfstring{Phases B2 and B3: Scaling of the baby Skyrmion action}{Phases B2 and B3: Scaling of the baby Skyrmion action}}\label{sub:B23}
In phases B2 and B3, the north-pole vacuum fixes the string orientation at world-sheet infinity. The south pole is metastable in phase B2 and unstable in phase B3.

To estimate its size by Derrick scaling~\cite{derrick1964}, choose a degree-one profile $\bmn_*(\boldsymbol s)$ in dimensionless coordinates $\boldsymbol s=(s^1,s^2)$ and define $\bmn_{\lambda_{\mathrm s}}(\tau,y)\equiv\bmn_*((\tau,y)/\lambda_{\mathrm s})$. Write $V_{\rm str}-V_{\min}=\Delta V\,\widehat V_B(\bmn)$ with a positive potential scale $\Delta V$. The shape integrals are $S_2\equiv G_B\int\dd^2s\,(\partial_\alpha\bmn_*)^2$, $C_4\equiv\int\dd^2s\,[\bmn_*\cdot(\partial_1\bmn_*\times\partial_2\bmn_*)]^2$, and $C_0\equiv\int\dd^2s\,\widehat V_B(\bmn_*)$. The action with the transverse string profile held fixed scales as
\begin{equation}
 \Delta S_{\rm str}(\lambda_{\mathrm s})=S_2+\frac{C_4\lambda_B}{\lambda_{\mathrm s}^2}
 +C_0\Delta V\lambda_{\mathrm s}^2+\cdots,\qquad C_0,C_4>0.
 \label{eq:babyscale}
\end{equation}
The kinetic term $S_2$ is scale invariant. The potential favors shrinking, and the baby Skyrme term favors spreading. This balance estimates the size for a fixed transverse profile, provided $\lambda_{\mathrm s}m_B\gg1$. Relaxation of the transverse width generates another term with the same $\lambda_{\mathrm s}^{-2}$ scaling (Appendix~\ref{app:coefficients}). In the benchmark of section~\ref{sec:merons}, the coefficient of this term has magnitude $|C_s|\simeq175$, compared with $\lambda_B\simeq3.14$ for the direct baby Skyrme term. Both terms are small compared with the two-derivative term at the meron scale, but that comparison does not control their balance with the potential in the size variation. Stability at finite size with the transverse profile allowed to relax therefore remains open.

\Needspace{5\baselineskip}
\subsection{Phases B1 and D1: Phase winding on a confined Nambu monopole}\label{sub:B1}
In gauge theories supporting non-Abelian vortices, confined 't~Hooft--Polyakov monopoles can be realized as kinks in the vortex world-sheet theory~\cite{HananyTong2004,ShifmanYung2004,TongMonopoles2003,NittaVinci2011}. In two-Higgs-doublet models, Nambu monopoles occur as junctions between two $Z$ strings~\cite{topmonopole,monodynamics,nambudyon}. For the inert Higgs interface, a kink between the two degenerate string orientations in phase B1 represents a confined Nambu monopole~\cite{interface2026}. To construct it, let $\alpha$ be the polar angle and $\chi$ the azimuthal angle of $\bmn$. The monopole centered at $y=y_0$ has
\begin{equation}
 \bmn=(\sin\alpha\cos\chi,\sin\alpha\sin\chi,\cos\alpha),\quad
 \alpha(y)=2\arctan e^{m_M(y-y_0)},\quad m_M\mathrel{\equiv}\sqrt{D/G_B}.
 \label{eq:monopole}
\end{equation}
For constant $\chi$, this field interpolates from $n_3=+1$ to $n_3=-1$ and follows a meridian of $S^2$. The angle $\chi$ is its internal $U(1)$ modulus. At $h=0$, the construction in phase D1 follows from that in phase B1 by interchanging the interface-field components $u_0\leftrightarrow u_3$ and the couplings $a\leftrightarrow b$.

Promoting $\chi$ to $\chi(\tau)$ sweeps this meridian over the orientational sphere. Using Eq.~\eqref{eq:stringlump}, the lump charge of the phase winding is
\begin{equation}
 Q_{\bmn}=\frac{1}{4\pi}\int_0^\pi\sin\alpha\,\dd\alpha\int\dd\chi
 =\frac{\Delta\chi}{2\pi}.
 \label{eq:monopolelump}
\end{equation}
At $g'=0$, a full turn $\Delta\chi=2\pi$ gives a unit-charge baby Skyrmion on the electroweak string world sheet. This configuration realizes a unit-charge Yang--Mills instanton in the bulk by Eqs.~\eqref{eq:suspensioncharge} and~\eqref{eq:QYM}. A higher-dimensional example is the Yang--Mills instanton on a monopole string inside a non-Abelian vortex sheet~\cite{nittaMonopoleStrings2013}. At $g'=0$, $\chi$ is a free collective coordinate at this order. Substitution of the monopole profile into the string effective action gives the Euclidean action for its winding relative to the static monopole on the same host string,
\begin{equation}
 \Delta S_\chi=\frac{\mathcal I_\chi}{2}\int\dd\tau\,(\chi')^2,
 \qquad \mathcal I_\chi=\frac{4G_B}{m_M}+\frac83\lambda_Bm_M.
 \label{eq:rotor}
\end{equation}
Here $\chi'\equiv\dd\chi/\dd\tau$, and the second contribution to the inertia $\mathcal I_\chi$ comes from the baby Skyrme term. On a Euclidean time circle of fixed period $T$, a uniform turn is stable within the phase theory at fixed winding. Its action in the unit-winding sector is $\Delta S_\chi=2\pi^2\mathcal I_\chi/T$. On the noncompact Euclidean time axis, the winding can spread indefinitely, lowering the action without changing the charge. Appendix~\ref{app:rotor} derives the compact-time minimum and gives a constrained representative of prescribed width on the line.

At $g'=0$, constant shifts of $\chi$ are a global symmetry of the string effective theory. At finite $g'$, the phase is gauged by $U(1)_{\rm EM}$, and a constant shift corresponds to an electromagnetic transformation approaching a constant at spatial infinity~\cite{nambudyon}. We fix the gauge convention at infinity. A time-dependent $\chi$ induces a temporal gauge field, and the finite-$g'$ correction to the inertia requires solving Gauss's law.

On the Euclidean string world sheet $(y,\tau)$, the monopole profile interpolates between the two string orientations as $y$ varies, while its internal phase $\chi$ winds by $2\pi$ as $\tau$ varies. Together these two variations form a unit-charge, two-dimensional domain-wall Skyrmion~\cite{josephson2012,jewels,ross2022,amari2023,jenningsSutcliffe2013} on the vortex world sheet. Table~\ref{tab:hierarchy} summarizes the construction in the bulk and on its successive hosts. Panel (c) shows the relation corresponding to Table~IV of Ref.~\cite{relations2022}. At $g'=0$, the $\mathbb{C}P^1$ kink on the electroweak string world sheet has an internal phase $\chi$ with a flat potential. A $2\pi$ winding of $\chi$ along the kink's Euclidean world line gives a unit-charge baby Skyrmion on the string world sheet.
\begin{table}[!htbp]
\centering\small
\renewcommand{\arraystretch}{1.18}
\medskip
\noindent\makebox[\linewidth][l]{\textbf{(a)} Soliton hierarchy from the bulk}\par\smallskip
\begin{tabular}{>{\raggedright\arraybackslash}p{3.12cm}|>{\raggedright\arraybackslash}p{3.06cm}>{\raggedright\arraybackslash}p{2.95cm}>{\raggedright\arraybackslash}p{3.59cm}}
\hline
Mother \textbackslash{} Daughter & Half-unit $Z$ string & Nambu monopole & YM instanton\\\hline
Interface & Kink & Global vortex & Skyrmion\\
Half-unit $Z$ string & --- & Kink & Baby Skyrmion/lump\\
Monopole & --- & --- & Internal-phase winding\\\hline
\end{tabular}\par
\medskip
\noindent\makebox[\linewidth][l]{\textbf{(b)} Soliton hierarchy within the interface}\par\smallskip
\begin{tabular}{>{\raggedright\arraybackslash}p{3.12cm}|>{\raggedright\arraybackslash}p{4.02cm}>{\raggedright\arraybackslash}p{5.87cm}}
\hline
Mother \textbackslash{} Daughter & Global vortex & 3D Skyrmion\\\hline
$\mathbb{C}P^1$ kink & $Z_2$ kink & Baby Skyrmion\\
Global vortex & --- & Internal-phase winding\\\hline
\end{tabular}\par
\medskip
\noindent\makebox[\linewidth][l]{\textbf{(c)} Soliton hierarchy on the string world sheet}\par\smallskip
\begin{tabular}{>{\raggedright\arraybackslash}p{3.12cm}|>{\raggedright\arraybackslash}p{10.32cm}}
\hline
Mother \textbackslash{} Daughter & Baby Skyrmion\\\hline
$\mathbb{C}P^1$ kink & Internal-phase winding\\\hline
\end{tabular}\par
\caption{Nested solitons in phases B1 and D1. (a) Starting from the bulk, the table follows the half-unit $Z$ string, the Nambu monopole, and the YM instanton through their successive descriptions on the interface, the string, and the monopole. (b) Starting from the interface theory, the table follows the global vortex and the three-dimensional Skyrmion onto defects within the interface. (c) Starting from the world-sheet theory of the half-unit $Z$ string, the table shows how the baby Skyrmion is realized by a $2\pi$ winding of the internal phase of the $\mathbb{C}P^1$ kink along its Euclidean world line. This winding carries one unit of instanton charge. YM denotes Yang--Mills.}\label{tab:hierarchy}
\end{table}

\Needspace{5\baselineskip}
\subsection{\texorpdfstring{Phase E: Internal-phase winding on a deconfined Nambu monopole}{Phase E: Internal-phase winding on a deconfined Nambu monopole}}\label{sub:E}
The phase-winding construction also applies to a deconfined Nambu monopole. In phase E, the interface vacuum is the neutral circle $|\eta|=1$, $c=0$. We can consider a global vortex on the interface, around which the phase of $\eta$ winds. This vortex corresponds to a deconfined Nambu monopole in the bulk~\cite{interface2026}. Its $Z$ flux remains confined to the interface and spreads radially within the wall, with axial symmetry about the wall normal through the monopole. The electromagnetic magnetic flux extends into the three-dimensional bulk. The charged component $c$ condenses in the vortex core. The vortex is therefore a bosonic superconducting string in the three-dimensional Euclidean interface theory.

For a monopole at a fixed position on the interface, the world line extends along the Euclidean time direction $\tau$. The phase of the charged core can wind as $\tau$ varies. In phase E, $h=0$ and $a=b=-\mu_E$ with $\mu_E>0$, so $V_E-V_{E,\min}=I\mu_E|c|^2$. At $g'=0$ the winding configuration can be written in polar coordinates $(r,\varphi)$ on the spatial interface as
\begin{equation}
 \eta=\sin F_E(r)\,e^{iq\varphi},\qquad
 c=\cos F_E(r)\,e^{i\chi(\tau)},\qquad
 F_E(0)=0,\quad F_E(\infty)=\frac{\pi}{2},
 \label{eq:Eansatz}
\end{equation}
where $q$ is the spatial vortex winding. With the spatial-vortex boundary condition held fixed, the magnitude of the relative interface Skyrmion charge is
\begin{equation}
 |k_{\rm rel}|=\frac{1}{2\pi^2}
 \left|\int_0^\infty\sin F_E\cos F_E\,F_E'(r)\,\dd r
 \int_0^{2\pi} q\,\dd\varphi\int\dd\chi\right|
 =\frac{|q\Delta\chi|}{2\pi}.
 \label{eq:Echarge}
\end{equation}
For a unit vortex of $\eta$, a full $2\pi$ winding of the charged core phase gives unit interface Skyrmion charge in magnitude. This realizes a unit-charge bulk Yang--Mills instanton through the gauge interpolation in Eq.~\eqref{eq:uplift}. Related constructions in which vortex-core winding carries Skyrmion charge have been studied in Skyrme models, principal chiral models, and low-energy QCD~\cite{incskyrmion,torii,beads,qiu,nittaFractionalPCM2015}.

At $g'=0$, the phase theory of the deconfined monopole also admits a uniform winding that minimizes the action at fixed winding number on a Euclidean time circle (Appendix~\ref{app:rotor}). At finite $g'$, the charged core phase is gauged by $U(1)_{\rm EM}$, as for the confined monopoles in section~\ref{sub:B1}. A time-dependent phase then couples to the temporal gauge field, whose response is governed by Gauss's law.

\subsection{\texorpdfstring{Phases B4, B5, and D5: Meron constituents on superconducting strings}{Phases B4, B5, and D5: Meron constituents on superconducting strings}}\label{sec:merons}
In phases B4, B5, and D5, the string world sheet supports merons because the vacuum orientations form a circle. With $D=-c_B<0$, completing the square in Eq.~\eqref{eq:orientpotential} gives
\begin{equation}
 V_{\rm str}-V_{\min}=c_B(n_3-s_*)^2,\qquad
 c_B\mathrel{\equiv}\frac{2Ia}{m_B}>0,\quad s_*\mathrel{\equiv}\frac{H}{2c_B}=\frac{\pi h}{2a}.
 \label{eq:latitude}
\end{equation}
The number $s_*$ is the vacuum value of $n_3$ on the orientational sphere. Phase B5 has $s_*=0$ and phase B4 has $0<s_*<1$. These expressions use $|a|,h\ll-b$, so the transverse kink profile is that of the reference string in phase B. Finite-$h$ changes of its asymptotic vacua are higher-order corrections to this approximation.

Phase B4 supports mixed $Z$--$W$ strings, while phases B5 and D5 support pure $W$ strings. The $W$-string configurations and the intermediate orientations between the $Z$ and $W$ strings were studied for bulk non-Abelian strings in two-Higgs-doublet models~\cite{eknstrings2018}. In the embedding of Eq.~\eqref{eq:Bembedding}, an easy-plane orientation has $n_1^2+n_2^2>0$, so the charged Higgs component $c$ condenses in the string core. The charged condensate supports bosonic superconductivity through the mechanism described by Witten~\cite{witten1985strings}. Its phase is the azimuthal angle on the circle of core orientations.

\begin{figure}[!htbp]
\centering
\includegraphics[width=.73\linewidth]{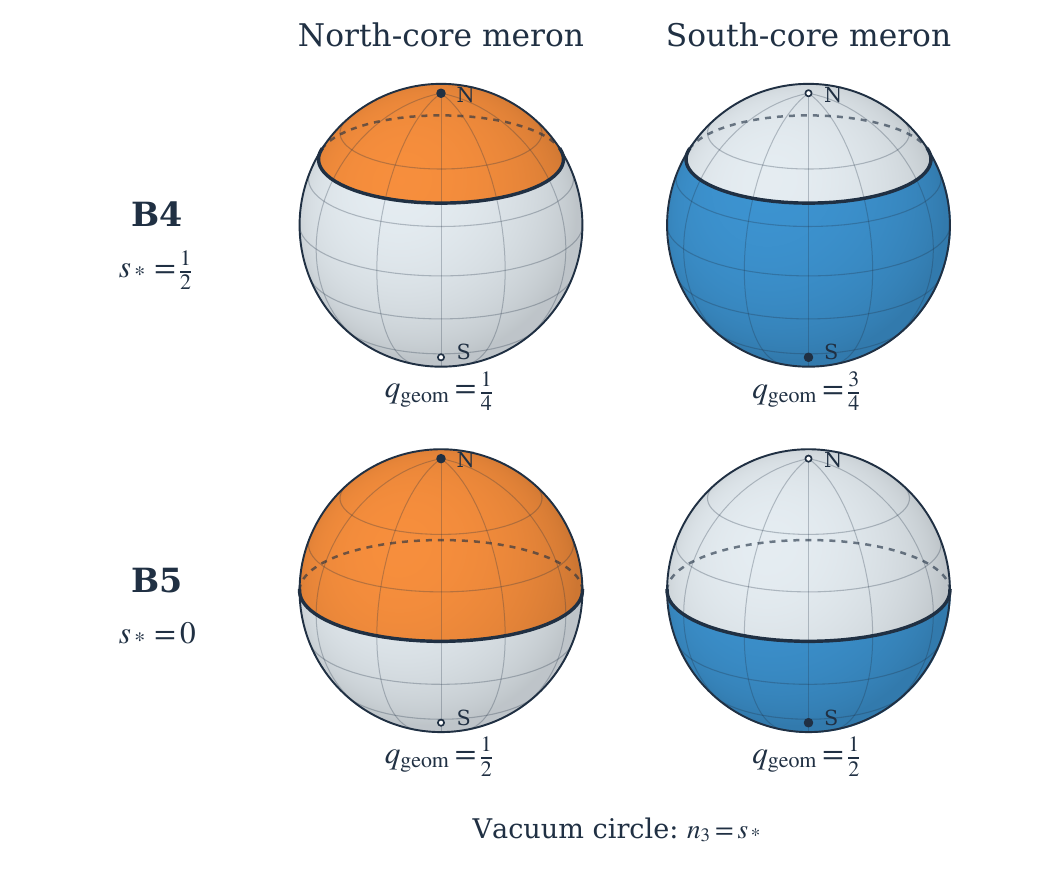}
\caption{Regions of the orientational sphere covered by north- and south-core merons. The thick curves mark the vacuum circle $n_3=s_*$. The shaded areas give the geometric charges in Eq.~\eqref{eq:cap}; phase B4 is illustrated at $s_*=1/2$, as in Table~\ref{tab:Bactions}.}\label{fig:meroncaps}
\end{figure}

A vortex of this phase on the Euclidean string world sheet is a meron. Far from the core, the field winds around the vacuum circle $n_3=s_*$, while at the core it reaches one of the two poles, $p\equiv n_3(0)=\pm1$. For vortex winding $\nu\in\mathbb Z$, the geometric charge is the oriented area of the corresponding spherical cap divided by $4\pi$,
\begin{equation}
 Q_{\rm geom}=\frac\nu2(p-s_*).\label{eq:cap}
\end{equation}
Figure~\ref{fig:meroncaps} shows the two caps in phases B4 and B5. The orientations $(p,\nu)=(+1,+1)$ and $(-1,-1)$ give complementary geometric charges $q_N\equiv(1-s_*)/2$ and $q_S\equiv(1+s_*)/2$. In the ungauged easy-plane baby Skyrme model, unit-charge baby Skyrmions have been constructed as bound states of two half-charge constituents~\cite{jaykkaSpeight2010,kobayashiNitta2013vortices}.

The azimuthal phase of the charged core condensate couples to the electromagnetic gauge field. On the world sheet the connection is $a\equiv a_\alpha\dd x^\alpha$, with $f\equiv\dd a$ and generator $\bm e_3=(0,0,1)$ for rotations of $\bmn$. The string effective action used to compute the meron profiles is
\begin{equation}
 S_{\rm loc}=\int\dd y\dd\tau\left[G_B(D_\alpha\bmn)^2+c_B(n_3-s_*)^2+
 \frac{f_{\alpha\beta}f_{\alpha\beta}}{4e_{\rm ws}^2}\right],
 \quad D_\alpha\bmn\equiv\partial_\alpha\bmn-a_\alpha\bm e_3\times\bmn.
 \label{eq:gaugedB}
\end{equation}
For the numerical profiles, we approximate the electromagnetic kinetic term by a local Maxwell term on the world sheet, with $e_{\rm ws}$ treated as an effective parameter. The orientational part retains the two-derivative term and the potential. The gauge field screens the azimuthal gradient in the local world-sheet theory. The profiles and actions below refer to this theory. Calculating the electromagnetic field outside the string and its contribution to the action requires the bulk gauge dynamics. Bogomolny solitons were studied in the gauged $O(3)$ nonlinear sigma model~\cite{schroers1995} and in gauged linear sigma models~\cite{schroers1996}. In the gauged $\mathbb{C}P^1$ model, a lump can decompose into two local vortices with north- and south-pole cores~\cite{vinci}, corresponding to the two types of meron here.

\begin{figure}[!htbp]
\centering
\includegraphics[width=\linewidth]{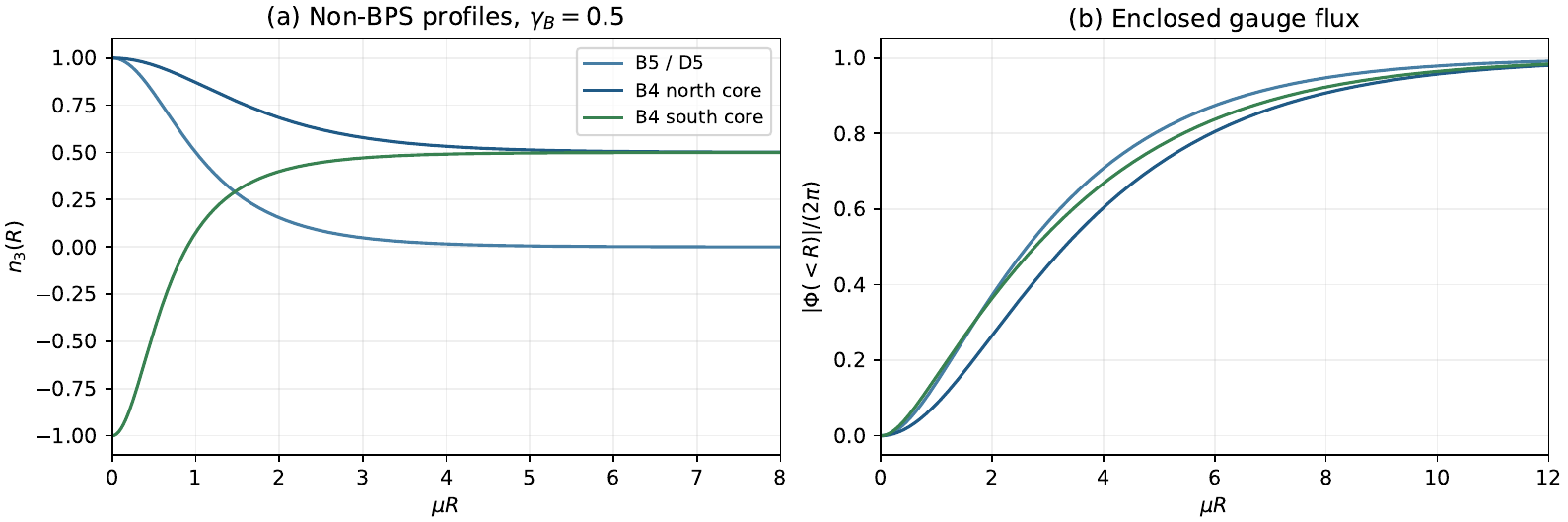}
\caption{Individual radial meron solutions at $\gamma_B=0.5$. (a) Orientational component $n_3$ for phases B5 and D5 and for the two cores in phase B4 at $s_*=1/2$. (b) Magnitudes of the corresponding enclosed magnetic fluxes, normalized by $2\pi$. The horizontal axis is the dimensionless radius $\rho$.}
\label{fig:meronprofiles}
\end{figure}

Use polar coordinates $(R,\varphi)$ on the Euclidean world sheet, with $R=\sqrt{y^2+\tau^2}$ and $\varphi=\arg(y+i\tau)$. An axially symmetric meron is
\begin{equation}
 \bmn=(\sin F\cos\nu\varphi,\sin F\sin\nu\varphi,\cos F),
 \qquad a=\nu\mathcal A(R)\dd\varphi.\label{eq:meronansatz}
\end{equation}
North and south cores have $F(0)=0$ and $\pi$, respectively, while $F(\infty)=\arccos s_*$. In both cases $\mathcal A(0)=0$ and $\mathcal A(\infty)=1$. For $|\nu|=1$, define
\begin{equation}
 \rho\mathrel{\equiv}\mu R,\quad \mu\mathrel{\equiv}\sqrt{c_B/G_B},\qquad
 \gamma_B\mathrel{\equiv}\frac{\sqrt{2e_{\rm ws}^2G_B}}\mu.
\end{equation}
The scale $\mu$ sets the inverse meron radius, $\rho$ is the dimensionless radius, and $\gamma_B$ measures the local gauge coupling. Subscripts $\rho$ denote derivatives with respect to this radius. The vacuum gauge-field mass is $m_A=\gamma_B\mu\sqrt{1-s_*^2}$. For either $\nu=+1$ or $-1$, the radial action is
\begin{align}
 \frac{S_{\rm loc}}{2\pi G_B}
 &=\int_0^\infty\rho\dd\rho\left[F_\rho^2+
 \frac{(1-\mathcal A)^2\sin^2F}{\rho^2}+(\cos F-s_*)^2+
 \frac{\mathcal A_\rho^2}{\gamma_B^2\rho^2}\right].\label{eq:Bactionrad}
\end{align}
\begin{align}
 F_{\rho\rho}+\frac{F_\rho}\rho-
 \frac{(1-\mathcal A)^2\sin F\cos F}{\rho^2}
 +\sin F(\cos F-s_*)&=0,\\
 \mathcal A_{\rho\rho}-\frac{\mathcal A_\rho}\rho+
 \gamma_B^2(1-\mathcal A)\sin^2F&=0.\label{eq:Beom}
\end{align}
Varying $F$ and $\mathcal A$ gives the two radial equations above. We solve them with the north and south boundary conditions separately. Figure~\ref{fig:meronprofiles} shows the resulting orientational profiles and enclosed fluxes at $\gamma_B=0.5$.
 At the critical coupling $\gamma_B=1$, the radial action can be written as a sum of squares plus a boundary term. Solutions of the first-order Bogomolny equations saturate the bound, giving the north- and south-core actions
\begin{equation}
 S_N=4\pi G_B(1-s_*),\qquad S_S=4\pi G_B(1+s_*).
 \label{eq:Bbps}
\end{equation}
Here $S_N$ and $S_S$ are the north- and south-core actions. The other values in Table~\ref{tab:Bactions} are obtained from the second-order equations away from this BPS coupling.
\begin{table}[tb]\centering
\begin{tabular}{rrrr}\toprule
$\gamma_B$ & Phases B5 and D5, each core & Phase B4, north core & Phase B4, south core\\\midrule
0.5 & 2.52034154 & 1.30605986 & 3.57252040\\
1 & 2.00000000 & 1.00000000 & 3.00000000\\
2 & 1.58407386 & 0.77217189 & 2.46785199\\\bottomrule
\end{tabular}
\caption{Meron actions $S_{\rm loc}/(2\pi G_B)$. The example in phase B4 has $s_*=1/2$. Symmetric cores in phases B5 and D5 have the same action.}\label{tab:Bactions}
\end{table}
For a normalization example, the reference wall at $g=2$ has $K_g=0.35355339$ and $\kappa_{\rm YM}=0.01473139$. With $m_B=0.1$, Eq.~\eqref{eq:GB} gives $G_B=7.10249478$ and $\lambda_B=3.14269681$. Choosing $\mu=0.005$ gives $b=-0.03$, $a=7.53333\times10^{-5}$, and $c_B=1.77562\times10^{-4}$. The example in phase B4 then has $h=2.39793\times10^{-5}$. For these parameters, the ratios $\mu/m_B=0.05$, $\lambda_B\mu^2/G_B=1.11\times10^{-5}$, and $|C_s|\mu^2/G_B=6.17\times10^{-4}$ quantify scales within the retained expansion.

A north-core and a south-core meron with opposite vortex windings have zero net flux and unit total lump charge. By section~\ref{sub:string}, the pair carries one unit of bulk Yang--Mills instanton charge. Appendix~\ref{app:charge} gives the relation between the geometric and Yang--Mills instanton charges of the individual gauged merons.
In Table~\ref{tab:Botherhierarchy}(b), the $\mathbb{C}P^1$ kink is the domain wall in the interface theory, and a three-dimensional Skyrmion trapped on it appears as a baby Skyrmion on the wall~\cite{nittaWallSkyrmion2013,nittaMatryoshka2013,gudnasonWallSkyrmions2014,incskyrmion,etoNittaNonAbelianSG2015}. In Table~\ref{tab:Botherhierarchy}(c), the mother vortex and the daughter baby Skyrmion have the same dimension on the Euclidean electroweak string world sheet. Each meron carries a fraction of the baby Skyrmion charge, and its core polarity is represented by an Ising spin. See Appendix~A.2 of Ref.~\cite{relations2022}.
\begin{table}[!htbp]
\centering\small
\renewcommand{\arraystretch}{1.18}
\medskip
\noindent\makebox[\linewidth][l]{\textbf{(a)} Soliton hierarchy from the bulk}\par\smallskip
\begin{tabular}{>{\raggedright\arraybackslash}p{3.12cm}|>{\raggedright\arraybackslash}p{2.4cm}>{\raggedright\arraybackslash}p{7.48cm}}
\hline
Mother \textbackslash{} Daughter & EW string & YM instanton\\\hline
Interface & Kink & Skyrmion\\
EW string & --- & Baby Skyrmion (B2, B3) or meron (B4, B5, D5)\\\hline
\end{tabular}\par
\medskip
\noindent\makebox[\linewidth][l]{\textbf{(b)} Soliton hierarchy within the interface}\par\smallskip
\begin{tabular}{>{\raggedright\arraybackslash}p{3.12cm}|>{\raggedright\arraybackslash}p{10.32cm}}
\hline
Mother \textbackslash{} Daughter & 3D Skyrmion\\\hline
$\mathbb{C}P^1$ kink & Baby Skyrmion (B2, B3) or meron (B4, B5, D5)\\\hline
\end{tabular}\par
\medskip
\noindent\makebox[\linewidth][l]{\textbf{(c)} Soliton hierarchy on the string world sheet}\par\smallskip
\begin{tabular}{>{\raggedright\arraybackslash}p{3.12cm}|>{\raggedright\arraybackslash}p{10.32cm}}
\hline
Mother \textbackslash{} Daughter & Baby Skyrmion\\\hline
Local vortex & \leavevmode{\begin{tabular}[t]{@{}l@{\hspace{0.4em}}l@{}}Ising spin: & Symmetric fractions $1/2,\,1/2$ (B5, D5)\\ & Asymmetric fractions $q_N,\,q_S$ with $q_N+q_S=1$ (B4)\end{tabular}}\\\hline
\end{tabular}\par
\caption{String configurations in phases B2--B5 and D5. The bulk string is fractional $Z$ in phases B2 and B3, mixed $Z$--$W$ in phase B4, and pure $W$ in phases B5 and D5.  In (b), the $\mathbb{C}P^1$ kink is the domain wall that hosts the Skyrmion. In (c), the mother vortex and the daughter baby Skyrmion have the same codimension on the string world sheet. The fractions in (c) are geometric baby Skyrmion charges. EW denotes electroweak.}\label{tab:Botherhierarchy}
\end{table}

\FloatBarrier
\Needspace{5\baselineskip}
\section{Instantons on superconducting-wall vortices: C phase}\label{sec:C}\suppressfloats[t]
In phase C, the charged condensate $c$ makes the interface superconducting. A vortex in this condensate carries electromagnetic flux, and its core is filled by the neutral component $\eta$. The phase of $\eta$ generally has a potential in the vortex core. In phases C1--C5, this potential supports stable instanton configurations of finite width in Euclidean time in the core effective theory. In the B1, D1, and E phases, the monopole-core phase at $g'=0$ has a flat potential. Its winding spreads on the noncompact Euclidean time axis and minimizes its action by winding uniformly on a circle of fixed period.
Section~\ref{sub:Cvortex} constructs the vortex on the superconducting interface. Section~\ref{sec:DSG} derives the double sine-Gordon theory of the neutral core phase and introduces the classification into subphases C1--C6. Section~\ref{sec:Cclassification} presents the corresponding kink solutions, computes their actions and topological charges, and tests selected solutions by relaxing the vortex profile.

\subsection{\texorpdfstring{The gauged vortex and its neutral core phase}{The gauged vortex and its neutral core phase}}\label{sub:Cvortex}
The phase of $c$ winds around a vortex on the superconducting interface and couples to the electromagnetic field. The neutral component $\eta$ fills the core, as in composite defects in gauged $O(N)$ models~\cite{EHN2023composite}. The superconducting condensate screens magnetic fields through the Meissner effect, while magnetic flux penetrates the vortex core. To describe this screening in the vortex calculation, we use a local Maxwell term on the interface with effective coupling $e_2$. The bulk on either side retains a massless photon and supports the stray field of the vortex, as in a superconducting film~\cite{Pearl1964}. The local Maxwell term models screening near the core. The bulk photon contributes a long-range response that is absent from this action. The Euclidean interface action is
\begin{equation}
 \begin{aligned}
 S_{w,\rm loc}&=\int\dd^3x\left[
 K_g\{(\partial u_0)^2+(\partial u_3)^2+|D c|^2\}
 +V_w+\frac{\mathfrak f_{ab}\mathfrak f_{ab}}{4e_2^2}\right],\\
 D_a c&\equiv(\partial_a-i\mathfrak a_a)c,\qquad
 \mathfrak f_{ab}\mathrel{\equiv}\partial_a\mathfrak a_b-\partial_b\mathfrak a_a .
 \end{aligned}\label{eq:Clocal}
\end{equation}
Here $a,b=0,1,2$ label the Euclidean interface coordinates, and $\mathfrak a=\mathfrak a_a\dd x^a$ is the electromagnetic connection restricted to the interface.

We first solve for the host vortex in phase C6, the symmetric point $a=b=a_0>0$ and $h=0$ of the interface potential in Eq.~\eqref{eq:wallpotential}. At this point the core has a continuous neutral phase $\beta$. In spatial polar coordinates $(r,\vartheta)$ on the interface, an axially symmetric vortex of winding $q$ is described by
\begin{equation}
 c=\cos F(r)\,e^{iq\vartheta},\qquad
 \eta=\sin F(r)\,e^{i\beta},\qquad
 \mathfrak a=q\,\mathcal A(\varrho)\dd\vartheta,\qquad
 m_0^2\mathrel{\equiv}\frac{Ia_0}{K_g},\quad
 \varrho\mathrel{\equiv}m_0r .
 \label{eq:Cansatz}
\end{equation}
The profile obeys $F(0)=\pi/2$, $F(\infty)=0$, $\mathcal A(0)=0$, and $\mathcal A(\infty)=1$. These conditions give a neutral condensate in the vortex core. A similar core structure occurs in local baryonic vortices in dense QCD, where the charged pion condensate winds around a core filled by the neutral pion field~\cite{qiu,mamedaNittaQiu2026}. At large radius the electromagnetic field screens the angular gradient of $c$. If $e$ is the charge of $c$ and $\mathfrak a=eA_{\rm em}$, the electromagnetic flux is $2\pi q/e$. The planar gauge mass $m_A$ and its ratio to $m_0$ are
\begin{equation}
 m_A=\sqrt{2e_2^2K_g},\qquad \gamma\equiv\frac{m_A}{m_0}.
\end{equation}
For a unit vortex ($q=1$), the energy is
\begin{equation}
 \frac{E_v}{2\pi K_g}=\int_0^\infty\dd\varrho\left[
 \varrho F_{\varrho}^2+
 \frac{(1-\mathcal A)^2\cos^2F}{\varrho}+\varrho\sin^2F+
 \frac{\mathcal A_{\varrho}^2}{\gamma^2\varrho}\right].
 \label{eq:CstaticE}
\end{equation}
Varying $F$ and $\mathcal A$ gives
\begin{align}
 F_{\varrho\varrho}+\frac{F_{\varrho}}\varrho+
 \frac{(1-\mathcal A)^2\sin F\cos F}{\varrho^2}-\sin F\cos F&=0,\\
 \mathcal A_{\varrho\varrho}-\frac{\mathcal A_\varrho}\varrho+
 \gamma^2\cos^2F(1-\mathcal A)&=0.
 \label{eq:Ceom}
\end{align}

\begin{figure}[tb]\centering
\includegraphics[width=.96\textwidth]{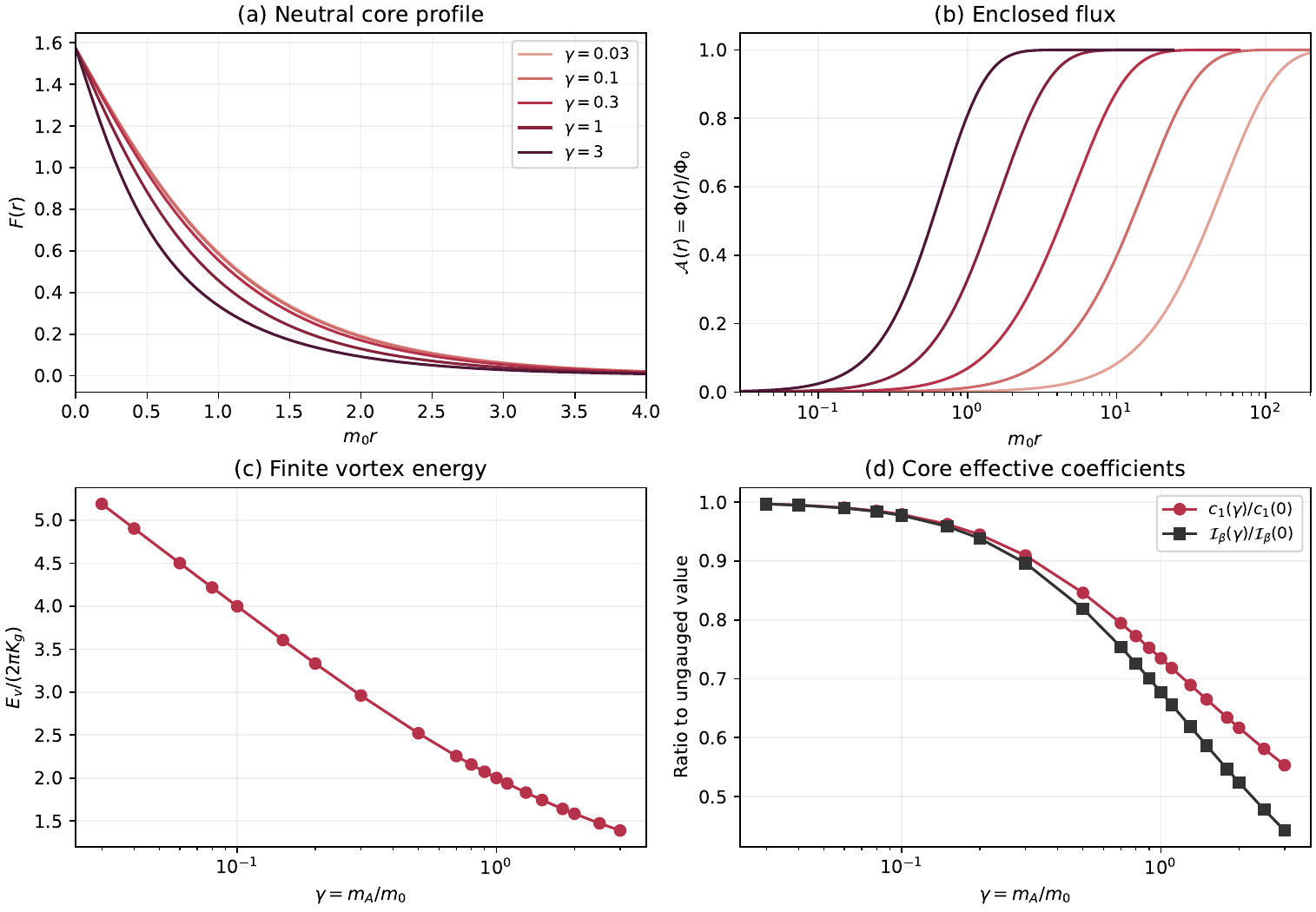}
\caption{Gauged vortex on the superconducting interface. The panels show the neutral core profile, the enclosed magnetic flux, the energy, and the core moments normalized by their $\gamma=0$ values.}\label{fig:Cstatic}
\end{figure}

The radial profiles determine the dimensionless core integrals
\begin{equation}
 c_n(\gamma)\equiv\int_0^\infty\varrho\,\sin^nF\,\dd\varrho,\qquad n=1,2,
 \label{eq:Cmoments}
\end{equation}
Table~\ref{tab:Cstatic} lists the resulting core integrals, and Fig.~\ref{fig:Cstatic} shows the radial profiles. For $\gamma>0$, the vortex has finite energy in the planar Maxwell approximation. The core and magnetic screening scales can differ greatly. At $\gamma=0.03$, the core radius defined by $F(r_c)=\pi/4$ is approximately $0.7405/m_0$, while the radius enclosing half of the magnetic flux is approximately $41.92/m_0$. At the critical coupling $\gamma=1$, the Bogomolny bound gives $E_v/(2\pi K_g)=2$. The $\gamma=0$ row provides the reference values used in the ungauged calculations of section~\ref{sec:Cnumerics}.

\begin{table}[tb]\centering\small
\begin{tabular}{rrrrrr}\toprule
$\gamma$ & $m_0r_c$ & $E_v/(2\pi K_g)$ & $c_1$ & $c_2$ & $c_1/c_2$\\\midrule
0 & 0.74174 & logarithmic & 1.36109 & 0.50000 & 2.72217\\
0.03 & 0.74054 & 5.190037 & 1.35702 & 0.49843 & 2.72262\\
0.1 & 0.73292 & 3.998131 & 1.33281 & 0.48872 & 2.72714\\
0.3 & 0.69901 & 2.959683 & 1.23779 & 0.44826 & 2.76136\\
1 & 0.59100 & 2.000000 & 1.00000 & 0.33865 & 2.95292\\
3 & 0.44105 & 1.388479 & 0.75304 & 0.22117 & 3.40481\\\bottomrule
\end{tabular}
\caption{Radial solutions for a unit vortex on the superconducting interface in the planar Maxwell approximation. The core radius satisfies $F(r_c)=\pi/4$. The core integrals $c_1$ and $c_2$ are defined in Eq.~\eqref{eq:Cmoments}. The $\gamma=0$ row gives the ungauged reference values.}\label{tab:Cstatic}
\end{table}

\subsection{\texorpdfstring{The double sine-Gordon theory and classification of core instantons}{The double sine-Gordon theory and classification of core instantons}}\label{sec:DSG}
At $a=b$ and $h=0$, the rotation $\eta\to e^{i\epsilon}\eta$ leaves $c$ and the electromagnetic profile fixed. To obtain the leading dynamics of the neutral core phase, let $\beta$ vary slowly with Euclidean time and integrate over the spatial vortex profile. The world-line Lagrangian and the inertia are
\begin{equation}
 \mathcal L_{\rm eff}=\frac{\mathcal I_\beta(\gamma)}{2}(\partial_\tau\beta)^2,
 \qquad
 \mathcal I_\beta(\gamma)=4\pi K_g\int_0^\infty r\,\dd r\,\sin^2F(r)
 =\frac{4\pi K_g}{m_0^2}c_2(\gamma).
 \label{eq:Ckinetic}
\end{equation}
The neutral variation $\delta\eta=i\sin F\,e^{i\beta}\delta\beta$ decays exponentially away from the core, so the inertia is finite. The phase motion generates no electromagnetic charge density at this order.

\begin{figure}[!tbp]\centering
\begingroup\small
\begin{tabular}{cl>{\raggedright\arraybackslash}p{4.2cm}>{\raggedright\arraybackslash}p{4.2cm}}\toprule
Region & Conditions & Core minima & Instanton structure\\\midrule
C1 & $A<0$, $B=0$ & $\pi/2$, $3\pi/2$ & Deconfined symmetric halves\\
C2 & $A<0$, $0<B<-4A$ & Two degenerate minima & Deconfined unequal fractions\\
C3 & $B>4|A|$ & One minimum modulo $2\pi$ & Single-peak unit event\\
C4 & $A>0$, $0<B<4A$ & True and higher local minimum & Confined two-peak unit molecule\\
C5 & $A>0$, $B=0$ & $0$, $\pi$ & Deconfined symmetric halves\\
C6 & $A=B=0$ & Full core circle & Free winding (uniform on a time circle)\\\bottomrule
\end{tabular}
\captionof{table}{Leading core DSG classification. ``Deconfined'' means that an isolated fractional event has finite relative action between degenerate static core states. In phase C4, the constituents form a unit molecule.}\label{tab:Cphases}
\par\endgroup
\medskip
\includegraphics[width=.98\textwidth]{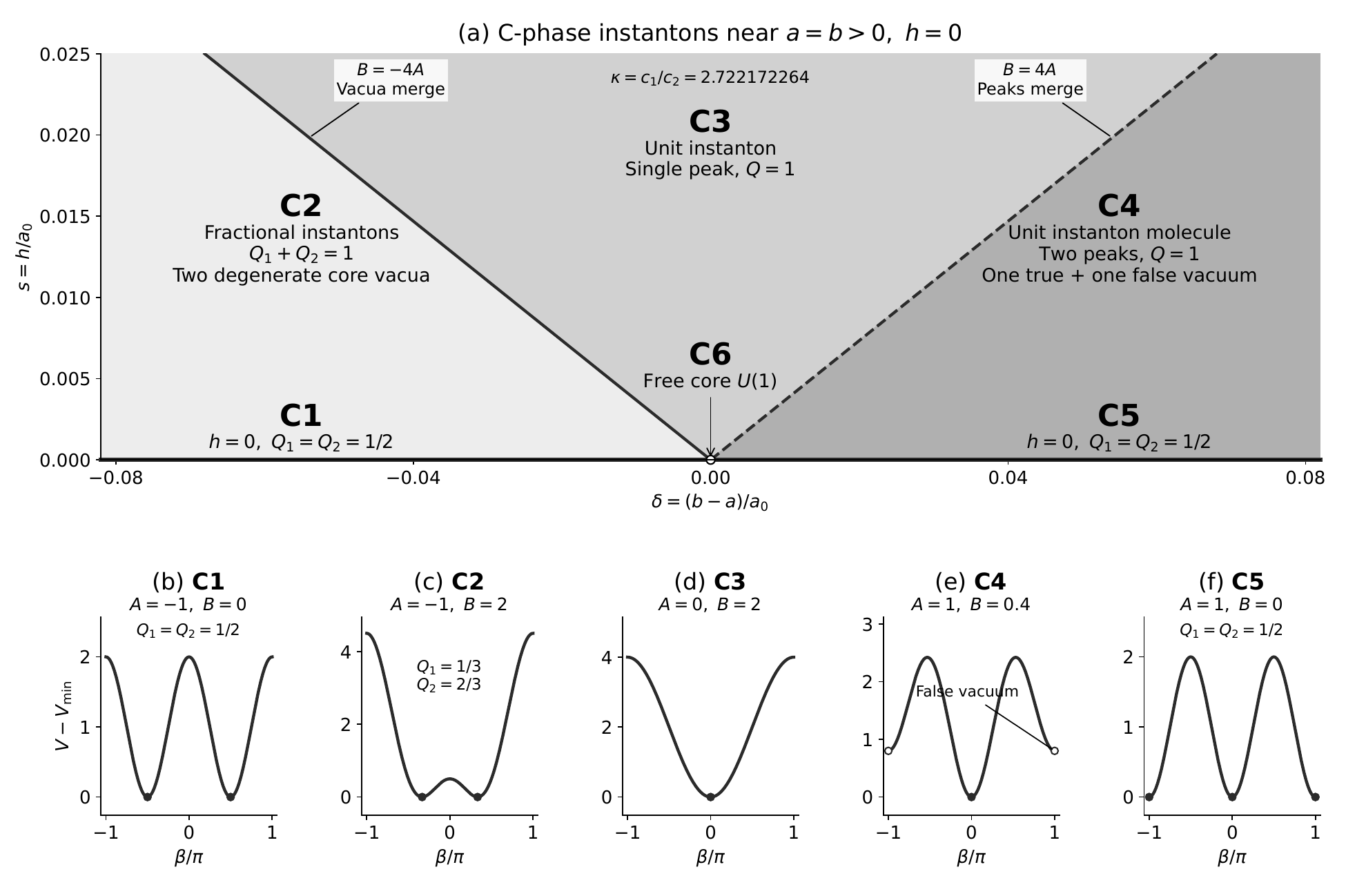}
\caption{C-phase core classification in the leading DSG theory. The upper plot uses $\delta=(b-a)/a_0$ and $s\mathrel{\equiv}h/a_0$. The origin is phase C6 with a free neutral core phase. The lower panels show representative potentials for subphases C1--C5, including both symmetric $h=0$ limits. Filled and open markers identify true and false minima. The solid boundary merges the two minima of phase C2; the dashed boundary merges the two density peaks of phase C4. Their straight lines use the ungauged reference moments and are replaced by Eq.~\eqref{eq:kappaC} when the planar vortex is gauged.}\label{fig:Cphase}
\end{figure}

Let $a=a_0$, $b=a_0(1+\delta)$, and $h=a_0s$, with $|\delta|,s\ll1$. The anisotropy $\delta$ and soft mixing $s$ lift the neutral core circle. Integrating their potential over the reference vortex gives the leading core action
\begin{equation}
\begin{aligned}
 \Delta S_{\rm core}&=\int\dd\tau\left[\frac{\mathcal I_\beta}{2}(\beta')^2+
 V(\beta)-V_{\min}\right],\\
 V&=A(1-\cos2\beta)+B(1-\cos\beta),\qquad B\geq0.
\end{aligned}
\label{eq:DSGaction}
\end{equation}
The coefficients of the second and first harmonics are
\begin{equation}
 A=\pi K_gc_2(\gamma)\delta,\qquad
 B=4\pi K_gc_1(\gamma)s.\label{eq:DSGcoeff}
\end{equation}
Here $\beta'\equiv\dd\beta/\dd\tau$ and $V_{\min}$ is the minimum of $V$. The relative action subtracts the static true-core vortex action. We choose the origin of $\beta$ so that $B\geq0$. The coefficients use the reference vortex at $a=b$ and $h=0$ to first order in the potential deformation. At finite $h$, the asymptotic vacuum of the spatial vortex is held fixed.\footnote{A trial configuration with this boundary condition is $\eta=s+[e^{i\beta(\tau)}-s]T(r)$, with $T(r)=[\sin F_h(r)-s]/(1-s)$ and $c=\sqrt{1-|\eta|^2}\,e^{i\vartheta}$. Here $\eta=\sin F_h(r)$ is a static real-positive core profile. Since $T(0)=1$ and $T(\infty)=0$, the core follows $e^{i\beta(\tau)}$ and the far field stays at $s$.}

In phase C6 ($A=B=0$), the potential vanishes and the vortex-core phase is free. On a Euclidean time circle of fixed period, uniform winding is stable in the phase theory because it minimizes the action at fixed winding. On the noncompact line, the winding spreads. Appendix~\ref{app:rotor} derives these results and gives a constrained configuration of prescribed temporal width.

The stationary condition is $V'=\sin\beta(4A\cos\beta+B)=0$. Table~\ref{tab:Cphases} and Fig.~\ref{fig:Cphase} summarize the resulting core minima and kink structures. The two symmetric half-instanton limits in phases C1 and C5 are exchanged by $\beta\mapsto\beta+\pi/2$ at $B=0$. Their continuation to $B>0$ is different and motivates keeping both labels.
In the original weak-deformation parameters, the boundaries between phases C2 and C3 and between phases C3 and C4 are $\delta=-\kappa_\gamma s$ and $\delta=+\kappa_\gamma s$, respectively, where
\begin{equation}
 \kappa_\gamma=\frac{c_1(\gamma)}{c_2(\gamma)}.\label{eq:kappaC}
\end{equation}
The first is a merger of two degenerate core vacua; the second is a merger of two kink-density peaks, with no massless true-vacuum mode. The diagram in Fig.~\ref{fig:Cphase} uses the ungauged reference $\kappa_0=2.722172264$; at $\gamma=0.3$ the coefficient is $2.76136$.

\subsection{\texorpdfstring{Kink solutions in the vortex core}{Kink solutions in the vortex core}}\label{sec:Cclassification}
The Euler--Lagrange equation and its first integral for a finite-action path between degenerate core minima are
\begin{equation}
 \mathcal I_\beta\beta''=2A\sin2\beta+B\sin\beta,\qquad
 \frac{\mathcal I_\beta}{2}(\beta')^2=V(\beta)-V_{\min}.
 \label{eq:DSGfirst}
\end{equation}
The first integral follows by multiplying the equation of motion by $\beta'$ and using $\beta'\to0$ and $V\to V_{\min}$ at both endpoints. It equates the kinetic and potential contributions along the kink. The same first-integral method was used for double sine-Gordon kinks on magnetic domain walls in Ref.~\cite{amari2023}. Changing the integration variable from $\tau$ to $\beta$ gives the action directly from the path in phase space
\begin{equation}
 S_{\rm path}=\int_{\rm path}|\dd\beta|\,
 \sqrt{2\mathcal I_\beta[V(\beta)-V_{\min}]}.\label{eq:DSGquad}
\end{equation}

With the asymptotic spatial vortex held fixed, the charge of the kink is the relative interface Skyrmion number $k_{\rm rel}$. For the reference vortex with $\beta=\beta(\tau)$, the pullback of the $S^3$ volume form gives
\begin{equation}
 k_{\rm rel}=\frac{q}{2\pi^2}\int r\dd r\dd\vartheta\dd\tau\,
 \frac{-F_r\sin F\cos F}{r}\beta'
 =\frac{q\Delta\beta}{2\pi}.\label{eq:Ccharge}
\end{equation}
The radial integral is $1/2$. A unit vortex and a full core turn therefore have $|k_{\rm rel}|=1$. The connection in Eq.~\eqref{eq:uplift} gives $Q_{\rm YM,rel}=-k_{\rm rel}$ at the ungauged reference.

At $g'=0$, exchanging $\eta$ and $c$ maps the winding construction at the symmetric point in phase C6 to that in the E phase given in Eq.~\eqref{eq:Eansatz}. At finite $g'$, the core phase is electrically neutral in the C phase and electrically charged in the E phase.

\paragraph{Symmetric halves in phases C1 and C5.}
Sine-Gordon kinks also describe Skyrmions on magnetic domain walls~\cite{ross2022}. In phase C5, $A>0$ and $B=0$, a half-period sine-Gordon kink is
\begin{equation}
 \beta(\tau)=2\arctan e^{2\sqrt{A/\mathcal I_\beta}(\tau-\tau_0)},
 \qquad S_{1/2}=4\sqrt{\mathcal I_\beta A}.\label{eq:Chalf}
\end{equation}
It connects $0$ to $\pi$, with center $\tau_0$. In phase C1, replace $A$ by $|A|$ and shift $\beta$ by $\pi/2$; the endpoints are then $\pi/2$ and $3\pi/2$. In each case $\Delta\beta=\pi$, giving $k_{\rm rel}=1/2$ for a unit spatial vortex by Eq.~\eqref{eq:Ccharge}. Two same-orientation half events supply a unit turn, but the first integral reaches the intermediate degenerate vacuum only at infinite time. The two halves are separated by an arbitrarily long interval in the intermediate vacuum.

\begin{figure}[tb]\centering
\includegraphics[width=.97\textwidth]{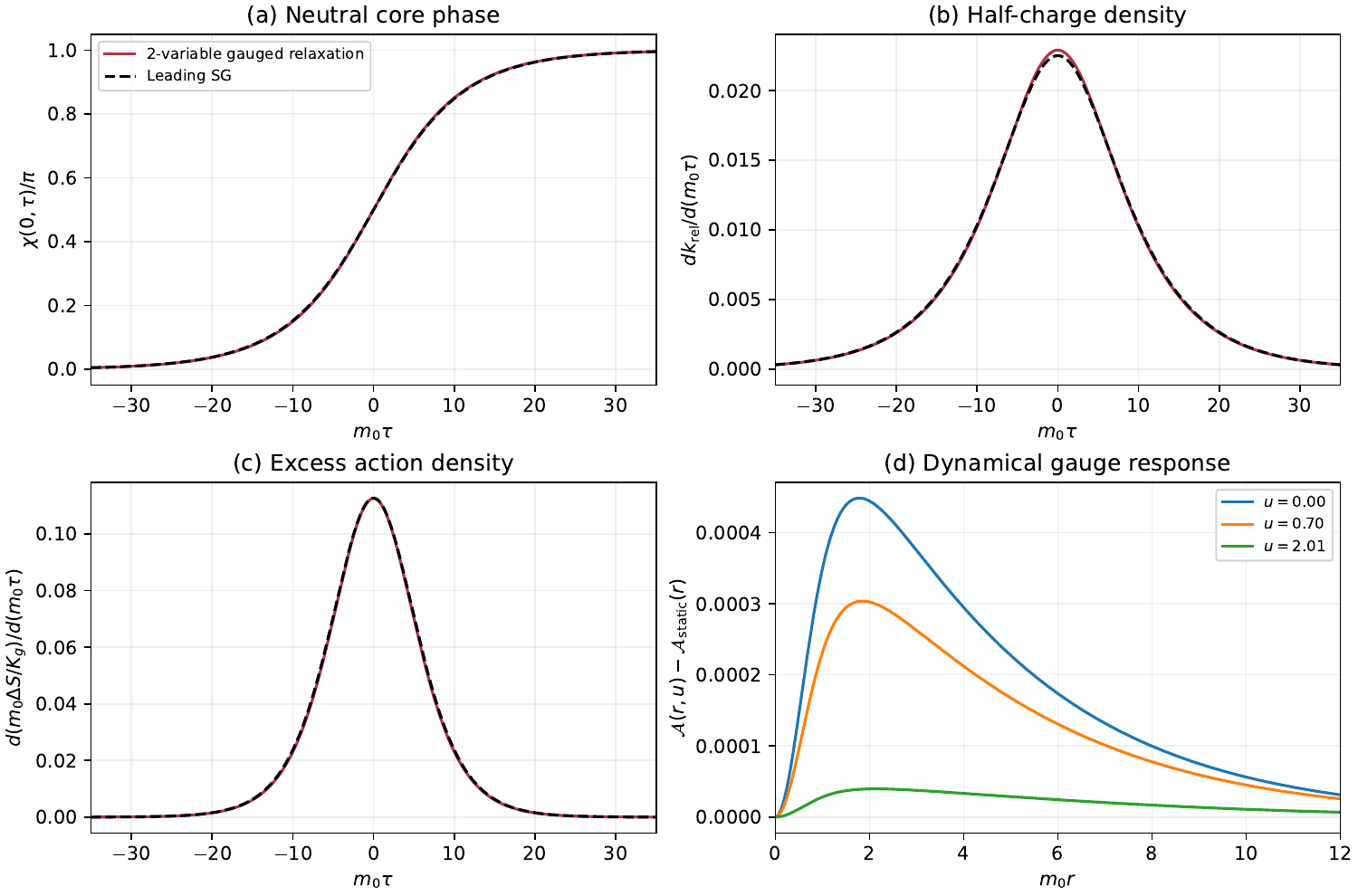}
\caption{Half event on a non-BPS local vortex in phase C at $\gamma=0.3$, $\delta=0.02$, $h=0$. The relaxed solution is compared with the SG theory using the gauged core coefficients. The panels show the core phase, relative degree density, relative action density, and departure from the static gauge profile. The displayed time coordinate is $u=\sqrt\delta\,m_0\tau$.}\label{fig:Cgaugedhalf}
\end{figure}

At $\gamma=0.3$, $h=0$, and $\delta=0.02$, we also solve for the half instanton in phase C5 with the vortex cross section and electromagnetic gauge profile allowed to vary in Euclidean time. In the local planar gauge theory, the relaxed action is $m_0\Delta S/K_g\simeq1.5740$, about $1.21\%$ below the sine-Gordon prediction $1.5932376$ computed with the corresponding core coefficients. The relative degree on the finest grid is $0.50000053$. Figure~\ref{fig:Cgaugedhalf} compares the profiles. Appendix~\ref{app:numerics} describes the relaxation.

\paragraph{Deconfined unequal fractions in phase C2.}
Set $C\equiv|A|$ and $k\equiv B/(4C)$, with $0<k<1$. The two minima are $\beta=\pm\beta_0$, $\beta_0=\arccos k$, and
\begin{equation}
 V-V_{\min}=2C(\cos\beta-k)^2.
\end{equation}
With $\mu_C=\sqrt{C/\mathcal I_\beta}\sqrt{1-k^2}$, the two increasing paths are
\begin{align}
 \beta_1&=2\arctan\left[\sqrt{\frac{1-k}{1+k}}\tanh\mu_C(\tau-\tau_1)\right],\\
 \beta_2&=\pi+2\arctan\left[\sqrt{\frac{1+k}{1-k}}\tanh\mu_C(\tau-\tau_2)\right],
 \label{eq:Cfractional}
\end{align}
connecting $-\beta_0$ to $\beta_0$ and $\beta_0$ to $2\pi-\beta_0$, respectively. Their arbitrary centers are $\tau_1$ and $\tau_2$. Evaluating Eq.~\eqref{eq:DSGquad} along the two arcs gives the actions below. The leading phase charges and actions are
\begin{align}
 Q_1^{(0)}&=\frac{\beta_0}\pi,\qquad Q_2^{(0)}=1-Q_1^{(0)},
\end{align}
\begin{equation}
 S_1=4\sqrt{\mathcal I_\beta C}(\sin\beta_0-k\beta_0),\qquad
 S_2=4\sqrt{\mathcal I_\beta C}[\sin\beta_0+k(\pi-\beta_0)].\label{eq:Cfracaction}
\end{equation}
For $k=1/2$, the leading charges are $1/3$ and $2/3$. The symmetry $\beta\mapsto-\beta$ keeps the endpoints degenerate at nonzero $h$. There is no false-core energy cost proportional to the separation of successive events. This is the sense in which the fractional events are deconfined.

\begin{figure}[tb]\centering
\includegraphics[width=.97\textwidth]{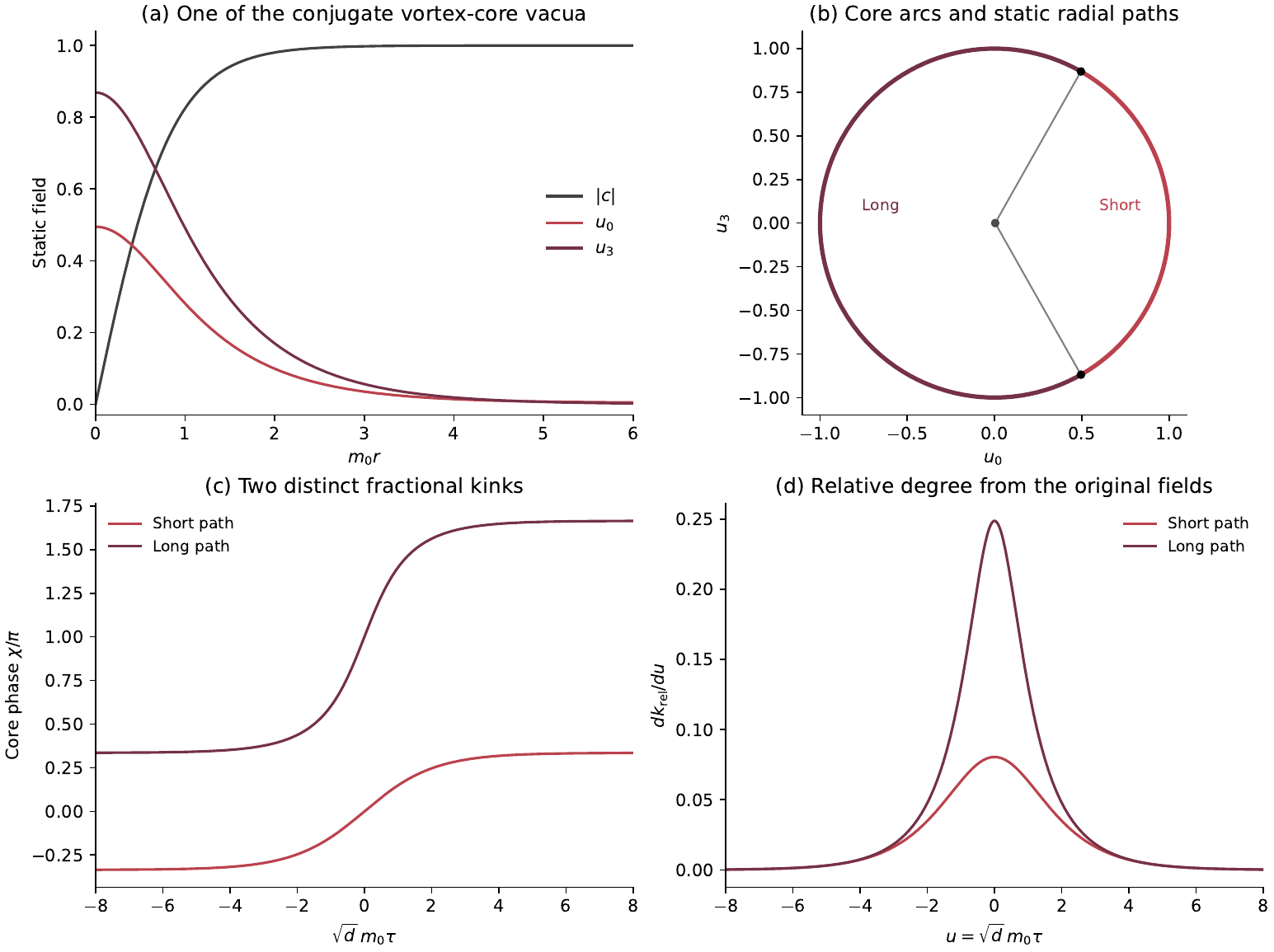}
\caption{Deconfined asymmetric fractional events in phase C2. The two-variable profiles use the ungauged parent theory at the representative point of Table~\ref{tab:Ctests}. The neutral core arc and static radial boundary paths both contribute to the relative degree. The integrated short and long values are approximately $0.33446$ and $0.66554$.}\label{fig:C2}
\end{figure}

At finite $h$, the static neutral-field profiles at the two temporal endpoints also contribute to the relative degree. Their radial paths contribute with opposite signs to the short and long events, so their charges still sum to one. For $\delta=-0.02$ and $s=0.003673536804$, the leading phase theory gives $1/3$ and $2/3$, while the radial boundary calculation gives
\begin{equation}
 k_{\rm short}=0.33445757,\qquad k_{\rm long}=0.66554243.
 \label{eq:Cfractionalnumbers}
\end{equation}
The correction in Eq.~\eqref{eq:Cboundary} follows from the area enclosed by the core arc and the two static radial paths. Figure~\ref{fig:C2} shows those paths and the charge densities obtained by relaxing the vortex cross section in the ungauged interface theory.

\paragraph{A unit event and its confined molecule in phases C3 and C4.}
For $B>0$ and $B+4A>0$, a unit-turn solution is
\begin{equation}
 \beta=\pi+2\arctan\left[\sqrt{\frac{B}{B+4A}}\sinh\Lambda(\tau-\tau_0)\right],
 \qquad \Lambda=\sqrt{\frac{B+4A}{\mathcal I_\beta}}.\label{eq:Cunit}
\end{equation}
It connects $0$ to $2\pi$. Both phases C3 and C4 have unit total relative degree for a unit spatial vortex by Eq.~\eqref{eq:Ccharge}. When $A>0$ and $B<4A$, the intermediate $\pi$ core is a higher local minimum. The two action-density peaks have separation
\begin{equation}
 d_\tau=\frac2\Lambda\operatorname{arsinh}\sqrt{\frac{4A}{B}-1}.
 \label{eq:Cseparation}
\end{equation}
A long interval in the higher core costs $2B\,d_\tau$ in the leading world-line action. Thus the constituents are confined into a unit molecule. Sending $B\to0^+$ removes this cost and makes the separation diverge, recovering phase C5. At $B=4A$ the peaks merge; beyond it the event in phase C3 has a single peak. The $A<0$, $B>-4A$ side also belongs to phase C3.

For the unit path, the vacuum-subtracted potential is $V-V_{\min}=2\sin^2(\beta/2)[B+4A\cos^2(\beta/2)]$. With $u=\cos(\beta/2)$, Eq.~\eqref{eq:DSGquad} gives
\[
 S_{\rm unit}=8\sqrt{\mathcal I_\beta}\int_0^1\sqrt{B+4Au^2}\,\dd u .
\]
For $A>0$ and $B>0$, this evaluates to
\begin{equation}
 S_{\rm unit}=4\sqrt{\mathcal I_\beta}\left[\sqrt{B+4A}+
 \frac{B}{2\sqrt A}\operatorname{arsinh}\frac{2\sqrt A}{\sqrt B}\right].
 \label{eq:Cunitaction}
\end{equation}
For $A=-C<0$ and $B>4C$, the same integral gives
\[
 S_{\rm unit}=4\sqrt{\mathcal I_\beta}\left[\sqrt{B-4C}+
 \frac{B}{2\sqrt C}\arcsin\frac{2\sqrt C}{\sqrt B}\right].
\]
Both expressions approach $8\sqrt{\mathcal I_\beta B}$ as $A\to0$. For $A>0$, the $B\to0^+$ limit is the sum of two half actions.
DSG kinks and their shape changes also occur in related wall systems~\cite{ekn,rossDSG,amarishort}; here both harmonics and the inertia follow from the inert-interface vortex.

\begin{figure}[tb]\centering
\includegraphics[width=.97\textwidth]{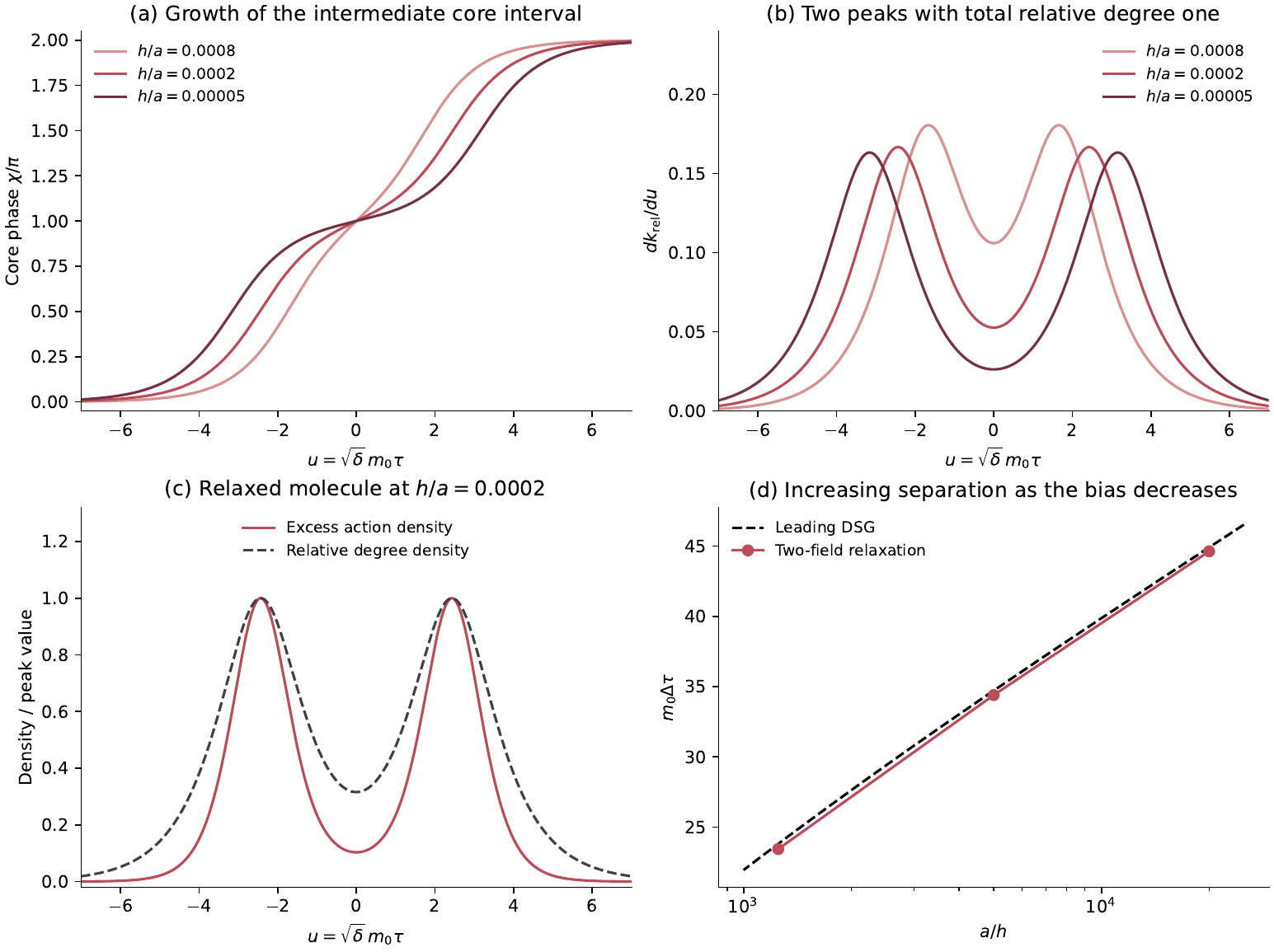}
\caption{Confined fractional-instanton molecule in phase C4. The two-variable parent-interface configurations have total relative degree one and retain two density peaks. Their separation increases as the soft mixing decreases. The intervening higher-energy vortex core supplies the linear separation cost; the $h\to0$ limit approaches the deconfined limit in phase C5.}\label{fig:C4}
\end{figure}

The two peaks in phase C4 persist when the cross section is relaxed in the ungauged interface theory. At $\delta=0.02$ and $s=0.0002$, their separation is $m_0d_\tau\simeq34.40$, compared with $34.72$ in the core theory. The peaks move farther apart as $s$ decreases toward phase C5, while their combined relative degree remains one. Figure~\ref{fig:C4} shows this behavior and Table~\ref{tab:C4scan} gives the separation and action at each test point.

At the boundary between phases C2 and C3, $A=-C$, $B=4C$, the unit event has an algebraic tail:
\begin{equation}
 \beta=\pi+2\arctan\left[2\sqrt{C/\mathcal I_\beta}(\tau-\tau_0)\right],
 \qquad S=4\pi\sqrt{\mathcal I_\beta C}.
\end{equation}
The small fractional charge tends to zero as the two minima of phase C2 merge.

The charge calculation identifies the successive descriptions collected in Table~\ref{tab:Chierarchy}: the same Euclidean configuration is a bulk instanton, an interface vortex Skyrmion, and a kink of the vortex-core phase. Related vortex Skyrmion constructions were studied in Skyrme and principal chiral models and in QCD~\cite{incskyrmion,torii,beads,qiu,nittaFractionalPCM2015}.
\begin{table}[!htbp]
\centering\small
\renewcommand{\arraystretch}{1.18}
\medskip
\noindent\makebox[\linewidth][l]{\textbf{(a)} Soliton hierarchy from the bulk}\par\smallskip
\begin{tabular}{>{\raggedright\arraybackslash}p{3.12cm}|>{\raggedright\arraybackslash}p{4.06cm}>{\raggedright\arraybackslash}p{5.82cm}}
\hline
Mother \textbackslash{} Daughter & EM magnetic flux & YM instanton\\\hline
Interface & Local vortex & Skyrmion\\
Local vortex & --- & Kink\\\hline
\end{tabular}\par
\medskip
\noindent\makebox[\linewidth][l]{\textbf{(b)} Soliton hierarchy within the interface}\par\smallskip
\begin{tabular}{>{\raggedright\arraybackslash}p{3.12cm}|>{\raggedright\arraybackslash}p{10.32cm}}
\hline
Mother \textbackslash{} Daughter & 3D Skyrmion\\\hline
Local vortex & Kink\\\hline
\end{tabular}\par
\caption{Soliton hierarchy in phase C. (a) The local vortex on the superconducting interface carries EM magnetic flux, and the bulk YM instanton appears as a Skyrmion on the interface. (b) On the local vortex, the three-dimensional interface Skyrmion appears as a kink of the neutral condensate phase along Euclidean time in phases C1--C5. In phase C6, it is a free winding of the same phase. EM denotes electromagnetic.}\label{tab:Chierarchy}
\end{table}

\FloatBarrier\section{Fractional instanton molecules: F and G phases}\label{sec:FG}\suppressfloats[t]
The vacuum spheres of phases F and G permit a different decomposition: an interface Skyrmion can separate into constituents with global-monopole structure. Section~\ref{sub:FGsphere} explains the vacuum geometry and the constituent core labels. Section~\ref{sub:FGfractions} then derives the symmetric half charges and their asymmetric deformation, relating the molecular construction to the bulk instanton charge.

\subsection{The vacuum sphere and monopole constituents}\label{sub:FGsphere}
The F and G phases have an $S^2$ vacuum set inside the $S^3$ target. Their instanton construction differs from a lump on a string or a core kink on a vortex: the three-dimensional Euclidean interface Skyrmion can split into global-monopole constituents. Fractional Skyrmions and their molecules in this type of potential were studied in Ref.~\cite{fractional2015}. We use that geometric mechanism with the potential and positive direct Skyrme contribution of the inert interface.

In phase F, define the local ratio $s=h/a$, with $0\leq s<1$. Subtracting the minimum gives
\begin{equation}
 V_F-V_{\min}=Ia(u_0-s)^2,\qquad b=0,\quad a>h.
 \label{eq:Fpotential}
\end{equation}
Phase F1 denotes $h=0$ and phase F2 denotes $0<h<a$. In phase G,
\begin{equation}
 V_G=Ib u_3^2,\qquad a=h=0,\quad b>0.
\end{equation}
In phase F1 the vacuum equator is $u_0=0$; in phase G it is $u_3=0$. Interchanging these target components and their coefficients maps the two symmetric constructions. A nonzero degree requires the cores to leave the vacuum sphere and approach opposite poles of $S^3$. A field restricted everywhere to the vacuum $S^2$ has zero pullback of the $S^3$ volume form.

The constituents are global monopoles of the interface orientation field in the three Euclidean coordinates $(\tau,x,y)$. These global monopoles and the bulk Nambu monopoles discussed in Section~\ref{sec:BD} are distinct types of defect. Their core polarity supplies an Ising label: the target field escapes toward the north or south pole. This same-dimensional decomposition does not add a new nested world-volume generation merely because the configuration has two constituents.

\subsection{Symmetric halves and asymmetric fractions}\label{sub:FGfractions}
The monopole number $m\in\pi_2(S^2)\simeq\mathbb Z$ measures the winding of the vacuum-sphere orientation around an interface point. In phase F1 the core polarity $\sigma=\pm1$ specifies $u_0=\pm1$; in phase G it specifies $u_3=\pm1$. With the orientation chosen consistently, the constituent charge is
\begin{equation}
 k_{\rm constituent}=\frac{m\sigma}{2}.
\end{equation}
The pair $(m,\sigma)=(+1,+1)$ and $(-1,-1)$ has vanishing net global-monopole number and unit total Skyrmion number. Each component contributes one half. The monopole sign and the core Ising spin are distinct labels.

In phase F2 the vacuum sphere shifts to $u_0=s$. Write $u_0=\cos f$ and the remaining components as $\sin f\,\bmN$, $\bmN\in S^2$. The vacuum latitude is $f_*=\arccos s$. A unit angular hedgehog gives the normalized cap charges
\begin{align}
 q_N&=\frac2\pi\int_0^{f_*}\sin^2 f\,\dd f
 =\frac{\arccos s-s\sqrt{1-s^2}}\pi,\\
 q_S&=1-q_N.\label{eq:Fcharges}
\end{align}
They reduce to $1/2+1/2$ at $s=0$ and approach $0+1$ as $s\to1^-$. At the endpoint the vacuum sphere collapses, so it is a limiting configuration of phase F2. Equation~\eqref{eq:Fcharges} is a geometric cap calculation; it is not a newly computed finite-$s$ molecular profile.

An isolated interface global monopole has an infrared-divergent action from its far-field angular gradient. Opposite monopole numbers allow a pair to return to a constant vacuum at infinity while retaining unit Skyrmion charge. The two- and four-derivative terms and the potential then provide the ingredients for a fractional Skyrmion molecule. The detailed equilibrium separation, action, and full-profile stability for the matched inert-interface coefficients have not been calculated here. We include the molecular charge construction as part of the phase organization without presenting it as a completed numerical solution.

For the reference gauge interpolation at $g'=0$, the degree-one molecule has a bulk Yang--Mills instanton charge of unit magnitude by Eq.~\eqref{eq:QYM}. Table~\ref{tab:FGhierarchy} summarizes how the global-monopole cores carry the Skyrmion charge in phases F and G.
\begin{table}[!htbp]
\centering\small
\renewcommand{\arraystretch}{1.18}
\medskip
\noindent\makebox[\linewidth][l]{\textbf{(a)} Soliton hierarchy from the bulk}\par\smallskip
\begin{tabular}{>{\raggedright\arraybackslash}p{3.12cm}|>{\raggedright\arraybackslash}p{10.32cm}}
\hline
Mother \textbackslash{} Daughter & YM instanton\\\hline
Interface & Skyrmion\\\hline
\end{tabular}\par
\medskip
\noindent\makebox[\linewidth][l]{\textbf{(b)} Mother and daughter of the same codimension}\par\smallskip
\begin{tabular}{>{\raggedright\arraybackslash}p{3.12cm}|>{\raggedright\arraybackslash}p{10.32cm}}
\hline
Mother \textbackslash{} Daughter & 3D Skyrmion\\\hline
Global monopole & \leavevmode{\begin{tabular}[t]{@{}l@{\hspace{0.4em}}l@{}}Ising spin: & Symmetric fractions $1/2,\,1/2$ (F1, G)\\ & Asymmetric fractions $q_N,\,q_S$ with $q_N+q_S=1$ (F2)\end{tabular}}\\\hline
\end{tabular}\par
\caption{Instanton charge in phases F and G. In (b), the global (anti-)monopole and the three-dimensional Skyrmion have the same codimension. Each fractional Skyrmion is represented by an Ising spin in the core of a global (anti-)monopole. See Ref.~\cite{fractional2015} and Appendix~A.3 of Ref.~\cite{relations2022}.}\label{tab:FGhierarchy}
\end{table}

\FloatBarrier\section{\texorpdfstring{Summary and discussion}{Summary and discussion}}\label{sec:discussion}\suppressfloats[t]
We have shown how electroweak instantons are realized on inert Higgs interfaces and on the defects they support. By constructing an explicit interpolation of the non-Abelian gauge connection, we have related the bulk Yang--Mills instanton charge to the Skyrmion charge on the Euclidean interface. We have derived the interface Skyrme term from the bulk Yang--Mills action and obtained the positive baby Skyrme term on the electroweak string by integrating over its transverse profile. These terms enter the Derrick estimates for the sizes of the interface Skyrmion and the baby Skyrmion on the string. In phases B6 and D6, the string orientation potential is flat and the full $S^2$ is degenerate.

\Needspace{3\baselineskip}
In the B1, D1, and E phases, a Nambu monopole has a free internal phase at $g'=0$. A full turn of this phase along Euclidean time carries one unit of Skyrmion charge on the interface. Spreading the winding over a longer time interval lowers its action while preserving its charge, so the winding spreads indefinitely on the noncompact Euclidean time axis. On a Euclidean time circle of fixed period, uniform winding minimizes the free-phase action in its winding sector. The same result holds for the free vortex-core phase in phase C6. Merons arise on the Euclidean world sheets of mixed $Z$--$W$ strings in phase B4 and pure $W$ strings in phases B5 and D5 (Table~\ref{tab:Botherhierarchy}). We have obtained individual radial meron solutions in the gauged string effective theory. In phase B4, the two merons have unequal geometric charges and actions. Including the magnetic-flux contribution for the explicit gauge interpolation gives each meron half a unit of Yang--Mills instanton charge in magnitude. The flux contributions cancel for a complementary pair, which carries unit total charge.

In phase C, we have derived a double sine-Gordon theory for the phase of the neutral condensate in the core of a local vortex. On the vortex, the interface Skyrmion appears as a kink of this phase in phases C1--C5 (Table~\ref{tab:Chierarchy}). The potential gives these kinks a finite width in Euclidean time. Phases C1 and C5 support symmetric half instantons, and phase C3 supports a single-peak unit instanton. In phase C2, the symmetry $\beta\mapsto-\beta$ preserves the degeneracy of the two core vacua at nonzero soft mixing, allowing two unequal fractional instantons to occur separately. In phase C4, the energy cost of the intermediate core grows with the separation of the constituents and binds them into a unit-charge molecule. We have obtained a half-instanton solution in phase C5 with the electromagnetic gauge profile allowed to relax. In phase C4, the two peaks persist when the vortex cross section is relaxed in the ungauged interface theory.

In phases F and G, the interface Skyrmion charge can be shared by two global-monopole constituents (Table~\ref{tab:FGhierarchy}). We have obtained equal geometric charges of $1/2$ in phases F1 and G and, in general, unequal geometric charges in phase F2. In each case the charges sum to one.

The nested configuration in phases B1 and D1 provides an electroweak realization of the domain-wall--vortex--monopole--instanton hierarchy of Ref.~\cite{relations2022}. The same Yang--Mills instanton appears as a winding of the Nambu monopole's internal phase along its Euclidean world line, as a baby Skyrmion on the electroweak string world sheet, and as a three-dimensional Skyrmion on the Euclidean interface (Table~\ref{tab:hierarchy}). The phase winding plays the role of the world-line kink in this hierarchy. The first steps of this electroweak construction were established in Ref.~\cite{interface2026}. A Nambu monopole appears as a kink on the electroweak string and as a global vortex on the interface. The electroweak string itself appears as a kink interpolating between two interface vacua. The present work completes the hierarchy with the instanton and extends it to include merons on electroweak strings, fractional instantons in local vortex cores, and Skyrmion charge shared by global-monopole constituents.
Table~\ref{tab:summary} summarizes the soliton configurations.
\par\medskip\noindent\begin{minipage}{\linewidth}\centering\small
\begin{tabular}{l>{\raggedright\arraybackslash}p{11.8cm}}\toprule
Phase & Interface or defect description\\\midrule
All phases & Skyrmion on the Euclidean interface\\
B1, D1 & Phase winding of a Nambu monopole confined by two half-unit $Z$-strings\\
B2, B3 & Baby Skyrmion on an electroweak string\\
B4 & Asymmetric merons on mixed $Z$--$W$ strings\\
B5, D5 & Symmetric merons on pure $W$ strings\\
B6, D6 & Lump on an electroweak string\\
C1, C5 & Half kinks in the core of a local vortex\\
C2 & Unequal fractional kinks in the core of a local vortex\\
C3 & Unit kink in the core of a local vortex\\
C4 & Two-peak kink in the core of a local vortex\\
C6 & Free neutral-phase winding in the core of a local vortex\\
E & Phase winding on a deconfined Nambu monopole\\
F1, G & Symmetric Skyrmion charge fractions on global monopoles ($1/2+1/2$)\\
F2 & Asymmetric Skyrmion charge fractions on global monopoles ($q_N+q_S=1$)\\\bottomrule
\end{tabular}
\captionof{table}{Soliton configurations carrying instanton charge in the interface and defect effective theories.}\label{tab:summary}
\end{minipage}\par\medskip

We conclude by discussing several directions for future work. A four-dimensional calculation with dynamical Higgs profiles and the electromagnetic field outside the interface would test the stability and localization of the configurations found here. In phase E, solving for the electromagnetic response at finite $g'$ would clarify the dynamics of the charged core phase. In phases F and G, computing the full molecular profiles would determine the equilibrium separation and action of the global-monopole pairs. The finite-width kinks in phase C also provide a starting point for calculating tunneling amplitudes between degenerate vortex-core states and assessing the validity of a dilute-instanton description.

A possible application is electroweak baryogenesis. If inert Higgs interfaces form during an electroweak phase transition, it would be interesting to determine whether the interfaces and their defects modify the rates of baryon-number-violating transitions relative to the surrounding plasma. An estimate of the resulting baryon asymmetry would require these rates at finite temperature, together with CP-violating sources and a treatment of transport and washout near a moving interface. The electromagnetic flux carried by the defects also raises the question of whether their formation and decay can contribute to the evolution of primordial magnetic fields.

The soliton configurations identified here provide a starting point for determining how spatial variations of the Higgs fields affect electroweak topological transitions.

\Needspace{7\baselineskip}
\acknowledgments
This work was supported in part by JSPS KAKENHI Grants No.~JP22H01221 and No.~JP23K22492, and by WPI-SKCM$^2$ at Hiroshima University.
The author used \mbox{OpenAI's} ChatGPT to help revise the manuscript and to assist with selected analytical derivations and numerical implementations. The author takes responsibility for the scientific content of the manuscript.

\appendix
\FloatBarrier\section{String coefficients and transverse-profile relaxation}\label{app:coefficients}\suppressfloats[t]
This appendix fixes the derivative and trace conventions used to obtain Eq.~\eqref{eq:GB}. For $U=\cos\theta+i\sin\theta\,\bmn\cdot\boldsymbol\sigma$ and $L=-\dd U U^{-1}$, the Pauli algebra gives
\begin{equation}
 -\Tr[L_x,L_\alpha]^2=8(\theta')^2\sin^2\theta(\partial_\alpha\bmn)^2,
 \qquad -\Tr[L_\tau,L_y]^2=8\sin^4\theta|\partial_\tau\bmn\times\partial_y\bmn|^2.
\end{equation}
The sum over ordered pairs doubles each independent pair. Using $\theta'=m_B\sin\theta$,
\begin{equation}
 \int\dd x\,\sin^2\theta=\frac2{m_B},\quad
 \int\dd x\,\sin^4\theta=\frac4{3m_B},\quad
 \int\dd x\,(\theta')^2\sin^2\theta=\frac{4m_B}3,
\end{equation}
gives $\Delta G_B=64\kappa_{\rm YM}m_B/3$ and $\lambda_B=64\kappa_{\rm YM}/(3m_B)$. Also $\int\sin\theta\dd x=\pi/m_B$, which fixes $H$ in Eq.~\eqref{eq:orientpotential}.

For the monopole profile, $\int\sin^2\alpha\dd y=2/m_M$ and $\int\alpha'^2\sin^2\alpha\dd y=4m_M/3$. Substitution in Eq.~\eqref{eq:stringaction} gives Eq.~\eqref{eq:rotor}. These are direct contributions within the stated truncation.

We also point out that allowing the transverse string width to relax generates an additional four-derivative term in the orientational action. To determine its coefficient, we consider a one-directional orientational gradient with locally constant $p\equiv\sum_\alpha(\partial_\alpha\bmn)^2$. Minimizing the transverse profile in the two-derivative interface action together with its direct Skyrme term gives the tension
\begin{equation}
 T(p)=2K_g\sqrt{m_B^2+p}\int_0^\pi\sin\theta
 \sqrt{1+\frac{16\kappa_{\rm YM}}{K_g}p\sin^2\theta}\,\dd\theta.
\end{equation}
For $\kappa_{\rm YM}=0$, minimizing the transverse width gives the Nambu--Goto-type square root $T(p)=4K_gm_B\sqrt{1+p/m_B^2}$ for a one-directional gradient of the internal orientation.
The expansion of the full expression is $T(p)=4K_gm_B+G_Bp+C_sp^2+O(p^3)$. It adds the four-derivative term
\begin{equation}
 C_s\left[\sum_\alpha(\partial_\alpha\bmn)^2\right]^2,
 \qquad C_s=-\frac{K_g}{2m_B^3}+\frac{32\kappa_{\rm YM}}{3m_B}
 -\frac{1024\kappa_{\rm YM}^2m_B}{15K_g},\label{eq:Cs}
\end{equation}
 This width minimization retains powers of $\kappa_{\rm YM}$ within the stated interface action. It gives the response of that action, rather than a complete higher-derivative matching of the bulk theory. The baby Skyrme operator and the width-induced operator have different tensor structures and independent coefficients. We use $C_s$ only to estimate the size of the width-relaxation correction in section~\ref{sec:merons}. The numerical meron profiles use the leading gauged world-sheet action in Eq.~\eqref{eq:gaugedB}, reduced to the radial ansatz in Eq.~\eqref{eq:Bactionrad}.

\FloatBarrier\section{Free phase winding and a constrained configuration}\label{app:rotor}\suppressfloats[t]
This appendix treats the winding of a free internal phase. The analysis applies to the Nambu-monopole phase $\chi$ in phases B1, D1, and E at $g'=0$, and to the vortex-core phase $\beta$ in phase C6, where $A=B=0$. For the latter, one replaces $\chi$ and $\mathcal I_\chi$ by $\beta$ and $\mathcal I_\beta$, respectively.

On a Euclidean time circle of period $T$, fix the winding number $n$ by $\chi(T)-\chi(0)=2\pi n$. The Cauchy--Schwarz inequality gives
\[
 \Delta S=\frac{\mathcal I_\chi}{2}\int_0^T\dd\tau\,(\chi')^2
 \geq \frac{\mathcal I_\chi}{2T}\left(\int_0^T\dd\tau\,\chi'\right)^2
 =\frac{2\pi^2\mathcal I_\chi n^2}{T}.
\]
The uniform winding $\chi(\tau)=\chi_0+2\pi n\tau/T$ saturates the bound and is stable within the phase theory at fixed winding number. A periodic fluctuation $\delta\chi$ changes the action by $(\mathcal I_\chi/2)\int_0^T\dd\tau\,(\partial_\tau\delta\chi)^2$, so a constant shift is a zero mode.
On a noncompact line, take a smooth dimensionless function $f$ with $f(-\infty)=0$, $f(+\infty)=2\pi$, a center $\tau_0$, and a width $\ell>0$. Then $\chi(\tau)=f((\tau-\tau_0)/\ell)$ gives $\Delta S=(\mathcal I_\chi/2\ell)\int(f')^2$, which tends to zero as $\ell\to\infty$.

A constrained variational problem can instead fix a width functional~\cite{affleck1981}:
\begin{equation}
 \mathcal C[\chi]\mathrel{\equiv}\int\dd\tau(1-\cos\chi)=4\ell,\qquad
 \delta\left[\frac{\mathcal I_\chi}{2}\int\chi'^2\dd\tau+
 \lambda(\mathcal C-4\ell)\right]=0.
\end{equation}
Here $\lambda$ is a Lagrange multiplier enforcing the chosen width $\ell$, independent of the Higgs quartic coupling. For endpoints $0$ and $2\pi$, the solution and its original relative action are
\begin{equation}
 \chi_\ell=4\arctan e^{(\tau-\tau_0)/\ell},\qquad
 \lambda=\frac{\mathcal I_\chi}{\ell^2},\qquad
 \Delta S[\chi_\ell]=\frac{4\mathcal I_\chi}{\ell}.
\end{equation}
The sine-Gordon equation here comes from a Lagrange multiplier, not an added physical potential. On the noncompact Euclidean time axis, removing the constraint restores spreading. At finite $g'$, a physical bulk constrained construction must also use gauge-compatible constraints and temporal fields. We use this example only to distinguish an explicit constrained solution from an unconstrained rotor winding. An embedded configuration can therefore be meaningful and carry the stated charge without being a stable finite-width saddle.

\FloatBarrier\section{Gauge completion of the meron charge}\label{app:charge}\suppressfloats[t]
In phase B4, the vacuum circle on the orientational sphere is displaced from the equator. The north- and south-core merons then cover spherical caps of different areas and carry geometric lump charges $(1-s_*)/2$ and $(1+s_*)/2$, respectively, in magnitude. Their actions are also unequal. Yet for the bulk gauge connection constructed below, each meron carries a Yang--Mills instanton charge of magnitude $1/2$. The charge of an individual meron in the bulk therefore stays fixed as the vacuum latitude changes.

The cancellation occurs at the boundary. Far from the core, the gauge field screens the winding of the orientation around the vacuum circle. For the radial ansatz below, the covariant lump density and the magnetic-flux term combine into a total derivative in the bulk charge integral. Screening makes the boundary contribution at infinity vanish, leaving the core polarity and the sign of the vortex winding to determine the charge. Equivalently, the flux term cancels the latitude-dependent shift in the geometric lump charge of each meron. For the complementary pair, the flux contributions cancel each other, and the total instanton charge has unit magnitude.

The world-sheet orientation is $\dd y\wedge\dd\tau$. Define
\begin{align}
 Q_{\rm cov}&\equiv\frac1{4\pi}\int\dd y\dd\tau\,\bmn\cdot(D_y\bmn\times D_\tau\bmn),\\
 Q_0&\equiv\frac1{4\pi}\int\dd y\dd\tau\left[\bmn\cdot(D_y\bmn\times D_\tau\bmn)+f_{y\tau}n_3\right],
 \qquad \Phi_{\rm ws}\equiv\int f=2\pi\nu.\label{eq:Q0}
\end{align}
For the radial ansatz,
\begin{equation}
 Q_0=\frac\nu2\int_0^\infty\dd R\left[(1-\mathcal A)\sin F F_R+
 \mathcal A_R\cos F\right]
 =-\frac\nu2[(1-\mathcal A)\cos F]_0^\infty=\frac{\nu p}{2}.
 \label{eq:Q0rad}
\end{equation}
Thus positive-charge north and south cores, with $\nu=p$, each have $Q_0=1/2$, even though their actions differ when $s_*\ne0$. 
A second gauge-invariant charge is obtained by replacing $n_3$ in the flux contribution by $n_3-s_*$:
\begin{equation}
 Q_{s_*}\equiv Q_0-\frac{s_*\Phi_{\rm ws}}{4\pi}
 =\frac1{4\pi}\int\dd y\dd\tau\left[\bmn\cdot(D_y\bmn\times D_\tau\bmn)+f_{y\tau}(n_3-s_*)\right].
 \label{eq:Qs}
\end{equation}
For Eq.~\eqref{eq:meronansatz}, $Q_{s_*}=\nu(p-s_*)/2$, agreeing with the ungauged cap value. For a pair of opposite windings, the total flux vanishes and both definitions give the same charge.

To identify the completion produced by a bulk connection, take $T=-i\sigma_3/2$ and a one-form $\mathcal B=-aT$ on the world sheet. In the reference embedding $U=\cos\theta\,\mathbf1+i\sin\theta\,\bmn\cdot\boldsymbol\sigma$, set
\begin{align}
 \mathcal L&\equiv U\mathcal B U^{-1}-\dd U U^{-1}-\mathcal B,\\
 \mathcal A&=\mathcal B+q_g(z)\mathcal L.\label{eq:gaugeduplift}
\end{align}
The connection interpolates between $\mathcal A_0\equiv\mathcal B$ and its gauge transform $\mathcal A_1\equiv U\mathcal B U^{-1}-\dd U U^{-1}$. Writing $q\equiv q_g(z)$, its curvature is
\begin{equation}
 \mathcal F=\dd q\wedge\mathcal L+(1-q)\mathcal F_0+q\mathcal F_1
 +q(q-1)\mathcal L\wedge\mathcal L,\qquad
 \mathcal F_1=U\mathcal F_0U^{-1}.
\end{equation}
The curvature $\mathcal F_0=-fT$ is supported on the string world sheet. Integrating the Yang--Mills topological charge density across $z$ and $x$, using the endpoints $q:0\to1$ and $\theta:0\to\pi$, gives
\begin{equation}
 \int\dd z\dd x\,\Tr\mathcal F\wedge\mathcal F
 =2\pi[\bmn\cdot(D_y\bmn\times D_\tau\bmn)+f_{y\tau}n_3]\,\dd y\wedge\dd\tau.
\end{equation}
The Yang--Mills instanton charge is therefore
\begin{equation}
 Q_{\rm YM}[\mathcal A]=-\frac1{4\pi}\int\dd y\dd\tau\left[
 \bmn\cdot(D_y\bmn\times D_\tau\bmn)+f_{y\tau}n_3\right]=-Q_0.
 \label{eq:BbulkQ}
\end{equation}

For the radial ansatz this gives $-Q_{\rm YM}=Q_{\rm geom}+s_*\Phi_{\rm ws}/(4\pi)$. The last term cancels for the pair.

This is a charge calculation for the specified interpolation and the reference transverse kink. At $z\to-\infty$ the curvature approaches $\mathcal F_0=-fT$, and at $z\to+\infty$ it approaches $\mathcal F_1=U\mathcal F_0U^{-1}$. For a meron with nonzero flux, the Yang--Mills action of this connection diverges in the transverse direction, while the integral of $\Tr\mathcal F\wedge\mathcal F$ is finite. A finite-action bulk realization requires an extension with appropriate transverse decay and, at finite $h$, the corresponding asymptotic Higgs orientations and profiles.

\FloatBarrier\section{Numerical methods and convergence}\label{app:numerics}\suppressfloats[t]
For the merons in phases B and D, we solve the radial boundary-value equations separately for the two core polarities. The solver residual is below $10^{-8}$ for the reported profiles, and the magnetic and potential terms satisfy the virial identity to better than $1.4\times10^{-10}$.

For the static planar vortex in phase C, we shoot from the regular core expansion $F=\pi/2-s_0\varrho+O(\varrho^3)$ and $\mathcal A=a_2\varrho^2+O(\varrho^4)$. The shooting parameters are fixed by the vacuum and flux boundary conditions at $\varrho_{\max}=\max(24,20/\gamma)$. Checks of the domain size and tolerance at $\gamma=0.03,0.3$ change the core moments by at most $1.1\times10^{-10}$.

\subsection{Relative charge at finite soft mixing}\label{sec:Ccharge}
At finite $h$, write the neutral fields as $\eta=N_0+iN_3$ and $c=N_c e^{i\vartheta}$, with $N_c^2+N_0^2+N_3^2=1$. Introduce dimensionless Euclidean time $t\mathrel{\equiv}m_0\tau$. For a unit spatial vortex, the relative Skyrmion charge in Eq.~\eqref{eq:Ccharge} can be written directly in terms of $N_0$ and $N_3$, allowing the radial profiles to vary with Euclidean time:
\begin{equation}
 k_{\rm rel}=-\frac1\pi\int\dd\varrho\dd t\,
 (\partial_\varrho N_0\partial_tN_3-\partial_tN_0\partial_\varrho N_3),
 \qquad t=m_0\tau.\label{eq:Cvolume}
\end{equation}
This is an oriented area in the neutral-field disk. Write the radial path of the positive static core in phase C2 as $\eta_+(\varrho)=P(\varrho)+iW(\varrho)$, with real functions $P,W$. Its core value is $e^{i\beta_c}$ and its asymptotic value is the real vacuum $s$. Applying Stokes' theorem to Eq.~\eqref{eq:Cvolume} expresses the two relative charges of Eq.~\eqref{eq:Cfractionalnumbers} in terms of the core arc and the static radial paths:
\begin{equation}
 k_{\rm short}=\frac1\pi\left[\beta_c+\int_0^\infty(PW'-WP')\dd\varrho\right],
 \qquad k_{\rm long}=1-k_{\rm short}.\label{eq:Cboundary}
\end{equation}
Here the prime differentiates the radial coordinate. The two conjugate radial boundary paths contribute oppositely to the short and long events and cancel in their sum. The full turn is integer. The individual fractions depend on the static profiles.

Equation~\eqref{eq:Cvolume} also measures the neutral relative degree in the planar gauged vortex sector.

\subsection{Numerical tests with a relaxing vortex}\label{sec:Cnumerics}
We test the core description by allowing the radial profile to depend on Euclidean time. The ungauged calculations use the two-derivative interface theory, axial symmetry in space, and time-reflection conditions for each chosen path. With the static vortex density $\mathcal E_{\rm host}$ subtracted on the same grid, the dimensionless parent action is
\begin{align}
 \frac{m_0\Delta S}{K_g}&=2\pi\int\dd t\int\varrho\dd\varrho\,
 \biggl[|\partial_\varrho\bmN|^2+|\partial_t\bmN|^2+\frac{N_c^2}{\varrho^2}
 +N_0^2+(1+\delta)N_3^2-2sN_0-\mathcal E_{\rm host}(\varrho)\biggr],\notag\\
 \bmN&=(N_c,N_0,N_3),\quad \bmN^2=1.\label{eq:Cparent}
\end{align}
The far field $(\sqrt{1-s^2},s,0)$ is held fixed at every time. In phase C2 the time endpoints are the two degenerate static vortex cores. In phases C3 and C4 they are the same true-core vortex. The core phase and cross section are varied together. We evaluate the degree from Eq.~\eqref{eq:Cvolume}, independently of the phase ansatz.

We discretize differences of the unit vector and subtract the independently minimized static reference on the same grid. The charge and action are also integrated using spline derivatives of the fields. In phase C3, a two-peak initial configuration relaxes to the same single-peak solution within the axially symmetric ansatz.

Table~\ref{tab:Ctests} collects representative comparisons for subphases C2--C4.
\begin{table}[tb]\centering\small
\begin{tabular}{lrrrr}\toprule
Configuration & Leading action & Relaxed action & Leading degree & Refined degree\\\midrule
C2 short & 0.6085446 & 0.6251 & $1/3$ & 0.33446\\
C2 long & 3.4000902 & 3.386 & $2/3$ & 0.66554\\
C3, $\delta=0$ & 3.708879 & 3.6937 & 1 & 1\\
C3, $\delta=0.002$ & 3.924843 & 3.9060 & 1 & 1\\
C4 molecule & 3.844414 & 3.7963 & 1 & 1\\\bottomrule
\end{tabular}
\caption{Two-variable ungauged parent-interface tests. Actions are $m_0\Delta S/K_g$. The phase C2 test uses $\delta=-0.02$, $s=0.003673536804$; the phase C3 test uses $s=0.002$; the phase C4 test uses $\delta=0.02$, $s=0.0002$. Relative degrees are computed from the original neutral fields.}\label{tab:Ctests}
\end{table}

Table~\ref{tab:C4scan} compares the peak separations of the relaxed and double sine-Gordon solutions as $s$ decreases at fixed $\delta=0.02$.

\begin{table}[tb]\centering
\begin{tabular}{rrrr}\toprule
$s=h/a_0$ & Relaxed action $m_0\Delta S/K_g$ & DSG separation & Relaxed separation\\\midrule
0.00080 & 4.3950 & 23.82 & 23.44\\
0.00020 & 3.7963 & 34.72 & 34.40\\
0.00005 & 3.5970 & 44.91 & 44.63\\\bottomrule
\end{tabular}
\caption{Approach to phase C5 from phase C4 at $\delta=0.02$. Separations use $m_0\Delta\tau$ and the relaxed values refer to the action-density peaks. The total relative degree is one.}\label{tab:C4scan}
\end{table}

For the gauged half instanton in phase C5, we relax $N_0$, $N_3$, and $\mathcal A(\varrho,t)$ together, at $\gamma=0.3$, $s=0$, and $\delta=0.02$. In the axial ansatz with real charged amplitude, the neutral motion does not source the temporal gauge potential. Relative to Eq.~\eqref{eq:Cparent}, the angular gradient is screened to $(1-\mathcal A)^2N_c^2/\varrho^2$, and the action includes $(\mathcal A_{\varrho}^2+\mathcal A_t^2)/(\gamma^2\varrho^2)$. At $\varrho_{\max}=24$ and $u_{\max}=10$, with $u=\sqrt\delta\,t$, the grid comparison is
\begin{equation}
\begin{array}{c|ccc}
 N_\varrho\times N_u & \hbox{discrete action} & \hbox{continuous action} & k_{\rm rel}\\\hline
 96\times128 & 1.53474667 & 1.57516124 & 0.50016545\\
 192\times256 & 1.56437498 & 1.57435416 & 0.50000727\\
 384\times512 & 1.57159255 & 1.57401137 & 0.50000053
\end{array}
\end{equation}
The fine-pair extrapolation gives $m_0\Delta S/K_g=1.573998$, reported as $1.5740$ in the text. At fixed grid spacing, enlarging the domain changes the discrete action by less than $2.2\times10^{-7}$.

\FloatBarrier


\begin{thebibliography}{999}
\bibitem{BPST} A. A. Belavin, A. M. Polyakov, A. S. Schwartz and Yu. S. Tyupkin, \emph{Pseudoparticle solutions of the Yang--Mills equations}, Phys. Lett. B \textbf{59}, 85--87 (1975), \href{https://doi.org/10.1016/0370-2693(75)90163-X}{doi:10.1016/0370-2693(75)90163-X}.
\bibitem{ADS} I.~Affleck, M.~Dine and N.~Seiberg, \emph{Dynamical supersymmetry breaking in supersymmetric QCD}, Nucl. Phys. B \textbf{241}, 493--534 (1984), \href{https://doi.org/10.1016/0550-3213(84)90058-0}{doi:10.1016/0550-3213(84)90058-0}.
\bibitem{SeibergVacua} N.~Seiberg, \emph{Exact results on the space of vacua of four-dimensional SUSY gauge theories}, Phys. Rev. D \textbf{49}, 6857--6863 (1994), \href{https://arxiv.org/abs/hep-th/9402044}{arXiv:hep-th/9402044}, \href{https://doi.org/10.1103/PhysRevD.49.6857}{doi:10.1103/PhysRevD.49.6857}.
\bibitem{SW} N.~Seiberg and E.~Witten, \emph{Electric-magnetic duality, monopole condensation, and confinement in $\mathcal N=2$ supersymmetric Yang--Mills theory}, Nucl. Phys. B \textbf{426}, 19--52 (1994) [Erratum: Nucl. Phys. B \textbf{430}, 485--486 (1994)], \href{https://arxiv.org/abs/hep-th/9407087}{arXiv:hep-th/9407087}, \href{https://doi.org/10.1016/0550-3213(94)90124-4}{doi:10.1016/0550-3213(94)90124-4}.
\bibitem{Nekrasov} N.~A.~Nekrasov, \emph{Seiberg--Witten prepotential from instanton counting}, Adv. Theor. Math. Phys. \textbf{7}, 831--864 (2003), \href{https://arxiv.org/abs/hep-th/0206161}{arXiv:hep-th/0206161}, \href{https://doi.org/10.4310/ATMP.2003.v7.n5.a4}{doi:10.4310/ATMP.2003.v7.n5.a4}.
\bibitem{HananyTong2003} A.~Hanany and D.~Tong, \emph{Vortices, Instantons and Branes}, JHEP \textbf{07}, 037 (2003), \href{https://arxiv.org/abs/hep-th/0306150}{arXiv:hep-th/0306150}, \href{https://doi.org/10.1088/1126-6708/2003/07/037}{doi:10.1088/1126-6708/2003/07/037}.
\bibitem{Auzzi2003} R.~Auzzi, S.~Bolognesi, J.~Evslin, K.~Konishi and A.~Yung, \emph{Nonabelian Superconductors: Vortices and Confinement in $\mathcal N=2$ SQCD}, Nucl. Phys. B \textbf{673}, 187--216 (2003), \href{https://arxiv.org/abs/hep-th/0307287}{arXiv:hep-th/0307287}, \href{https://doi.org/10.1016/j.nuclphysb.2003.09.029}{doi:10.1016/j.nuclphysb.2003.09.029}.
\bibitem{EtoModuli2006} M.~Eto, Y.~Isozumi, M.~Nitta, K.~Ohashi and N.~Sakai, \emph{Moduli Space of Non-Abelian Vortices}, Phys. Rev. Lett. \textbf{96}, 161601 (2006), \href{https://arxiv.org/abs/hep-th/0511088}{arXiv:hep-th/0511088}, \href{https://doi.org/10.1103/PhysRevLett.96.161601}{doi:10.1103/PhysRevLett.96.161601}.
\bibitem{higgsinstantons} M. Eto, Y. Isozumi, M. Nitta, K. Ohashi and N. Sakai, \emph{Instantons in the Higgs Phase}, Phys. Rev. D \textbf{72}, 025011 (2005), \href{https://arxiv.org/abs/hep-th/0412048}{arXiv:hep-th/0412048}, \href{https://doi.org/10.1103/PhysRevD.72.025011}{doi:10.1103/PhysRevD.72.025011}.
\bibitem{HananyTong2004} A.~Hanany and D.~Tong, \emph{Vortex Strings and Four-Dimensional Gauge Dynamics}, JHEP \textbf{04}, 066 (2004), \href{https://arxiv.org/abs/hep-th/0403158}{arXiv:hep-th/0403158}, \href{https://doi.org/10.1088/1126-6708/2004/04/066}{doi:10.1088/1126-6708/2004/04/066}.
\bibitem{ShifmanYung2004} M.~Shifman and A.~Yung, \emph{Non-Abelian String Junctions as Confined Monopoles}, Phys. Rev. D \textbf{70}, 045004 (2004), \href{https://arxiv.org/abs/hep-th/0403149}{arXiv:hep-th/0403149}, \href{https://doi.org/10.1103/PhysRevD.70.045004}{doi:10.1103/PhysRevD.70.045004}.
\bibitem{TongMonopoles2003} D.~Tong, \emph{Monopoles in the Higgs Phase}, Phys. Rev. D \textbf{69}, 065003 (2004), \href{https://arxiv.org/abs/hep-th/0307302}{arXiv:hep-th/0307302}, \href{https://doi.org/10.1103/PhysRevD.69.065003}{doi:10.1103/PhysRevD.69.065003}.
\bibitem{NittaVinci2011} M.~Nitta and W.~Vinci, \emph{Non-Abelian Monopoles in the Higgs Phase}, Nucl. Phys. B \textbf{848}, 121--154 (2011), \href{https://arxiv.org/abs/1012.4057}{arXiv:1012.4057}, \href{https://doi.org/10.1016/j.nuclphysb.2011.02.014}{doi:10.1016/j.nuclphysb.2011.02.014}.
\bibitem{Dorey1998} N.~Dorey, \emph{The BPS Spectra of Two-Dimensional Supersymmetric Gauge Theories with Twisted Mass Terms}, JHEP \textbf{11}, 005 (1998), \href{https://arxiv.org/abs/hep-th/9806056}{arXiv:hep-th/9806056}, \href{https://doi.org/10.1088/1126-6708/1998/11/005}{doi:10.1088/1126-6708/1998/11/005}.
\bibitem{DoreyHollowoodTong1999} N.~Dorey, T.~J.~Hollowood and D.~Tong, \emph{The BPS Spectra of Gauge Theories in Two and Four Dimensions}, JHEP \textbf{05}, 006 (1999), \href{https://arxiv.org/abs/hep-th/9902134}{arXiv:hep-th/9902134}, \href{https://doi.org/10.1088/1126-6708/1999/05/006}{doi:10.1088/1126-6708/1999/05/006}.
\bibitem{DoreyHollowoodLee2011} N.~Dorey, T.~J.~Hollowood and S.~Lee, \emph{Quantization of Integrable Systems and a 2d/4d Duality}, JHEP \textbf{10}, 077 (2011), \href{https://arxiv.org/abs/1103.5726}{arXiv:1103.5726}, \href{https://doi.org/10.1007/JHEP10(2011)077}{doi:10.1007/JHEP10(2011)077}.
\bibitem{ChenDorey2011} H.-Y.~Chen, N.~Dorey, T.~J.~Hollowood and S.~Lee, \emph{A New 2d/4d Duality via Integrability}, JHEP \textbf{09}, 040 (2011), \href{https://arxiv.org/abs/1104.3021}{arXiv:1104.3021}, \href{https://doi.org/10.1007/JHEP09(2011)040}{doi:10.1007/JHEP09(2011)040}.
\bibitem{FujimoriCounting2012} T.~Fujimori, T.~Kimura, M.~Nitta and K.~Ohashi, \emph{Vortex Counting from Field Theory}, JHEP \textbf{06}, 028 (2012), \href{https://arxiv.org/abs/1204.1968}{arXiv:1204.1968}, \href{https://doi.org/10.1007/JHEP06(2012)028}{doi:10.1007/JHEP06(2012)028}.
\bibitem{FujimoriKimuraNittaOhashi2015} T.~Fujimori, T.~Kimura, M.~Nitta and K.~Ohashi, \emph{2d Partition Function in $\Omega$-background and Vortex/Instanton Correspondence}, JHEP \textbf{12}, 110 (2015), \href{https://arxiv.org/abs/1509.08630}{arXiv:1509.08630}, \href{https://doi.org/10.1007/JHEP12(2015)110}{doi:10.1007/JHEP12(2015)110}.
\bibitem{eto2005} M. Eto, M. Nitta, K. Ohashi, D. Tong, \emph{Skyrmions from Instantons inside Domain Walls}, Phys. Rev. Lett. 95, 252003 (2005), \href{https://arxiv.org/abs/hep-th/0508130}{arXiv:hep-th/0508130}, \href{https://doi.org/10.1103/PhysRevLett.95.252003}{doi:10.1103/PhysRevLett.95.252003}.
\bibitem{earlywall} M.~Shifman, A.~Yung, \emph{Localization of Non-Abelian Gauge Fields on Domain Walls at Weak Coupling (D-brane Prototypes)}, Phys. Rev. D \textbf{70}, 025013 (2004), \href{https://arxiv.org/abs/hep-th/0312257}{arXiv:hep-th/0312257}, \href{https://doi.org/10.1103/PhysRevD.70.025013}{doi:10.1103/PhysRevD.70.025013}.
\bibitem{cloud} M.~Eto, T.~Fujimori, M.~Nitta, K.~Ohashi \emph{et al.}, \emph{Domain Walls with Non-Abelian Clouds}, Phys. Rev. D \textbf{77}, 125008 (2008), \href{https://arxiv.org/abs/0802.3135}{arXiv:0802.3135}, \href{https://doi.org/10.1103/PhysRevD.77.125008}{doi:10.1103/PhysRevD.77.125008}.
\bibitem{jwall} M.~Nitta, \emph{Josephson junction of non-Abelian superconductors and non-Abelian Josephson vortices}, Nucl. Phys. B \textbf{899}, 78 (2015), \href{https://arxiv.org/abs/1502.02525}{arXiv:1502.02525}, \href{https://doi.org/10.1016/j.nuclphysb.2015.07.027}{doi:10.1016/j.nuclphysb.2015.07.027}.
\bibitem{josephson} M. Nitta, \emph{Josephson instantons and Josephson monopoles in a non-Abelian Josephson junction}, \href{https://arxiv.org/abs/1503.02060}{arXiv:1503.02060}, \href{https://doi.org/10.1103/PhysRevD.92.045010}{doi:10.1103/PhysRevD.92.045010}.
\bibitem{amoeba2008} T.~Fujimori, M.~Nitta, K.~Ohta, N.~Sakai and M.~Yamazaki, \emph{Intersecting Solitons, Amoeba and Tropical Geometry}, Phys. Rev. D \textbf{78}, 105004 (2008), \href{https://arxiv.org/abs/0805.1194}{arXiv:0805.1194}, \href{https://doi.org/10.1103/PhysRevD.78.105004}{doi:10.1103/PhysRevD.78.105004}.
\bibitem{nittaMonopoleStrings2013} M.~Nitta, \emph{Instantons confined by monopole strings}, Phys. Rev. D \textbf{87}, 066008 (2013), \href{https://arxiv.org/abs/1301.3268}{arXiv:1301.3268}, \href{https://doi.org/10.1103/PhysRevD.87.066008}{doi:10.1103/PhysRevD.87.066008}.
\bibitem{incarnations} M. Nitta, \emph{Incarnations of Instantons}, Nucl. Phys. B \textbf{885}, 493 (2014), \href{https://arxiv.org/abs/1311.2718}{arXiv:1311.2718}, \href{https://doi.org/10.1016/j.nuclphysb.2014.05.032}{doi:10.1016/j.nuclphysb.2014.05.032}.
\bibitem{relations2022} M. Nitta, \emph{Relations among topological solitons}, Phys. Rev. D \textbf{105}, 105006 (2022), \href{https://arxiv.org/abs/2202.03929}{arXiv:2202.03929}, \href{https://doi.org/10.1103/PhysRevD.105.105006}{doi:10.1103/PhysRevD.105.105006}.
\bibitem{tHooft} G. 't Hooft, \emph{Symmetry Breaking through Bell--Jackiw Anomalies}, Phys. Rev. Lett. \textbf{37}, 8--11 (1976), \href{https://doi.org/10.1103/PhysRevLett.37.8}{doi:10.1103/PhysRevLett.37.8}.
\bibitem{affleck1981} I. Affleck, \emph{On constrained instantons}, Nucl. Phys. B \textbf{191}, 429--444 (1981), \href{https://doi.org/10.1016/0550-3213(81)90307-2}{doi:10.1016/0550-3213(81)90307-2}.
\bibitem{nambu} Y.~Nambu, \emph{String-like configurations in the Weinberg--Salam theory}, Nucl. Phys. B \textbf{130}, 505 (1977), \href{https://doi.org/10.1016/0550-3213(77)90252-8}{doi:10.1016/0550-3213(77)90252-8}.
\bibitem{vachaspati} T.~Vachaspati, \emph{Vortex solutions in the Weinberg--Salam model}, Phys. Rev. Lett. \textbf{68}, 1977 (1992) [Erratum: Phys. Rev. Lett. \textbf{69}, 216 (1992)], \href{https://doi.org/10.1103/PhysRevLett.68.1977}{doi:10.1103/PhysRevLett.68.1977}.
\bibitem{review} A.~Achucarro, T.~Vachaspati, \emph{Semilocal and Electroweak Strings}, Phys. Rept. \textbf{327}, 347 (2000), \href{https://arxiv.org/abs/hep-ph/9904229}{arXiv:hep-ph/9904229}, \href{https://doi.org/10.1016/S0370-1573(99)00103-9}{doi:10.1016/S0370-1573(99)00103-9}.
\bibitem{james1992} M.~James, L.~Perivolaropoulos, T.~Vachaspati, \emph{Stability of electroweak strings}, Phys. Rev. D \textbf{46}, R5232 (1992), \href{https://doi.org/10.1103/PhysRevD.46.R5232}{doi:10.1103/PhysRevD.46.R5232}.
\bibitem{james} M.~James, L.~Perivolaropoulos, T.~Vachaspati, \emph{Detailed Stability Analysis of Electroweak Strings}, Nucl. Phys. B \textbf{395}, 534 (1993), \href{https://arxiv.org/abs/hep-ph/9212301}{arXiv:hep-ph/9212301}, \href{https://doi.org/10.1016/0550-3213(93)90046-R}{doi:10.1016/0550-3213(93)90046-R}.
\bibitem{goodband} M.~Goodband, M.~Hindmarsh, \emph{Instabilities of Electroweak Strings}, Phys. Lett. B \textbf{363}, 58 (1995), \href{https://arxiv.org/abs/hep-ph/9505357}{arXiv:hep-ph/9505357}, \href{https://doi.org/10.1016/0370-2693(95)01198-Y}{doi:10.1016/0370-2693(95)01198-Y}.
\bibitem{baryogenesis} R.~H.~Brandenberger, A.~Davis, \emph{Electroweak Baryogenesis with Electroweak Strings}, Phys. Lett. B \textbf{308}, 79 (1993), \href{https://arxiv.org/abs/astro-ph/9206001}{arXiv:astro-ph/9206001}, \href{https://doi.org/10.1016/0370-2693(93)90604-G}{doi:10.1016/0370-2693(93)90604-G}.
\bibitem{magnetic} T.~Vachaspati, \emph{Magnetic fields from cosmological phase transitions}, Phys. Lett. B \textbf{265}, 258 (1991), \href{https://doi.org/10.1016/0370-2693(91)90051-Q}{doi:10.1016/0370-2693(91)90051-Q}.
\bibitem{helicity} T.~Vachaspati, \emph{Estimate of the primordial magnetic field helicity}, Phys. Rev. Lett. \textbf{87}, 251302 (2001), \href{https://arxiv.org/abs/astro-ph/0101261}{arXiv:astro-ph/0101261}, \href{https://doi.org/10.1103/PhysRevLett.87.251302}{doi:10.1103/PhysRevLett.87.251302}.
\bibitem{vachaspatiwatkins} T.~Vachaspati, R.~Watkins, \emph{Bound States Can Stabilize Electroweak Strings}, Phys. Lett. B \textbf{318}, 163 (1993), \href{https://arxiv.org/abs/hep-ph/9211284}{arXiv:hep-ph/9211284}, \href{https://doi.org/10.1016/0370-2693(93)91800-3}{doi:10.1016/0370-2693(93)91800-3}.
\bibitem{garriga} J.~Garriga, X.~Montes, \emph{Stability of $Z$ Strings in Strong Magnetic Fields}, Phys. Rev. Lett. \textbf{75}, 2268 (1995), \href{https://arxiv.org/abs/hep-ph/9505424}{arXiv:hep-ph/9505424}, \href{https://doi.org/10.1103/PhysRevLett.75.2268}{doi:10.1103/PhysRevLett.75.2268}.
\bibitem{nagasawa} M.~Nagasawa, R.~Brandenberger, \emph{Stabilization of the Electroweak Z String in the Early Universe}, Phys. Rev. D \textbf{67}, 043504 (2003), \href{https://arxiv.org/abs/hep-ph/0207246}{arXiv:hep-ph/0207246}, \href{https://doi.org/10.1103/PhysRevD.67.043504}{doi:10.1103/PhysRevD.67.043504}.
\bibitem{forgacs} P.~Forg\'acs, A.~Luk\'acs, \emph{Electroweak strings with dark scalar condensates and their (in)stability}, Phys. Rev. D \textbf{102}, 023009 (2020), \href{https://arxiv.org/abs/1909.07447}{arXiv:1909.07447}, \href{https://doi.org/10.1103/PhysRevD.102.023009}{doi:10.1103/PhysRevD.102.023009}.
\bibitem{naculich} S.~G.~Naculich, \emph{Fermions Destabilize Electroweak Strings}, Phys. Rev. Lett. \textbf{75}, 998 (1995), \href{https://arxiv.org/abs/hep-ph/9501388}{arXiv:hep-ph/9501388}, \href{https://doi.org/10.1103/PhysRevLett.75.998}{doi:10.1103/PhysRevLett.75.998}.
\bibitem{etohamada2024} M.~Eto, Y.~Hamada, R.~Jinno, M.~Nitta, M.~Yamada, \emph{Neutrino zeromodes on electroweak strings in light of topological insulators}, JHEP \textbf{06}, 062 (2024), \href{https://arxiv.org/abs/2402.19417}{arXiv:2402.19417}, \href{https://doi.org/10.1007/JHEP06(2024)062}{doi:10.1007/JHEP06(2024)062}.
\bibitem{DrogoszOscillons2026} Z.~Drogosz, E.~I.~Sfakianakis, K.~Slawinska and A.~Wereszczynski, \emph{Standard Model Effective Field Theory and Oscillons}, \href{https://arxiv.org/abs/2607.23616}{arXiv:2607.23616}.
\bibitem{2hdmreview} G.~C.~Branco, P.~M.~Ferreira, L.~Lavoura, M.~N.~Rebelo, M.~Sher, J.~P.~Silva, \emph{Theory and phenomenology of two-Higgs-doublet models}, Phys. Rept. \textbf{516}, 1 (2012), \href{https://arxiv.org/abs/1106.0034}{arXiv:1106.0034}, \href{https://doi.org/10.1016/j.physrep.2012.02.002}{doi:10.1016/j.physrep.2012.02.002}.
\bibitem{vacuumtopology} R.~A.~Battye, G.~D.~Brawn, A.~Pilaftsis, \emph{Vacuum Topology of the Two Higgs Doublet Model}, JHEP \textbf{08}, 020 (2011), \href{https://arxiv.org/abs/1106.3482}{arXiv:1106.3482}, \href{https://doi.org/10.1007/JHEP08(2011)020}{doi:10.1007/JHEP08(2011)020}.
\bibitem{earlyvortex} H.~La, \emph{Vortex Solutions in Two-Higgs Systems and $\tan\beta$}, \href{https://arxiv.org/abs/hep-ph/9302220}{arXiv:hep-ph/9302220} (1993).
\bibitem{twohiggsvortex} G.~Bimonte, G.~Lozano, \emph{Vortex Solutions in Two-Higgs-Doublet Systems}, Phys. Lett. B \textbf{326}, 270 (1994), \href{https://arxiv.org/abs/hep-ph/9401313}{arXiv:hep-ph/9401313}, \href{https://doi.org/10.1016/0370-2693(94)91321-8}{doi:10.1016/0370-2693(94)91321-8}.
\bibitem{fluxtube} G.~R.~Dvali, G.~Senjanovic, \emph{Topologically Stable Electroweak Flux Tube}, Phys. Rev. Lett. \textbf{71}, 2376 (1993), \href{https://arxiv.org/abs/hep-ph/9305278}{arXiv:hep-ph/9305278}, \href{https://doi.org/10.1103/PhysRevLett.71.2376}{doi:10.1103/PhysRevLett.71.2376}.
\bibitem{ekn} M. Eto, M. Kurachi and M. Nitta, \emph{Constraints on two Higgs doublet models from domain walls}, Phys. Lett. B \textbf{785}, 447--453 (2018), \href{https://arxiv.org/abs/1803.04662}{arXiv:1803.04662}, \href{https://doi.org/10.1016/j.physletb.2018.09.002}{doi:10.1016/j.physletb.2018.09.002}.
\bibitem{eknstrings2018} M. Eto, M. Kurachi and M. Nitta, \emph{Non-Abelian strings and domain walls in two Higgs doublet models}, JHEP \textbf{08}, 195 (2018), \href{https://arxiv.org/abs/1805.07015}{arXiv:1805.07015}, \href{https://doi.org/10.1007/JHEP08(2018)195}{doi:10.1007/JHEP08(2018)195}.
\bibitem{topmonopole} M.~Eto, Y.~Hamada, M.~Kurachi, M.~Nitta, \emph{Topological Nambu monopole in two Higgs doublet models}, Phys. Lett. B \textbf{802}, 135220 (2020), \href{https://arxiv.org/abs/1904.09269}{arXiv:1904.09269}, \href{https://doi.org/10.1016/j.physletb.2020.135220}{doi:10.1016/j.physletb.2020.135220}.
\bibitem{monodynamics} M.~Eto, Y.~Hamada, M.~Kurachi, M.~Nitta, \emph{Dynamics of Nambu monopole in two Higgs doublet models -- Cosmological Monopole Collider --}, JHEP \textbf{07}, 004 (2020), \href{https://arxiv.org/abs/2003.08772}{arXiv:2003.08772}, \href{https://doi.org/10.1007/JHEP07(2020)004}{doi:10.1007/JHEP07(2020)004}.
\bibitem{nambudyon} M. Eto, Y. Hamada, and M. Nitta, \emph{Topological structure of a Nambu monopole in two-Higgs-doublet models: Fiber bundle, Dirac's quantization, and a dyon}, Phys. Rev. D \textbf{102}, 105018 (2020), \href{https://arxiv.org/abs/2007.15587}{arXiv:2007.15587}, \href{https://doi.org/10.1103/PhysRevD.102.105018}{doi:10.1103/PhysRevD.102.105018}.
\bibitem{molecule} M.~Eto, Y.~Hamada, M.~Nitta, \emph{Stable $Z$-strings with topological polarization in two Higgs doublet model}, JHEP \textbf{02}, 099 (2022), \href{https://arxiv.org/abs/2111.13345}{arXiv:2111.13345}, \href{https://doi.org/10.1007/JHEP02(2022)099}{doi:10.1007/JHEP02(2022)099}.
\bibitem{membrane} C.~Bachas, T.~N.~Tomaras, \emph{Membranes in the Two-Higgs-Doublet Standard Model}, Phys. Rev. Lett. \textbf{76}, 356 (1996), \href{https://arxiv.org/abs/hep-ph/9508395}{arXiv:hep-ph/9508395}, \href{https://doi.org/10.1103/PhysRevLett.76.356}{doi:10.1103/PhysRevLett.76.356}.
\bibitem{cpsoliton} A.~Riotto, O.~Tornkvist, \emph{CP-violating solitons in the minimal supersymmetric standard model}, Phys. Rev. D \textbf{56}, 3917 (1997), \href{https://arxiv.org/abs/hep-ph/9704371}{arXiv:hep-ph/9704371}, \href{https://doi.org/10.1103/PhysRevD.56.3917}{doi:10.1103/PhysRevD.56.3917}.
\bibitem{chargedwall} K.~H.~Law, A.~Pilaftsis, \emph{Charged and CP-violating Kink Solutions in the Two Higgs Doublet Model}, Phys. Rev. D \textbf{105}, 056007 (2022), \href{https://arxiv.org/abs/2110.12550}{arXiv:2110.12550}, \href{https://doi.org/10.1103/PhysRevD.105.056007}{doi:10.1103/PhysRevD.105.056007}.
\bibitem{wallsim} R.~A.~Battye, A.~Pilaftsis, D.~G.~Viatic, \emph{Simulations of Domain Walls in Two Higgs Doublet Models}, JHEP \textbf{01}, 105 (2021), \href{https://arxiv.org/abs/2006.13273}{arXiv:2006.13273}, \href{https://doi.org/10.1007/JHEP01(2021)105}{doi:10.1007/JHEP01(2021)105}.
\bibitem{wallconstraint} R.~A.~Battye, A.~Pilaftsis, D.~G.~Viatic, \emph{Domain Wall Constraints on Two Higgs Doublet Models with $Z_2$ Symmetry}, Phys. Rev. D \textbf{102}, 123536 (2020), \href{https://arxiv.org/abs/2010.09840}{arXiv:2010.09840}, \href{https://doi.org/10.1103/PhysRevD.102.123536}{doi:10.1103/PhysRevD.102.123536}.
\bibitem{collapse} N.~Chen, T.~Li, Z.~Teng, Y.~Wu, \emph{Collapsing domain walls in the two-Higgs-doublet model and deep insights from the EDM}, JHEP \textbf{10}, 081 (2020), \href{https://arxiv.org/abs/2006.06913}{arXiv:2006.06913}, \href{https://doi.org/10.1007/JHEP10(2020)081}{doi:10.1007/JHEP10(2020)081}.
\bibitem{wallclassification} R.~A.~Battye, S.~J.~Cotterill, A.~K.~Thomasson, \emph{Complete Classification of Domain Wall Solutions in the $\mathbb{Z}_2$-symmetric 2HDM}, Phys. Rev. D \textbf{112}, 115015 (2025), \href{https://arxiv.org/abs/2509.23332}{arXiv:2509.23332}, \href{https://doi.org/10.1103/p1sq-38cc}{doi:10.1103/p1sq-38cc}.
\bibitem{superconducting2hdm} R.~A.~Battye, S.~J.~Cotterill, \emph{Superconducting strings in the two-Higgs doublet model}, Phys. Lett. B \textbf{868}, 139714 (2025), \href{https://arxiv.org/abs/2410.03300}{arXiv:2410.03300}, \href{https://doi.org/10.1016/j.physletb.2025.139714}{doi:10.1016/j.physletb.2025.139714}.
\bibitem{percolation} R.~A.~Battye, S.~J.~Cotterill, E.~Sabater~Andres, A.~K.~Thomasson, \emph{Percolation of Domain Walls in the Two-Higgs Doublet Model}, Phys. Lett. B \textbf{862}, 139311 (2025), \href{https://arxiv.org/abs/2502.03209}{arXiv:2502.03209}, \href{https://doi.org/10.1016/j.physletb.2025.139311}{doi:10.1016/j.physletb.2025.139311}.
\bibitem{kinkyvortons} R.~A.~Battye, S.~J.~Cotterill, A.~K.~Thomasson, \emph{Kinky vortons in the 2HDM}, Phys. Rev. D \textbf{113}, 115035 (2026), \href{https://arxiv.org/abs/2603.22040}{arXiv:2603.22040}, \href{https://doi.org/10.1103/yf92-llgw}{doi:10.1103/yf92-llgw}.
\bibitem{globalmonopoles} R.~A.~Battye, S.~J.~Cotterill, D.~G.~Viatic, \emph{Global monopoles in the two-Higgs-doublet-model}, Phys. Lett. B \textbf{844}, 138091 (2023), \href{https://arxiv.org/abs/2309.04400}{arXiv:2309.04400}, \href{https://doi.org/10.1016/j.physletb.2023.138091}{doi:10.1016/j.physletb.2023.138091}.
\bibitem{hopffibration} R.~A.~Battye, S.~J.~Cotterill, \emph{Spontaneous Hopf fibration in the two-Higgs-doublet model}, Phys. Rev. Lett. \textbf{132}, 061601 (2024), \href{https://arxiv.org/abs/2401.06567}{arXiv:2401.06567}, \href{https://doi.org/10.1103/PhysRevLett.132.061601}{doi:10.1103/PhysRevLett.132.061601}.
\bibitem{deshpande} N.~G.~Deshpande, E.~Ma, \emph{Pattern of symmetry breaking with two Higgs doublets}, Phys. Rev. D \textbf{18}, 2574 (1978), \href{https://doi.org/10.1103/PhysRevD.18.2574}{doi:10.1103/PhysRevD.18.2574}.
\bibitem{barbieri} R.~Barbieri, L.~J.~Hall, V.~S.~Rychkov, \emph{Improved Naturalness with a Heavy Higgs: An Alternative Road to LHC Physics}, Phys. Rev. D \textbf{74}, 015007 (2006), \href{https://arxiv.org/abs/hep-ph/0603188}{arXiv:hep-ph/0603188}, \href{https://doi.org/10.1103/PhysRevD.74.015007}{doi:10.1103/PhysRevD.74.015007}.
\bibitem{idmdm} L.~Lopez~Honorez, E.~Nezri, J.~F.~Oliver, M.~H.~G.~Tytgat, \emph{The Inert Doublet Model: an Archetype for Dark Matter}, JCAP \textbf{02}, 028 (2007), \href{https://arxiv.org/abs/hep-ph/0612275}{arXiv:hep-ph/0612275}, \href{https://doi.org/10.1088/1475-7516/2007/02/028}{doi:10.1088/1475-7516/2007/02/028}.
\bibitem{aokicollider} M.~Aoki, S.~Kanemura, H.~Yokoya, \emph{Reconstruction of Inert Doublet Scalars at the International Linear Collider}, Phys. Lett. B \textbf{725}, 302 (2013), \href{https://arxiv.org/abs/1303.6191}{arXiv:1303.6191}, \href{https://doi.org/10.1016/j.physletb.2013.07.011}{doi:10.1016/j.physletb.2013.07.011}.
\bibitem{idmprecision} S.~Kanemura, M.~Kikuchi, K.~Sakurai, \emph{Testing the dark matter scenario in the inert doublet model by future precision measurements of the Higgs boson couplings}, Phys. Rev. D \textbf{94}, 115011 (2016), \href{https://arxiv.org/abs/1605.08520}{arXiv:1605.08520}, \href{https://doi.org/10.1103/PhysRevD.94.115011}{doi:10.1103/PhysRevD.94.115011}.
\bibitem{idmlimits} A.~Ilnicka, M.~Krawczyk, T.~Robens, \emph{Inert Doublet Model in light of LHC Run I and astrophysical data}, Phys. Rev. D \textbf{93}, 055026 (2016), \href{https://arxiv.org/abs/1508.01671}{arXiv:1508.01671}, \href{https://doi.org/10.1103/PhysRevD.93.055026}{doi:10.1103/PhysRevD.93.055026}.
\bibitem{idmewpt} J.~M.~Cline, K.~Kainulainen, \emph{Improved Electroweak Phase Transition with Subdominant Inert Doublet Dark Matter}, Phys. Rev. D \textbf{87}, 071701(R) (2013), \href{https://arxiv.org/abs/1302.2614}{arXiv:1302.2614}, \href{https://doi.org/10.1103/PhysRevD.87.071701}{doi:10.1103/PhysRevD.87.071701}.
\bibitem{gil} G.~Gil, P.~Chankowski, M.~Krawczyk, \emph{Inert Dark Matter and Strong Electroweak Phase Transition}, Phys. Lett. B \textbf{717}, 396 (2012), \href{https://arxiv.org/abs/1207.0084}{arXiv:1207.0084}, \href{https://doi.org/10.1016/j.physletb.2012.09.052}{doi:10.1016/j.physletb.2012.09.052}.
\bibitem{idmperturbativity} S.~Kanemura, Y.~Mura, \emph{On the criterion of perturbativity with the mass-dependent beta function in extended Higgs models}, Phys. Rev. D \textbf{110}, 075016 (2024), \href{https://arxiv.org/abs/2310.15622}{arXiv:2310.15622}, \href{https://doi.org/10.1103/PhysRevD.110.075016}{doi:10.1103/PhysRevD.110.075016}.
\bibitem{kanemuraewbg} S.~Kanemura, Y.~Okada, E.~Senaha, \emph{Electroweak baryogenesis and quantum corrections to the triple Higgs boson coupling}, Phys. Lett. B \textbf{606}, 361 (2005), \href{https://arxiv.org/abs/hep-ph/0411354}{arXiv:hep-ph/0411354}, \href{https://doi.org/10.1016/j.physletb.2004.12.004}{doi:10.1016/j.physletb.2004.12.004}.
\bibitem{aks} M.~Aoki, S.~Kanemura, O.~Seto, \emph{Neutrino mass, Dark Matter and Baryon Asymmetry via TeV-Scale Physics without Fine-Tuning}, Phys. Rev. Lett. \textbf{102}, 051805 (2009), \href{https://arxiv.org/abs/0807.0361}{arXiv:0807.0361}, \href{https://doi.org/10.1103/PhysRevLett.102.051805}{doi:10.1103/PhysRevLett.102.051805}.
\bibitem{aksewbg} M.~Aoki, K.~Enomoto, S.~Kanemura, \emph{Electroweak baryogenesis in the three-loop neutrino mass model with dark matter}, Phys. Rev. D \textbf{107}, 115022 (2023), \href{https://arxiv.org/abs/2212.14786}{arXiv:2212.14786}, \href{https://doi.org/10.1103/PhysRevD.107.115022}{doi:10.1103/PhysRevD.107.115022}.
\bibitem{ferreira} P.~M.~Ferreira, B.~Swiezewska, \emph{One-loop contributions to neutral minima in the inert doublet model}, JHEP \textbf{04}, 099 (2016), \href{https://arxiv.org/abs/1511.02879}{arXiv:1511.02879}, \href{https://doi.org/10.1007/JHEP04(2016)099}{doi:10.1007/JHEP04(2016)099}.
\bibitem{ginzburg} I.~Ginzburg, K.~Kanishev, M.~Krawczyk, D.~Sokolowska, \emph{Evolution of Universe to the present inert phase}, Phys. Rev. D \textbf{82}, 123533 (2010), \href{https://arxiv.org/abs/1009.4593}{arXiv:1009.4593}, \href{https://doi.org/10.1103/PhysRevD.82.123533}{doi:10.1103/PhysRevD.82.123533}.
\bibitem{aokimultistep} M.~Aoki, T.~Komatsu, H.~Shibuya, \emph{Possibility of multi-step electroweak phase transition in the two Higgs doublet models}, Prog. Theor. Exp. Phys. \textbf{2022}, 063B05 (2022), \href{https://arxiv.org/abs/2106.03439}{arXiv:2106.03439}, \href{https://doi.org/10.1093/ptep/ptac068}{doi:10.1093/ptep/ptac068}.
\bibitem{aoki} M.~Aoki, H.~Shibuya, \emph{Electroweak baryogenesis between broken phases in multi-step phase transition}, Phys. Lett. B \textbf{843}, 138041 (2023), \href{https://arxiv.org/abs/2302.11551}{arXiv:2302.11551}, \href{https://doi.org/10.1016/j.physletb.2023.138041}{doi:10.1016/j.physletb.2023.138041}.
\bibitem{fabian} S.~Fabian, F.~Goertz, Y.~Jiang, \emph{Dark Matter and Nature of Electroweak Phase Transition with an Inert Doublet}, JCAP \textbf{09}, 011 (2021), \href{https://arxiv.org/abs/2012.12847}{arXiv:2012.12847}, \href{https://doi.org/10.1088/1475-7516/2021/09/011}{doi:10.1088/1475-7516/2021/09/011}.
\bibitem{aokicharged} M.~Aoki, L.~Biermann, C.~Borschensky, I.~P.~Ivanov, M.~Mühlleitner, H.~Shibuya, \emph{Intermediate Charge-Breaking Phases and Symmetry Non-Restoration in the 2-Higgs-Doublet Model}, JHEP \textbf{02}, 232 (2024), \href{https://arxiv.org/abs/2308.04141}{arXiv:2308.04141}, \href{https://doi.org/10.1007/JHEP02(2024)232}{doi:10.1007/JHEP02(2024)232}.
\bibitem{interface2026} M. Nitta, \emph{Stable Electroweak Strings and Nambu Monopoles Nested at Inert-Higgs Interfaces}, \href{https://arxiv.org/abs/2610.39173}{arXiv:2610.39173v1} (2026).
\bibitem{fractional2015} S. B. Gudnason and M. Nitta, \emph{Fractional Skyrmions and their molecules}, Phys. Rev. D \textbf{91}, 085040 (2015), \href{https://arxiv.org/abs/1502.06596}{arXiv:1502.06596}, \href{https://doi.org/10.1103/PhysRevD.91.085040}{doi:10.1103/PhysRevD.91.085040}.
\bibitem{symmetry2012} A.~Pilaftsis, \emph{On the Classification of Accidental Symmetries of the Two Higgs Doublet Model Potential}, Phys. Lett. B \textbf{706}, 465 (2012), \href{https://arxiv.org/abs/1109.3787}{arXiv:1109.3787}, \href{https://doi.org/10.1016/j.physletb.2011.11.047}{doi:10.1016/j.physletb.2011.11.047}.
\bibitem{symmetry2020} N.~Darvishi, A.~Pilaftsis, \emph{Classifying Accidental Symmetries in Multi-Higgs Doublet Models}, Phys. Rev. D \textbf{101}, 095008 (2020), \href{https://arxiv.org/abs/1912.00887}{arXiv:1912.00887}, \href{https://doi.org/10.1103/PhysRevD.101.095008}{doi:10.1103/PhysRevD.101.095008}.
\bibitem{maxsym} P.~S.~Bhupal~Dev, A.~Pilaftsis, \emph{Maximally Symmetric Two Higgs Doublet Model with Natural Standard Model Alignment}, JHEP \textbf{12}, 024 (2014) [Erratum: JHEP \textbf{11}, 147 (2015)], \href{https://arxiv.org/abs/1408.3405}{arXiv:1408.3405}, \href{https://doi.org/10.1007/JHEP12(2014)024}{doi:10.1007/JHEP12(2014)024}, \href{https://doi.org/10.1007/JHEP11(2015)147}{doi:10.1007/JHEP11(2015)147}.
\bibitem{manton} N.~S.~Manton, \emph{A remark on the scattering of BPS monopoles}, Phys. Lett. B \textbf{110}, 54 (1982), \href{https://doi.org/10.1016/0370-2693(82)90950-9}{doi:10.1016/0370-2693(82)90950-9}.
\bibitem{solitoneft} M.~Eto, Y.~Isozumi, M.~Nitta, K.~Ohashi \emph{et al.}, \emph{Manifestly supersymmetric effective Lagrangians on BPS solitons}, Phys. Rev. D \textbf{73}, 125008 (2006), \href{https://arxiv.org/abs/hep-th/0602289}{arXiv:hep-th/0602289}, \href{https://doi.org/10.1103/PhysRevD.73.125008}{doi:10.1103/PhysRevD.73.125008}.
\bibitem{derrick1964} G. H. Derrick, \emph{Comments on Nonlinear Wave Equations as Models for Elementary Particles}, J. Math. Phys. \textbf{5}, 1252--1254 (1964), \href{https://doi.org/10.1063/1.1704233}{doi:10.1063/1.1704233}.
\bibitem{pietteMultisolitons} B. M. A. G. Piette, B. J. Schroers and W. J. Zakrzewski, \emph{Multisolitons in a Two-Dimensional Skyrme Model}, Z. Phys. C \textbf{65}, 165--174 (1995), \href{https://arxiv.org/abs/hep-th/9406160}{arXiv:hep-th/9406160}, \href{https://doi.org/10.1007/BF01571317}{doi:10.1007/BF01571317}.
\bibitem{pietteDynamics} B. M. A. G. Piette, B. J. Schroers and W. J. Zakrzewski, \emph{Dynamics of Baby Skyrmions}, Nucl. Phys. B \textbf{439}, 205--235 (1995), \href{https://arxiv.org/abs/hep-ph/9410256}{arXiv:hep-ph/9410256}, \href{https://doi.org/10.1016/0550-3213(95)00011-G}{doi:10.1016/0550-3213(95)00011-G}.
\bibitem{nittaWallSkyrmion2013} M. Nitta, \emph{Correspondence between Skyrmions in 2+1 and 3+1 Dimensions}, Phys. Rev. D \textbf{87}, 025013 (2013), \href{https://arxiv.org/abs/1210.2233}{arXiv:1210.2233}, \href{https://doi.org/10.1103/PhysRevD.87.025013}{doi:10.1103/PhysRevD.87.025013}.
\bibitem{nittaMatryoshka2013} M. Nitta, \emph{Matryoshka Skyrmions}, Nucl. Phys. B \textbf{872}, 62--71 (2013), \href{https://arxiv.org/abs/1211.4916}{arXiv:1211.4916}, \href{https://doi.org/10.1016/j.nuclphysb.2013.03.003}{doi:10.1016/j.nuclphysb.2013.03.003}.
\bibitem{gudnasonWallSkyrmions2014} S. B. Gudnason and M. Nitta, \emph{Domain wall Skyrmions}, Phys. Rev. D \textbf{89}, 085022 (2014), \href{https://arxiv.org/abs/1403.1245}{arXiv:1403.1245}, \href{https://doi.org/10.1103/PhysRevD.89.085022}{doi:10.1103/PhysRevD.89.085022}.
\bibitem{incskyrmion} S. B. Gudnason, M. Nitta, \emph{Incarnations of Skyrmions}, Phys. Rev. D \textbf{90}, 085007 (2014), \href{https://arxiv.org/abs/1407.7210}{arXiv:1407.7210}, \href{https://doi.org/10.1103/PhysRevD.90.085007}{doi:10.1103/PhysRevD.90.085007}.
\bibitem{etoNittaNonAbelianSG2015} M.~Eto and M.~Nitta, \emph{Non-Abelian Sine-Gordon Solitons: Correspondence between $SU(N)$ Skyrmions and $\mathbb{C}P^{N-1}$ Lumps}, Phys. Rev. D \textbf{91}, 085044 (2015), \href{https://arxiv.org/abs/1501.07038}{arXiv:1501.07038}, \href{https://doi.org/10.1103/PhysRevD.91.085044}{doi:10.1103/PhysRevD.91.085044}.
\bibitem{etoQCDwall} M. Eto, K. Nishimura and M. Nitta, \emph{Domain-Wall Skyrmion Phase in Dense QCD at Strong Magnetic Fields Using Leading-Order Chiral Perturbation Theory}, Phys. Rev. Lett. \textbf{134}, 181902 (2025), \href{https://arxiv.org/abs/2304.02940}{arXiv:2304.02940}, \href{https://doi.org/10.1103/PhysRevLett.134.181902}{doi:10.1103/PhysRevLett.134.181902}.
\bibitem{etoQCDchain} M. Eto, K. Nishimura and M. Nitta, \emph{Phase diagram of QCD matter with magnetic field: domain-wall Skyrmion chain in chiral soliton lattice}, JHEP \textbf{12}, 032 (2023), \href{https://arxiv.org/abs/2311.01112}{arXiv:2311.01112}, \href{https://doi.org/10.1007/JHEP12(2023)032}{doi:10.1007/JHEP12(2023)032}.
\bibitem{etoQCDrotation} M. Eto, K. Nishimura and M. Nitta, \emph{Domain-wall Skyrmion phase in a rapidly rotating QCD matter}, JHEP \textbf{03}, 019 (2024), \href{https://arxiv.org/abs/2310.17511}{arXiv:2310.17511}, \href{https://doi.org/10.1007/JHEP03(2024)019}{doi:10.1007/JHEP03(2024)019}.
\bibitem{amariQCDgauge} Y. Amari, M. Eto and M. Nitta, \emph{Domain-wall Skyrmion phase of QCD in magnetic field: gauge field dynamics}, JHEP \textbf{05}, 037 (2025), \href{https://arxiv.org/abs/2409.08841}{arXiv:2409.08841}, \href{https://doi.org/10.1007/JHEP05(2025)037}{doi:10.1007/JHEP05(2025)037}.
\bibitem{copingerFermionicWalls2026} P.~Copinger, M.~Eto, M.~Nitta and Z.~Qiu, \emph{Fermionic domain-wall Skyrmions of QCD in a magnetic field}, JHEP \textbf{05}, 063 (2026), \href{https://arxiv.org/abs/2512.22023}{arXiv:2512.22023}, \href{https://doi.org/10.1007/JHEP05(2026)063}{doi:10.1007/JHEP05(2026)063}.
\bibitem{copingerWallFermiGas2026} P.~Copinger, M.~Eto and M.~Nitta, \emph{Fermi gas of domain-wall Skyrmions in QCD in a strong magnetic field}, \href{https://arxiv.org/abs/2608.26608}{arXiv:2608.26608}.
\bibitem{josephson2012} M. Nitta, \emph{Josephson vortices and the Atiyah--Manton construction}, Phys. Rev. D \textbf{86}, 125004 (2012), \href{https://arxiv.org/abs/1207.6958}{arXiv:1207.6958}, \href{https://doi.org/10.1103/PhysRevD.86.125004}{doi:10.1103/PhysRevD.86.125004}.
\bibitem{jewels} M.~Kobayashi and M.~Nitta, \emph{Sine-Gordon kinks on a domain wall ring}, Phys. Rev. D \textbf{87}, 085003 (2013), \href{https://arxiv.org/abs/1302.0989}{arXiv:1302.0989}, \href{https://doi.org/10.1103/PhysRevD.87.085003}{doi:10.1103/PhysRevD.87.085003}.
\bibitem{ross2022} C.~Ross and M.~Nitta, \emph{Domain-wall skyrmions in chiral magnets}, Phys. Rev. B \textbf{107}, 024422 (2023), \href{https://arxiv.org/abs/2205.11417}{arXiv:2205.11417}, \href{https://doi.org/10.1103/PhysRevB.107.024422}{doi:10.1103/PhysRevB.107.024422}.
\bibitem{amari2023} Y.~Amari, C.~Ross and M.~Nitta, \emph{Domain-wall skyrmion chain and domain-wall bimerons in chiral magnets}, Phys. Rev. B \textbf{109}, 104426 (2024), \href{https://arxiv.org/abs/2311.05174}{arXiv:2311.05174}, \href{https://doi.org/10.1103/PhysRevB.109.104426}{doi:10.1103/PhysRevB.109.104426}.
\bibitem{jenningsSutcliffe2013} P.~Jennings and P.~Sutcliffe, \emph{The dynamics of domain wall Skyrmions}, J. Phys. A \textbf{46}, 465401 (2013), \href{https://arxiv.org/abs/1305.2869}{arXiv:1305.2869}, \href{https://doi.org/10.1088/1751-8113/46/46/465401}{doi:10.1088/1751-8113/46/46/465401}.
\bibitem{torii} S. B. Gudnason, M. Nitta, \emph{Baryonic torii: Toroidal baryons in a generalized Skyrme model}, Phys. Rev. D \textbf{91}, 045027 (2015), \href{https://arxiv.org/abs/1410.8407}{arXiv:1410.8407}, \href{https://doi.org/10.1103/PhysRevD.91.045027}{doi:10.1103/PhysRevD.91.045027}.
\bibitem{beads} S. B. Gudnason, M. Nitta, \emph{Skyrmions confined as beads on a vortex ring}, Phys. Rev. D \textbf{94}, 025008 (2016), \href{https://arxiv.org/abs/1606.00336}{arXiv:1606.00336}, \href{https://doi.org/10.1103/PhysRevD.94.025008}{doi:10.1103/PhysRevD.94.025008}.
\bibitem{qiu} Z. Qiu, M. Nitta, \emph{Baryonic Vortex Phase and Magnetic Field Generation in QCD with Isospin and Baryon Chemical Potentials}, JHEP \textbf{06}, 139 (2024), \href{https://arxiv.org/abs/2403.07433}{arXiv:2403.07433}, \href{https://doi.org/10.1007/JHEP06(2024)139}{doi:10.1007/JHEP06(2024)139}.
\bibitem{nittaFractionalPCM2015} M.~Nitta, \emph{Fractional instantons and bions in the principal chiral model on $\mathbb R^2\times S^1$ with twisted boundary conditions}, JHEP \textbf{08}, 063 (2015), \href{https://arxiv.org/abs/1503.06336}{arXiv:1503.06336}, \href{https://doi.org/10.1007/JHEP08(2015)063}{doi:10.1007/JHEP08(2015)063}.
\bibitem{witten1985strings} E. Witten, \emph{Superconducting strings}, Nucl. Phys. B \textbf{249}, 557--592 (1985), \href{https://doi.org/10.1016/0550-3213(85)90022-7}{doi:10.1016/0550-3213(85)90022-7}.
\bibitem{jaykkaSpeight2010} J. J\"aykk\"a and M. Speight, \emph{Easy plane baby skyrmions}, Phys. Rev. D \textbf{82}, 125030 (2010), \href{https://arxiv.org/abs/1010.2217}{arXiv:1010.2217}, \href{https://doi.org/10.1103/PhysRevD.82.125030}{doi:10.1103/PhysRevD.82.125030}.
\bibitem{kobayashiNitta2013vortices} M. Kobayashi and M. Nitta, \emph{Fractional vortex molecules and vortex polygons in a baby Skyrme model}, Phys. Rev. D \textbf{87}, 125013 (2013), \href{https://arxiv.org/abs/1307.0242}{arXiv:1307.0242}, \href{https://doi.org/10.1103/PhysRevD.87.125013}{doi:10.1103/PhysRevD.87.125013}.
\bibitem{schroers1995} B. J. Schroers, \emph{Bogomol'nyi solitons in a gauged $O(3)$ sigma model}, Phys. Lett. B \textbf{356}, 291--296 (1995), \href{https://arxiv.org/abs/hep-th/9506004}{arXiv:hep-th/9506004}, \href{https://doi.org/10.1016/0370-2693(95)00833-7}{doi:10.1016/0370-2693(95)00833-7}.
\bibitem{schroers1996} B. J. Schroers, \emph{The Spectrum of Bogomol'nyi Solitons in Gauged Linear Sigma Models}, Nucl. Phys. B \textbf{475}, 440--468 (1996), \href{https://arxiv.org/abs/hep-th/9603101}{arXiv:hep-th/9603101}, \href{https://doi.org/10.1016/0550-3213(96)00348-3}{doi:10.1016/0550-3213(96)00348-3}.
\bibitem{vinci} M. Nitta and W. Vinci, \emph{Decomposing Instantons in Two Dimensions}, J. Phys. A 45, 175401 (2012), \href{https://arxiv.org/abs/1108.5742}{arXiv:1108.5742}, \href{https://doi.org/10.1088/1751-8113/45/17/175401}{doi:10.1088/1751-8113/45/17/175401}.
\bibitem{EHN2023composite} M.~Eto, Y.~Hamada and M.~Nitta, \emph{Composite topological solitons consisting of domain walls, strings, and monopoles in $O(N)$ models}, JHEP \textbf{08}, 150 (2023), \href{https://arxiv.org/abs/2304.14143}{arXiv:2304.14143}, \href{https://doi.org/10.1007/JHEP08(2023)150}{doi:10.1007/JHEP08(2023)150}.
\bibitem{Pearl1964} J. Pearl, \emph{Current distribution in superconducting films carrying quantized fluxoids}, Appl. Phys. Lett. \textbf{5}, 65--66 (1964), \href{https://doi.org/10.1063/1.1754056}{doi:10.1063/1.1754056}.
\bibitem{mamedaNittaQiu2026} K. Mameda, M. Nitta and Z. Qiu, \emph{Baryonic vortices in rotating nuclear matter}, JHEP~(to~appear), \href{https://arxiv.org/abs/2603.29325}{arXiv:2603.29325}.
\bibitem{rossDSG} C. Ross, N. Sakai and M. Nitta, \emph{Exact Ground States and Domain Walls in One Dimensional Chiral Magnets}, JHEP \textbf{12}, 163 (2021), \href{https://arxiv.org/abs/2012.08800}{arXiv:2012.08800}, \href{https://doi.org/10.1007/JHEP12(2021)163}{doi:10.1007/JHEP12(2021)163}.
\bibitem{amarishort} Y.~Amari, M.~Nitta, \emph{Skyrmion crystal phase on a magnetic domain wall in chiral magnets}, Phys. Rev. B \textbf{111}, 134441 (2025), \href{https://arxiv.org/abs/2409.07943}{arXiv:2409.07943}, \href{https://doi.org/10.1103/PhysRevB.111.134441}{doi:10.1103/PhysRevB.111.134441}.
\end{thebibliography}
\end{document}